\documentclass[11pt]{article}
\usepackage[margin=1in]{geometry}
\usepackage{amsmath,amssymb,amsthm,bm}
\usepackage[authoryear,round]{natbib}
\usepackage{booktabs,array,multirow}
\usepackage{enumitem}
\usepackage{setspace}
\usepackage{caption}
\usepackage{placeins}
\usepackage{graphicx}
\graphicspath{{./}{figures/}{source/figures/}}
\usepackage{xcolor}
\definecolor{citered}{RGB}{200,16,46}
\definecolor{refblue}{RGB}{0,51,204}
\definecolor{xdocblue}{RGB}{0,160,230}
\usepackage[colorlinks=true,linkcolor=refblue,citecolor=citered,urlcolor=refblue,filecolor=xdocblue]{hyperref}
\newtheorem{assumption}{Assumption}
\newtheorem{theorem}{Theorem}
\newtheorem{lemma}{Lemma}
\newtheorem{proposition}{Proposition}
\newtheorem{corollary}{Corollary}
\theoremstyle{definition}
\newtheorem{remark}{Remark}

\newcommand{\R}{\mathbb{R}}
\newcommand{\E}{\mathbb{E}}
\newcommand{\1}{\mathbf{1}}
\newcommand{\weak}{\Rightarrow}
\newcommand{\convp}{\xrightarrow{\;p\;}}
\newcommand{\convd}{\xrightarrow{\;d\;}}

\DeclareMathOperator*{\argmax}{arg\,max}
\newcommand{\Lt}{L_{J}}
\newcommand{\lt}{\ell}
\newcommand{\F}{\mathcal{F}}
\newcommand{\Wald}{\mathrm{Wald}}


\numberwithin{equation}{section}

\title{Threshold Regression with a Nearly Integrated Threshold Variable\thanks{The views expressed in this paper are those of the authors and do not necessarily reflect the views of the Magyar Nemzeti Bank.}}
\author{Chaoyi Chen\thanks{%
Magyar Nemzeti Bank (Central Bank of Hungary),  Budapest, 1054, Hungary; Email: \texttt{chenc@mnb.hu}.}\  \and Thanasis Stengos\thanks{Department of Economics and Finance, University of Guelph, Guelph, Ontario, Canada; Email: \texttt{tstengos@uoguelph.ca} } } 
\date{This version: October 2026}

\begin{document}
\maketitle

\begin{abstract}
\noindent
This paper considers threshold regression with an integrated or local-to-unity threshold variable and a threshold fixed in levels. We study the least squares threshold estimator and establish its asymptotic theory under both fixed and diminishing threshold effects. Under a fixed effect, the threshold estimator has a compound Cox argmin limit, whereas under a diminishing effect, its limiting distribution is characterized by a two-sided Brownian argmax with a random scale. We also develop likelihood ratio inference, which is asymptotically pivotal under a diminishing effect and conservative under Gaussian errors for a fixed effect, and show that conventional $t$-ratios for the slope coefficients remain asymptotically standard normal. A sup-Wald test is proposed to detect threshold effects that may be too small for consistent threshold localization. We further extend the framework to allow for integrated slope regressors and for contemporaneous endogeneity of the threshold variable, showing that the leading endogenous contribution cancels over shrinking threshold windows. Monte Carlo simulations examine the finite-sample performance of the proposed methods. An illustration using public debt and economic growth highlights the relevance of allowing for a persistent threshold variable.


\end{abstract}

\newpage
\section{Introduction}\label{sec:intro}

Threshold regression allows the coefficients of a relationship to change when an observed variable crosses an unknown level. Conventional threshold models typically assume that both the threshold variable and the slope regressors are stationary. Under a fixed threshold effect, \citet{Chan1993} shows that the least squares estimator of the threshold parameter is $n$-consistent and converges to the argmin of a compound Poisson process. Under a diminishing threshold effect, \citet{Hansen2000} derives a limiting distribution based on a two-sided Brownian criterion with drift and develops an asymptotically pivotal likelihood ratio statistic.\footnote{\citet{YuPhillips2018} link these two limits by showing that a suitably rescaled compound Poisson criterion approaches the Brownian criterion as the threshold effect becomes small.} More recent work allows for integrated slope regressors while maintaining a stationary threshold variable \citep{GonzaloPitarakis2006,Chen2015,ChenStengos2022}, and for endogenous slope regressors or threshold variables \citep{CanerHansen2004,KourtellosStengosTan2016,YuLiaoPhillips2024}.

The aforementioned studies maintain a stationary threshold variable, which can be restrictive in many applications. Policy interest rates, inflation, debt ratios, and financial valuation ratios are often highly persistent, and persistent predictors are commonly modelled as local-to-unity processes \citep{Phillips2015}. A growing literature considers parametric nonlinear models with nonstationary state variables. For null-recurrent threshold autoregressions, \citet{PhamChanTong1991} establish strong consistency of least squares. \citet{LiuLingShao2011} study a self-exciting threshold autoregression with a nonstationary process and a \textit{known} threshold. \citet{GaoTjostheimYin2013} establish asymptotic results for the coefficient estimators, discuss estimation of an \textit{unknown} threshold, and leave the distribution theory of the threshold estimator for future research. \citet{CaiGaoTjostheim2017} consider bivariate threshold cointegration with a nonstationary regime. In continuous time, threshold diffusions with a drift or volatility threshold are estimated from continuous or high-frequency records by \citet{Kutoyants2012}, \citet{SuChan2015} and \citet{LejayPigato2018}. Related questions have also been studied in nonstationary binary choice models \citep{ParkPhillips2000,HuPhillips2004,PhillipsJinHu2007} and in smooth-transition or latent-factor regime-switching models \citep{SaikkonenChoi2004,ChangChoiPark2017}. Recently, \citet{PengZhouSong2026} study a threshold cointegration model with an integrated threshold variable and develop tests for cointegration and threshold effects.

However, none of these studies establishes the asymptotic distribution of the least squares threshold estimator when the threshold variable is nonstationary. In this paper, we therefore fill this gap in the literature by considering the following model:
\[
y_t=\beta'x_t+\delta_n'x_t\1\{q_t\le\gamma_0\}+u_t,
\]
where $q_t$ is integrated or local to unity \citep{Phillips1987b,Phillips1988,ChanWei1987} and $\gamma_0$ is fixed in its original units. In the baseline model, we assume ${x_t}$, ${q_t}$ and ${u_t}$ are mutually independent sequences.\footnote{See Assumption~\ref{ass:x}(iii). Section~\ref{sec:endog} considers a specified form of contemporaneous error--innovation dependence.} Our contribution to the literature is threefold.

First, we establish the asymptotic properties of the least squares threshold estimator under both fixed and diminishing threshold effects. Unlike a stationary threshold variable, an integrated or local-to-unity $q_t$ makes only order $\sqrt{n}$ visits to a fixed neighbourhood of the threshold. These local visits determine the amount of information available for estimating $\gamma_0$ and, consequently, the rate and limiting distribution of the threshold estimator. Under a fixed threshold effect, we show that the least squares estimator of $\gamma_0$ is $\sqrt n$-consistent and converges to the argmin of a compound Cox process, that is, a compound Poisson process with \textit{random} intensity. The result is driven by the local time of the limiting diffusion and builds on the local-time-directed Poisson perspective of \citet{BrutscheRohde2026} and \citet{BrutscheHahnRohde2026}. For a volatility threshold in a null-recurrent diffusion observed at time intervals $n^{-a}$, $a\in[0,1)$, \citet{BrutscheHahnRohde2026} obtain the convergence rate $n^{(1+a)/2}$ and a limit driven by local time that does not depend on $a$. Therefore, our $\sqrt n$-consistent rate under the fixed threshold effect is not new in itself.\footnote{The low-frequency case of $a=0$ in \citet{BrutscheHahnRohde2026} gives the $\sqrt n$ rate that also arises here.} The differences lie in the regression setting: the threshold shifts the conditional mean of $y_t$, the slope coefficients are profiled out, and each visit to the threshold neighbourhood carries a random loss increment determined by the regressors and the regression error, so the limiting criterion has random marks and no drift between arrivals.\footnote{Technically, our proof uses Laplace functionals and exponential tilting, together with occupation-time methods from nonlinear cointegration \citep{ParkPhillips1999,ParkPhillips2001,WangPhillips2009a,WangPhillips2009b,Wang2015}. The infill experiment of \citet{BrutscheRohde2026} has rate $n$ on the event of positive local time at the true threshold. Their diffusion likelihood and our regression criterion require different proofs.}

We also derive the asymptotic distribution under a diminishing threshold effect, in the spirit of \citet{Hansen2000}. We show that, for a threshold effect $\delta_n=n^{-\alpha}\delta_0$, the least squares estimator of $\gamma_0$ converges at rate $n^{1/2-2\alpha}$ to a \textit{randomly} scaled argmax of a two-sided Brownian motion with linear drift.\footnote{In \citet{Hansen2000}, the scale factor is an \textit{unknown} but \textit{nonrandom} constant that depends on the threshold effect, the threshold density, and conditional moments evaluated at the true threshold.} Therefore, the intuition behind the transition from a fixed to a diminishing threshold effect is closely related to the connection between compound Poisson and Brownian limits established by \citet{YuPhillips2018}: as the threshold effect becomes small, many small loss increments transform the compound Cox criterion into a Brownian criterion with drift. In our setting, this argument applies conditionally on the diffusion path, so the local time remains in the limit as a random scale. The analysis also reveals a stricter consistency boundary for least squares estimation: the threshold effect must diminish more slowly than $n^{-1/4}$, rather than $n^{-1/2}$ as in the stationary case.\footnote{Section~\ref{rem:compare} illustrates the connection through two simple examples.}

Our second contribution concerns likelihood ratio inference for the threshold parameter. Under a diminishing threshold effect, local time cancels from the limiting likelihood ratio statistic. Thus, the critical values of \citet{Hansen2000} remain applicable despite the nonstationarity of the threshold variable. Under a fixed threshold effect and Gaussian errors, the same critical values yield asymptotically conservative confidence sets. The slope coefficients are estimated from whole regimes rather than from boundary visits, and conventional $t$-ratios for them remain asymptotically standard normal. We also distinguish between detecting a threshold effect and consistently locating the threshold. When the regressors, errors and threshold innovations are mutually independent, a quantile-trimmed sup-Wald test can detect effects over a wider range than that in which the least squares threshold estimator is consistent.

Third, we extend the model in two directions, allowing for independent integrated slope regressors and for contemporaneous dependence between the regression error and the innovation to the threshold variable. With integrated slope regressors, we derive the asymptotic properties of the least squares estimator of $\gamma_0$ under both fixed and diminishing threshold effects. The convergence rates remain the same after the appropriate normalization. However, the information governing the limiting distribution becomes path dependent and is determined by the values of the slope process when the threshold variable visits the threshold. With \textit{contemporaneous} threshold endogeneity of the form considered in the paper, we show that the least squares estimator of $\gamma_0$ remains consistent and, under a diminishing threshold effect, retains the same limiting distribution and likelihood ratio inference as in the exogenous case. The key insight is that aggregation over the shrinking windows relevant for threshold estimation removes the leading endogenous contribution. This intuition is related to the integrated-difference argument of \citet{YuPhillipsEndogeneity2018}, and also relates to the zero-energy functionals arising in structural nonparametric cointegrating regression \citep{WangPhillips2009b,WangPhillips2011}.

Empirically, we apply the model to the US debt and growth data of \citet{Hansen2017}. Persistence diagnostics provide strong evidence of persistence in the full sample and mixed evidence in the post-war sample. Comparing the threshold estimates across the two samples illustrates how persistence and the frequency of visits near the threshold can affect threshold estimation.

The rest of the paper is organised as follows. Section~\ref{sec:model} introduces the model and estimator. Section~\ref{sec:results} states the regularity assumptions and develops the large-sample theory with stationary slope regressors. Sections~\ref{sec:test}, \ref{sec:endog} and \ref{sec:nonstat} consider testing for a threshold effect, an endogenous threshold variable and nonstationary slope regressors, respectively. Section~\ref{sec:mc} reports the Monte Carlo evidence, Section~\ref{sec:app} presents the empirical application, and Section~\ref{sec:conclusion} concludes. Appendix~\ref{app:A} states the lemmas and Appendix~\ref{app:B} proves the main results; the Online Appendix proves the lemmas and illustrates the mechanism behind the endogeneity result.

Throughout the paper, $\1\{\cdot\}$ denotes the indicator function, $\|\cdot\|$ the Euclidean norm, $[a]$ the integer part of $a$, and $|A|$ the length of an interval $A$. The symbols $\convp$, $\convd$ and $\weak$ denote convergence in probability, convergence in distribution and weak convergence of stochastic processes, respectively. $\mathbb D[a,b]$ denotes the space of c\`adl\`ag functions on $[a,b]$, equipped with the Skorokhod $J_1$ topology. Point-process convergence is understood with respect to the vague topology.

\section{Model and estimator}\label{sec:model}

\subsection{The model}

We observe $\{(y_t,x_t,q_t)\}_{t=1}^n$ generated by the threshold regression
\begin{equation}\label{eq:model}
y_t=\beta'x_t+\delta_n'x_t\1\{q_t\le\gamma_0\}+u_t,\qquad t=1,\ldots,n,
\end{equation}
where $x_t\in\R^d$ is a vector of slope regressors that may contain a constant, $u_t$ is the regression error, $\gamma_0$ is the threshold and $\delta_n$ is the threshold effect. The threshold variable is nearly integrated,
\begin{equation}\label{eq:q}
q_t=\rho_nq_{t-1}+w_t,\qquad \rho_n=\exp(c/n),\qquad q_0=0,
\end{equation}
with a fixed localising coefficient $c\in\R$. This specification covers a unit root ($c=0$), local mean reversion ($c<0$) and local explosiveness ($c>0$). Following \citet{Hansen2000}, the threshold effect is either fixed or diminishing,
\begin{equation}\label{eq:delta}
\delta_n=n^{-\alpha}\delta_0,\qquad\delta_0\neq0,\qquad0\le\alpha<1/4,
\end{equation}
where $\alpha=0$ corresponds to a fixed threshold effect. The upper bound $1/4$ replaces the bound $1/2$ of the stationary theory.\footnote{Theorem~\ref{thm:cons}(b) shows that this bound is sharp for least squares.} If the effect diminishes faster, the observations near the threshold provide too little information to locate it consistently.

\begin{remark}
The threshold $\gamma_0$ is fixed in the original units of $q_t$ and does not vary with $n$, as when $\gamma_0$ represents an economically meaningful level such as an inflation rate or a debt ratio. Together with $q_0=0$, this implies that, after diffusion scaling, both the initial condition and the threshold correspond to the origin, where the limiting process $J_c$ in \eqref{eq:FCLT} accumulates positive local time $\lt$, defined in \eqref{eq:ellmu} below. Consequently, the number of visits to any fixed neighbourhood of $\gamma_0$ diverges in probability. This feature is important for consistent threshold estimation. If instead $\gamma_{0n}=\sqrt n\,g_0$ with $g_0\neq0$, or $q_0=\sqrt n\,\bar q$ with $\bar q\neq0$, the relevant local time is evaluated at a nonzero level that the limiting path need not visit. The local time is then zero with positive probability, and on those paths the threshold cannot be estimated consistently. A diffusion-scale initial condition is natural when $c<0$ and the process has run for a long time. In that case we expect the results below to hold on the event $\{\lt>0\}$, with $\lt$ the local time at zero of the shifted limit $\bar qe^{cs}+J_c(s)$; a formal statement is left for future work.
\end{remark}

\subsection{The least squares estimator}

Estimation proceeds by profiling out the slope coefficients and then searching over the threshold parameter. For a given candidate threshold $\gamma$, the model is linear in $\theta_n=(\beta',\delta_n')'$. Let
$X_t(\gamma)=(x_t',x_t'\1\{q_t\leq\gamma\})'$ and denote by $\hat\theta(\gamma)$ the least squares estimator from the regression of $y_t$ on $X_t(\gamma)$. Substituting $\hat\theta(\gamma)$ back into the objective function gives the concentrated sum of squared residuals
\begin{equation}\label{eq:Sn}
S_n(\gamma)=\sum_{t=1}^n\bigl(y_t-\hat\theta(\gamma)'X_t(\gamma)\bigr)^2 .
\end{equation}
The threshold estimator is then obtained by minimising $S_n(\gamma)$ over the admissible values of $\gamma$.

For later use, let $X$ and $X_\gamma$ be the $n\times d$ matrices with rows $x_t'$ and $x_t'\1\{q_t\leq\gamma\}$, respectively. Let $u$ and $y$ stack $u_t$ and $y_t$, and let $P_\gamma$ denote the orthogonal projection onto the column space of $[X_\gamma,X-X_\gamma]$. Then, $S_n(\gamma)=y'(I-P_\gamma)y$, which is well defined for every $\gamma$.

Let $\Gamma=[\underline\gamma,\bar\gamma]$ be a compact parameter space, fixed in the units of $q_t$. We restrict the threshold search to the feasible set
\begin{equation}\label{eq:feasible}
\hat\Gamma=
\bigl\{\gamma\in\Gamma:
\mathrm{rank}(X_\gamma)=\mathrm{rank}(X-X_\gamma)=d,\ 
n_{\min}\leq\#\{t:q_t\leq\gamma\}\leq n-n_{\min}
\bigr\},
\end{equation}
where $n_{\min}\geq d$ is a fixed minimum regime size. The rank conditions ensure that the coefficient estimates in both regimes are uniquely defined, while the minimum-size restriction rules out splits with too few observations in either regime. The least squares estimator of the threshold is the smallest minimiser of the concentrated criterion,
\begin{equation}\label{eq:gammahat}
\hat\gamma=
\min\Bigl\{
\gamma\in\hat\Gamma:
S_n(\gamma)=\min_{\gamma'\in\hat\Gamma}S_n(\gamma')
\Bigr\}.
\end{equation}

In practice, the minimisation in \eqref{eq:gammahat} is implemented by a grid search. Both the criterion and the feasibility conditions change only when $\gamma$ crosses an observed value of $q_t$, so $S_n(\gamma)$ is constant on each interval $[q_{(k)},q_{(k+1)})\cap\Gamma$, where $q_{(1)}\le\cdots\le q_{(n)}$ are the order statistics of $q_1,\ldots,q_n$. It is therefore sufficient to evaluate the criterion at the corresponding observed threshold values in $\Gamma$, together with the lower boundary when relevant. The smallest-minimiser rule makes the estimator unique and also resolves ties between distinct optimal splits, which may arise with discrete errors.

If $\hat\Gamma=\varnothing$, we set $\hat\gamma=\underline\gamma$. In either case, define $\hat\sigma^2=S_n(\hat\gamma)/n$. This convention has no asymptotic effect because $P(\hat\Gamma=\Gamma)\to1$. Both regimes have positive limiting proportions uniformly over the fixed parameter space, so the rank and minimum-regime-size restrictions affect finite-sample implementation but not the asymptotic theory.\footnote{The regime proportion $\mu$ is defined in \eqref{eq:ellmu}. See Lemma~\ref{lem:orders} for the full-rank and minimum-size argument. Section~\ref{sec:mc} examines their practical effect.} 

We maintain a threshold parameter space fixed in levels rather than using quantile trimming. For a nearly integrated threshold variable, the fraction of observations below any fixed level remains random in the limit. Consequently, quantile trimming would exclude the true threshold $\gamma_0$ with a probability that does not vanish. For a random walk, the limiting fraction has the arcsine distribution, so trimming 15\% of the observations at each end excludes $\gamma_0$ with probability $1-\frac2\pi\{\arcsin\sqrt{0.85}-\arcsin\sqrt{0.15}\}\approx0.51$, and trimming 5\% excludes it with probability about $0.29$.\footnote{A simulation with $n=2{,}000$ and 100{,}000 replications gives $0.506$ and $0.286$.} In practice, $\Gamma$ should be chosen on economic grounds as the range of levels at which a regime change is plausible, wide enough to contain $\gamma_0$ in its interior. Asymptotically, a wide $\Gamma$ costs nothing: by Theorem~\ref{thm:cons} the normalised criterion increases linearly in the distance from $\gamma_0$, so distant thresholds are strongly separated. In finite samples, the minimum regime size $n_{\min}$, rather than $\Gamma$, determines the feasible set, and parts of $\Gamma$ that the sample path rarely visits cannot be rejected. In the design of Section~\ref{sec:mc} with $c=0$ and $n=500$, doubling the width of $\Gamma$ changes median errors by less than 20\% and coverage by less than 0.01, but lengthens the confidence sets by about 30\%.\footnote{An asymptotic theory for a parameter space that expands with $n$ would require uniform occupation limits over growing ranges and is left for future work.} 

\section{Asymptotic theory with stationary slope regressors}\label{sec:results}

In this section, we establish the asymptotic theory for the least squares estimator of $\gamma_0$ in model~\eqref{eq:model} with stationary slope regressors. We first state the regularity assumptions, then establish consistency and the least squares consistency boundary, and finally derive the limiting distribution of the threshold estimator and likelihood ratio inference. We also provide two examples to relate these results to the familiar stationary threshold regression framework.

\subsection{Assumptions}

Let $\F_t=\sigma(\eta_s:s\le t)$ denote the filtration generated by the innovations of the threshold variable.

\begin{assumption}[Threshold variable]\label{ass:q}
$w_t=\sum_{j\ge0}\varphi_j\eta_{t-j}$ with $\varphi_0=1$, $\sum_{j\ge0}j|\varphi_j|<\infty$, and $\varphi:=\sum_{j\ge0}\varphi_j\neq0$. The innovations $\{\eta_t\}_{t\in\mathbb Z}$ are \textit{i.i.d.}\ with $\E\eta_t=0$, $\E\eta_t^2=\sigma_\eta^2>0$, and $\E|\eta_t|^p<\infty$ for some $p>4$. Their characteristic function, $\psi_\eta(\lambda)=\E e^{i\lambda\eta_t}$, satisfies $\int_{\R}(1+|\lambda|)|\psi_\eta(\lambda)|\,d\lambda<\infty$.
\end{assumption}

\begin{assumption}[Slope regressors and errors]\label{ass:x}\leavevmode
\begin{enumerate}[label=(\roman*),leftmargin=*]
\item $\{x_t\}$ is strictly stationary and strongly mixing with mixing coefficients $\alpha_x(m)\le C\varsigma^m$ for some $C<\infty$ and $\varsigma\in(0,1)$, $\E\|x_t\|^r<\infty$ for some $r>4$, and $D=\E(x_tx_t')$ is positive definite.
\item $\{u_t\}$ is \textit{i.i.d.}\ with $\E u_t=0$, $\E u_t^2=\sigma^2>0$, and $\E|u_t|^r<\infty$.
\item The sequences $\{x_t\}$, $\{u_t\}$, and $\{\eta_t\}$ are mutually independent.
\end{enumerate}
\end{assumption}

\begin{assumption}[Parameters]\label{ass:par}
$\gamma_0$ lies in the interior of $\Gamma$, $\beta$ and $\delta_0\neq0$ are fixed, and $\delta_n$ satisfies \eqref{eq:delta}.
\end{assumption}

Assumption~\ref{ass:q} follows the nonlinear cointegration framework of \citet{WangPhillips2009b}. The weighted integrability condition on the characteristic function provides the density regularity needed for the local limit theory. In particular, the innovation density is bounded and Lipschitz continuous, ruling out lattice-valued innovations. As the prediction horizon increases, the conditional density of $q_t$ becomes more diffuse and locally flat. These properties underlie the occupation-time approximation. Under Assumption~\ref{ass:q},
\begin{equation}\label{eq:FCLT}
X_n(s):=n^{-1/2}q_{[ns]}\weak J_c(s)=\omega\int_0^se^{c(s-r)}d\mathbb W(r),\qquad s\in[0,1],
\end{equation}
on $\mathbb D[0,1]$, where $\mathbb W$ is a standard Brownian motion and $\omega=\sigma_\eta|\varphi|$ \citep{Phillips1987b,PhillipsSolo1992}.

Assumption~\ref{ass:x}(i) allows for serial dependence in the slope regressors. The whole-path independence imposed in part (iii) allows us to condition on the threshold path, so that visits to a given window can be treated as a fixed set of observation indices. It also rules out dependence between the regression error and future threshold innovations, which a martingale difference condition alone would permit and which can alter the null distribution of the sup-Wald statistic (Section~\ref{sec:test}). Section~\ref{sec:endog} considers contemporaneous dependence between the regression error and the threshold innovation, while Remark~\ref{rem:het} allows for conditional heteroskedasticity. The nearly integrated threshold variable itself is not required to be mixing.

Whole-path independence is the binding restriction of the paper. It excludes the threshold variable itself and lagged dependent variables as slope regressors, as well as regressors correlated with the threshold innovations. The local averaging argument of Section~\ref{sec:endog} motivates investigating stationary regressors that depend on those innovations, but this would require additional dependence and exogeneity conditions and a joint occupation theory, and is not attempted here. Lagged dependent variables require particular care because they also inherit the regime indicator $\1\{q_{t-1}\le\gamma_0\}$.

Assumption~\ref{ass:par} places the true threshold away from the boundaries of the parameter space and ensures a nonzero regime contrast. The shrinking-effect restriction determines how much information about the threshold remains as the sample size increases.

\subsection{Asymptotic results}

Because $\gamma_0/\sqrt n\to0$, every threshold fixed in levels corresponds to the origin after diffusion scaling. This explains why boundary visits and regime proportions are described by the limiting path at zero. Let $\Lt(s,a)$ denote the chronological local time of the limiting process $J_c$ in \eqref{eq:FCLT}\footnote{Chronological local time is the occupation density at level $a$ up to time $s$, defined by $\Lt(s,a)=\lim_{\varepsilon\downarrow0}(2\varepsilon)^{-1}\int_0^s\mathbf{1}\{|J_c(r)-a|<\varepsilon\}\,dr$ almost surely. The integral measures the time spent in $(a-\varepsilon,a+\varepsilon)$, which is divided by the band width. Local time can be positive even though the diffusion spends zero time at the exact level $a$. See \citet{ParkPhillips2001}.}, and define
\begin{equation}\label{eq:ellmu}
\lt:=\Lt(1,0),\qquad
\mu:=\int_0^1\1\{J_c(s)\le0\}\,ds.
\end{equation}
Local time satisfies the occupation formula $\int_0^1g(J_c(s))\,ds=\int_{\R}g(a)\Lt(1,a)\,da$ \citep{ParkPhillips2001}. The corresponding sample occupation limit implies that the number of observations in a fixed bounded interval $A$ is approximately $\sqrt n\,\lt\,|A|$, compared with order $n$ under stationary sampling when $A$ has positive probability. Since $J_c$ starts at zero and has a nonzero diffusion coefficient, $\lt>0$ and $0<\mu<1$ almost surely. Every fixed threshold has the same limiting regime proportion $\mu$. For $c=0$, $\mu$ has the arcsine distribution and $\lt$ is distributed as $|N(0,1)|/\omega$.

The limits are joint with the threshold path.\footnote{Local time also enters the threshold limits of \citet{BrutscheRohde2026,BrutscheHahnRohde2026}. Here, we adapt the occupation-time arguments of \citet{ParkPhillips2001} to establish weighted occupation and score limits uniformly over moving and shrinking indicator windows. See Lemma~\ref{lem:local}(b)--(d).} As in \citet{ParkPhillips2001}, statements such as $Y_n\convp\lt Y$ are made on a Skorokhod representation of \eqref{eq:FCLT} carrying the independent regressor and error sequences. They imply joint convergence in distribution in the original experiment. Let $\kappa_0=\delta_0'D\delta_0>0$.

\begin{theorem}[Consistency]\label{thm:cons}
Let Assumptions~\ref{ass:q}--\ref{ass:par} hold.
\begin{enumerate}[label=(\alph*),leftmargin=*]
\item If $0\le\alpha<1/4$, then, uniformly in $\gamma\in\Gamma$,
\begin{equation}\label{eq:globalSn}
n^{2\alpha-1/2}\big[S_n(\gamma)-S_n(\gamma_0)\big]\convp\lt\,\kappa_0\,|\gamma-\gamma_0|.
\end{equation}
Consequently $\hat\gamma\convp\gamma_0$ and $\hat\sigma^2\convp\sigma^2$.
\item If instead \eqref{eq:delta} holds with $1/4\le\alpha<1/2$ in place of $0\le\alpha<1/4$, then
\[
n^{\alpha-1/4}\big[S_n(\gamma)-S_n(\gamma_0)\big]\weak\1\{\alpha=1/4\}\,\lt\,\kappa_0|\gamma-\gamma_0|-2\sigma(\lt\,\kappa_0)^{1/2}\mathcal B(\gamma-\gamma_0)
\]
on $\Gamma$, where $\mathcal B$ is a two-sided standard Brownian motion independent of $J_c$. The estimator $\hat\gamma-\gamma_0$ converges in distribution to the minimiser of the right-hand side over $\Gamma-\gamma_0$, which differs from zero almost surely.
\end{enumerate}
\end{theorem}

Theorem~\ref{thm:cons} characterises the global least squares criterion and establishes the consistency boundary. When $\alpha<1/4$, the normalised criterion converges uniformly to $\ell\kappa_0|\gamma-\gamma_0|$, so the objective increases linearly away from $\gamma_0$ with a random slope proportional to local time. Since $\ell>0$ almost surely, $\hat\gamma$ is consistent. The boundary follows from a simple signal--noise comparison: a fixed displacement of the threshold misclassifies order $\sqrt n$ observations, generating a signal of order $n^{1/2-2\alpha}$ and noise of order $n^{1/4-\alpha}$. The signal therefore dominates only when $\alpha<1/4$. With a stationary threshold variable, the corresponding neighbourhood contains order $n$ observations, yielding the familiar boundary $\alpha<1/2$. This boundary is specific to least squares under the stated assumptions; an impossibility result covering all estimators would require additional restrictions.\footnote{For example, with $x_t=1$, $0<\delta_n<1$, and $u_t\in\{-1,1\}$, the four possible outcomes identify the regime of each observation once all four appear. The nearest observations in the two regimes then bracket the threshold even when $\alpha>1/4$. An impossibility result would therefore require additional restrictions on the error distribution.}

To state the distribution results, we introduce two limiting processes. First, let $N_1$ and $N_2$ be conditionally independent Poisson processes on $[0,\infty)$ with intensity $\lt$, given $J_c$. Let $\{(x_{1i},u_{1i})\}_{i\ge1}$ and $\{(x_{2i},u_{2i})\}_{i\ge1}$ be \textit{i.i.d.}\ sequences, independent of $(N_1,N_2,J_c)$, with distribution $P_x\otimes P_u$, where $P_x$ and $P_u$ denote the distributions of $x_t$ and $u_t$. Define
\begin{equation}\label{eq:Dcal}
\mathcal D(v)=
\begin{cases}
\displaystyle\sum_{i=1}^{N_1(v)} z_{1i}, & v>0,\\[1mm]
0, & v=0,\\[1mm]
\displaystyle\sum_{i=1}^{N_2(-v)} z_{2i}, & v<0,
\end{cases}
\qquad
\begin{aligned}
z_{1i}&=(\delta_0'x_{1i})^2-2\delta_0'x_{1i}u_{1i},\\
z_{2i}&=(\delta_0'x_{2i})^2+2\delta_0'x_{2i}u_{2i},
\end{aligned}
\end{equation}
where $N_1(r)$ counts points in $(0,r]$ and $N_2(r)$ in $(0,r)$, so that $\mathcal D$ is right-continuous and matches the assignment rule $q_t\leq\gamma$. Conditional on $J_c$, $\mathcal D$ is a two-sided compound Poisson process with intensity $\lt$. Unconditionally, it is a compound Cox process because $\lt$ is random. Let $\mathcal V$ denote the left endpoint of the interval on which $\mathcal D$ attains its minimum. When $u_t$ has a continuous distribution and $P(\delta_0'x_t=0)=0$, the levels of $\mathcal D$ on its intervals of constancy are almost surely distinct, so this minimizing interval is unique and its left endpoint corresponds to the selection rule in \eqref{eq:gammahat}. 

Next, let $\mathcal B_1$ and $\mathcal B_2$ be independent standard Brownian motions on $[0,\infty)$, independent of $J_c$, and define
\begin{equation}\label{eq:Tcal}
T(r)=
\begin{cases}
\mathcal B_1(r)-\frac12|r|, & r\geq0,\\
\mathcal B_2(-r)-\frac12|r|, & r<0,
\end{cases}
\qquad
\mathcal T=\argmax_{r\in\R}T(r),
\end{equation}
where the distribution of $\mathcal T$ is given by \citet{Bai1997} and \citet{Hansen2000}.

\begin{theorem}[Asymptotic distribution of $\hat\gamma$]\label{thm:dist}
Let Assumptions~\ref{ass:q}--\ref{ass:par} hold and let $a_n=n^{1/2-2\alpha}$.
\begin{enumerate}[label=(\alph*),leftmargin=*]
\item Fixed threshold effect. If $\alpha=0$, then $a_n(\hat\gamma-\gamma_0)=\sqrt n(\hat\gamma-\gamma_0)=O_p(1)$. If, in addition, $u_t$ has a continuous distribution and $P(\delta_0'x_t=0)=0$, then $\sqrt n(\hat\gamma-\gamma_0)\convd\mathcal V$.
\item Diminishing threshold effect. If $0<\alpha<1/4$, then
\begin{equation}\label{eq:shrink}
a_n(\hat\gamma-\gamma_0)\convd\lambda_\ell\,\mathcal T,\qquad \lambda_\ell=\frac{\sigma^2}{\lt\,\kappa_0},
\end{equation}
where $\mathcal T$ is independent of $\lt$.
\end{enumerate}
\end{theorem}

Theorem~\ref{thm:dist} establishes the limiting distribution of the least-squares estimator of $\gamma_0$ under fixed and diminishing threshold effects.  Part~(a) has the sparse-arrival mechanism of \citet{BrutscheHahnRohde2026}: a window of width $n^{-1/2}$ receives finitely many visits in the limit, with a Poisson arrival rate determined by local time. We adapt that mechanism by attaching the regression-loss increment to each visit, producing the compound Cox criterion in \eqref{eq:Dcal}. Its sample paths are constant between arrivals, while its expected loss rises with distance from the threshold.\footnote{Specifically, $\E\{\mathcal D(v)\mid J_c\}=\lt\kappa_0|v|$.}

Part~(b) is the nearly integrated counterpart of \citet{Hansen2000}. The wider window of width $a_n^{-1}$ contains many small contributions whose sum approaches a Brownian criterion. Paths with fewer visits have a larger scale $\lambda_\ell$ and therefore less precise threshold estimates. The random scale also has heavy tails: for $c=0$, $\lt$ is distributed as $|N(0,1)|/\omega$, so $1/\lt$ has no finite mean, and neither do the limits $\mathcal V$ and $\lambda_\ell\mathcal T$, which scale with $1/\lt$. Medians and other quantiles are therefore the appropriate summaries of the estimation error, and Section~\ref{sec:mc} reports them.

The slope coefficients are not affected by the slow localisation of the threshold. Let $\hat\beta_1$ and $\hat\beta_2$ be the least squares estimators of $\beta_1=\beta+\delta_n$ and $\beta_2=\beta$ from the observations with $q_t\le\hat\gamma$ and $q_t>\hat\gamma$, respectively, so that $\hat\theta(\hat\gamma)=(\hat\beta_2',\hat\beta_1'-\hat\beta_2')'$, and let $\hat H_1=\sum_tx_tx_t'\1\{q_t\le\hat\gamma\}$ and $\hat H_2=\sum_tx_tx_t'\1\{q_t>\hat\gamma\}$.

\begin{corollary}[Slope coefficients]\label{cor:slope}
Let Assumptions~\ref{ass:q}--\ref{ass:par} hold with $0\le\alpha<1/4$. Then, jointly with \eqref{eq:FCLT},
\[
\sqrt n(\hat\beta_1-\beta_1)\convd\mathrm{MN}\big(0,\sigma^2D^{-1}/\mu\big),\qquad
\sqrt n(\hat\beta_2-\beta_2)\convd\mathrm{MN}\big(0,\sigma^2D^{-1}/(1-\mu)\big),
\]
where $\mathrm{MN}(0,\Sigma)$ denotes the mixed normal distribution whose conditional distribution given $J_c$ is $N(0,\Sigma)$, the two limits are independent conditionally on $J_c$, and
\[
\hat\sigma^{-1}\big(\hat H_1^{1/2}(\hat\beta_1-\beta_1),\ \hat H_2^{1/2}(\hat\beta_2-\beta_2)\big)\convd N(0,I_{2d}).
\]
\end{corollary}

Corollary~\ref{cor:slope} establishes mixed normal limiting distributions for the least squares slope estimators.\footnote{Replacing $\gamma_0$ by $\hat\gamma$ misclassifies $O_p(n^{2\alpha})$ observations and changes the standardised slope estimators by $O_p(n^{\alpha-1/2})=o_p(1)$.} In the stationary framework of \citet{Hansen2000}, regime proportions converge to nonrandom constants, yielding normal limits for the slope estimators. With a nearly integrated threshold variable, the limiting regime proportion $\mu$ remains random, so the asymptotic precision varies with the threshold path. Conventional standard errors capture this variation through the estimated regime moment matrices. Studentisation therefore yields standard normal and chi-square limits for $t$-ratios and Wald statistics, respectively.

\subsection{Heuristic comparison: stationary and nearly integrated thresholds}\label{rem:compare}

In this subsection, we use two simple examples to relate Theorem~\ref{thm:dist} to the familiar asymptotic theory for stationary threshold regression. The first considers an error-free step model and isolates the arrival mechanism around the threshold. The second adds a regression error and allows the threshold effect to shrink, which connects the compound Cox limit in Theorem~\ref{thm:dist}(a) to the Brownian limit in Theorem~\ref{thm:dist}(b), borrowing the argument of \citet{YuPhillips2018}.

\emph{Example 1: an error-free step.} Consider
\begin{equation}\label{eq:step_example}
y_t=\1\{q_t\leq\gamma_0\},\qquad t=1,\ldots,n.
\end{equation}
With probability tending to one, the smallest least squares minimiser is
$\hat\gamma=\max\{q_t:q_t\leq\gamma_0\}$, so estimation of $\gamma_0$ reduces to the spacing between the threshold and the nearest observation below it.

Suppose first that $q_t$ is \textit{i.i.d.}\ with distribution function $F$ and a density $f$ that is continuous and positive at $\gamma_0$. The event $\{n(\gamma_0-\hat\gamma)>v\}$ occurs when no observation falls in the window $(\gamma_0-v/n,\gamma_0]$. Each observation falls there independently with the same probability $F(\gamma_0)-F(\gamma_0-v/n)\approx f(\gamma_0)v/n$, so
\begin{equation*}
P\{n(\gamma_0-\hat\gamma)>v\}\approx\Bigl\{1-\frac{f(\gamma_0)v}{n}\Bigr\}^n\longrightarrow
e^{-f(\gamma_0)v},\qquad v>0.
\end{equation*}
Thus, the estimator is $n$-consistent, and local arrivals follow a Poisson process with fixed intensity $f(\gamma_0)$. This arrival mechanism underlies the compound Poisson limit for the least squares estimator in \citet{Chan1993}.

Next, with a nearly integrated $q_t$, a fixed neighbourhood of $\gamma_0$ contains only order $\sqrt n$ observations \citep{BrutscheHahnRohde2026}, so the nearest observation below $\gamma_0$ is at a distance of order $n^{-1/2}$. The event $\{\sqrt n(\gamma_0-\hat\gamma)>v\}$ now occurs when no observation falls in the window $(\gamma_0-v/\sqrt n,\gamma_0]$, but the probability of falling there differs across observations. Let $q_t=z_t+\eta_t$, where $z_t=\rho_nq_{t-1}+\sum_{j\ge1}\varphi_j\eta_{t-j}$ is known at time $t-1$, and let $f_\eta$ denote the density of $\eta_t$. Given the past, $q_t$ has density $f_\eta(\cdot-z_t)$. Therefore, observation $t$ falls in the window with probability
\begin{equation*}
P\bigl(\gamma_0-v/\sqrt n<q_t\le\gamma_0\bigm|\F_{t-1}\bigr)
=\int_{\gamma_0-v/\sqrt n}^{\gamma_0}f_\eta(y-z_t)\,dy
\approx f_\eta(\gamma_0-z_t)\,\frac{v}{\sqrt n}.
\end{equation*}
In contrast with the fixed probability $f(\gamma_0)v/n$ of the stationary case, this probability is random and varies over time with the position of the path. It is negligible when the path is far from $\gamma_0$, which is most of the time, and of order $n^{-1/2}$ when the path is within $O(1)$ of $\gamma_0$, which happens in only order $\sqrt n$ periods. The probabilities therefore add up to a random quantity of order one. Since $f_\eta$ integrates to one, the occupation-time argument behind the count approximation $\sqrt n\,\lt\,|A|$ above gives
\begin{equation*}
\sum_{t=1}^nP\bigl(\gamma_0-v/\sqrt n<q_t\le\gamma_0\bigm|\F_{t-1}\bigr)
\approx\frac{v}{\sqrt n}\sum_{t=1}^nf_\eta(\gamma_0-z_t)
\convp v\lt .
\end{equation*}
The local time is thus the accumulated arrival probability per unit of $v$, and it plays the role of $f(\gamma_0)$. Because each probability is small, the Poisson approximation for rare events applies path by path: conditionally on $J_c$, the number of observations in the window is asymptotically Poisson with mean $v\lt$, and the window is empty with probability $e^{-\lt v}$. Hence
\begin{equation*}
\bigl(X_n,\sqrt n(\gamma_0-\hat\gamma)\bigr)
\weak (J_c,\mathcal E),
\qquad
P(\mathcal E>v\mid J_c)=e^{-\lt v},\quad v>0.
\end{equation*}

The convergence rate thus falls from $n$ to $\sqrt n$, while the fixed density $f(\gamma_0)$ in the stationary case is replaced by the random local time $\lt$, which illustrates the compound Cox result in Theorem~\ref{thm:dist}(a). A Poisson process whose intensity is itself random is a Cox, or doubly stochastic Poisson, process. Given the threshold path, arrivals are Poisson, but their intensity varies across paths and, in calendar time, is concentrated at boundary visits.\footnote{Lemma~\ref{lem:cox} establishes the joint limit in calendar time, location and marks; the arrival probabilities up to date $[nr]$ converge to $v\Lt(r,0)$.} The error is negative because the smallest minimiser in \eqref{eq:gammahat} is the left end-point of the minimising interval. Here the nearest observation below $\gamma_0$. The mid-point of that interval, proposed by \citet{Yu2012} for stationary threshold regression, has error $(\mathcal E_1-\mathcal E_2)/2$ with $\mathcal E_1,\mathcal E_2$ conditionally independent exponentials of rate $\lt$, which halves the median absolute error. The same convention could be used here.

\emph{Example 2: a diminishing threshold effect with regression error.} Now consider
\begin{equation}\label{eq:step_error_example}
y_t=\delta_n\1\{q_t\leq\gamma_0\}+u_t,
\qquad
\delta_n=n^{-\alpha}\delta_0,\quad \delta_0\neq0,
\end{equation}
where $\E u_t=0$ and $\E u_t^2=\sigma^2$. For a  \textit{i.i.d.}\ threshold variable with density $f(\gamma_0)>0$, a fixed effect gives the familiar compound Poisson criterion. \citet{YuPhillips2018} show that, as the effect diminishes, the suitably rescaled compound Poisson criterion converges to a two-sided Brownian criterion with linear drift. In the present scalar example,
\begin{equation*}
n^{1-2\alpha}(\hat\gamma-\gamma_0)
\convd
\frac{\sigma^2}{f(\gamma_0)\delta_0^2}\,\mathcal T,
\end{equation*}
provided $0<\alpha<1/2$. This gives the usual stationary threshold rate and consistency boundary.

The same argument explains the nearly integrated result. Conditional on $J_c$, the fixed-effect criterion $\mathcal D$ is compound Poisson with intensity $\lt$, so the small-effect argument of \citet{YuPhillips2018} applies path by path. As $\delta_n$ shrinks, the many small loss increments generate the Brownian approximation
\begin{equation*}
\lt\delta_0^2|v|
-2\sigma(\lt\delta_0^2)^{1/2}\mathcal B(v),
\end{equation*}
where $\mathcal B$ is a two-sided standard Brownian motion. Its minimiser has distribution
$\lambda_\ell\mathcal T$, with $\lambda_\ell=\sigma^2/(\lt\delta_0^2)$. Hence
\begin{equation*}
n^{1/2-2\alpha}(\hat\gamma-\gamma_0)
\convd
\lambda_\ell\mathcal T,
\end{equation*}
which is the scalar version of Theorem~\ref{thm:dist}(b). This argument is heuristic as the theorem establishes the joint limit directly rather than by first taking the fixed-effect limit and then shrinking the threshold effect.

The two examples make the comparison with stationary threshold regression clear. The effective number of observations per unit interval around $\gamma_0$ is $nf(\gamma_0)$ in the stationary case but $\sqrt n\,\lt$ when the threshold variable is nearly integrated. The localisation rate therefore falls from $n^{1-2\alpha}$ to $n^{1/2-2\alpha}$, and the consistency boundary changes from $\alpha<1/2$ to $\alpha<1/4$. The Brownian argmax $\mathcal T$ remains the same, while nonstationarity enters through the random local-time scale $\lambda_\ell$. This also anticipates why local time affects the precision of threshold estimation but drops out of the likelihood ratio limit below.

\subsection{Likelihood ratio inference}

The limit in Theorem~\ref{thm:dist}(a) depends on the distribution of the jumps, and the scale of the limit in part (b) is random and depends on $c$. Following \citet{Hansen2000}, we construct confidence sets for $\gamma_0$ by inverting the likelihood ratio statistic
\[
LR_n(\gamma)=n\,\frac{S_n(\gamma)-S_n(\hat\gamma)}{S_n(\hat\gamma)} .
\]
In practice, if the denominator is zero, define $LR_n(\gamma)=0$ when the numerator is nonpositive and $LR_n(\gamma)=+\infty$ otherwise. This convention covers finite-sample events permitted by discrete errors and does not affect the limits because $\hat\sigma^2\convp\sigma^2>0$.

Let $\xi=\sup_{r\in\R}\{2\mathcal B(r)-|r|\}$, where $\mathcal B$ is a two-sided standard Brownian motion. Then $P(\xi\le z)=(1-e^{-z/2})^2$ for $z\ge0$ \citep{Hansen2000}. For $p\in(0,1)$ let $c_p=-2\log(1-\sqrt p)$, so that $c_{0.95}\approx7.35$, and $C_n(p)=\{\gamma\in\Gamma:LR_n(\gamma)\le c_p\}$.

\begin{theorem}[Likelihood ratio]\label{thm:LR}
Let Assumptions~\ref{ass:q}--\ref{ass:par} hold.
\begin{enumerate}[label=(\alph*),leftmargin=*]
\item \emph{Diminishing threshold effect.} If $0<\alpha<1/4$, then $LR_n(\gamma_0)\convd\xi$, and hence $P\{\gamma_0\in C_n(p)\}\to p$.

\item \emph{Fixed threshold effect with Gaussian errors.} If $\alpha=0$ and $u_t\sim N(0,\sigma^2)$, then, for every $z>0$,
\[
\limsup_{n\to\infty}P\{LR_n(\gamma_0)>z\}
\leq 1-(1-e^{-z/2})^2,
\]
and hence $\liminf_{n\to\infty}P\{\gamma_0\in C_n(p)\}\geq p$.

\item \emph{Fixed threshold effect.} If $\alpha=0$, then
\[
LR_n(\gamma_0)\convd-\sigma^{-2}\min_{v\in\R}\mathcal D(v).
\]
\end{enumerate}
\end{theorem}

Theorem~\ref{ass:par} characterizes the asymptotic behaviour of the likelihood ratio statistic under diminishing and fixed threshold effects. Part (a) permits likelihood ratio inference without estimating $c$, $\alpha$, or local time. The same path-dependent intensity enters the drift and variance of the limiting criterion, so it rescales the minimizer but leaves the distribution of its minimum unchanged. Part (b) gives the fixed-effect counterpart of \citet{Hansen2000} using an exponential martingale bound. Part (c) gives the general fixed-effect limit, which depends on the jump distribution of the compound Cox criterion but not on $\lt$ or $c$: rescaling time by $\lt$ leaves the law of the minimum of a compound Poisson process unchanged. Unlike the argmin in Theorem~\ref{thm:dist}(a), the minimum value is a continuous functional of the jump configuration even when ties occur, so part~(c) requires neither a continuous error distribution nor $P(\delta_0'x_t=0)=0$. The small-effect approximation of the resulting limit approaches $\xi$, as suggested by Section~\ref{rem:compare}.\footnote{\label{rem:nongauss}For example, let $x_t=1$, $\delta_0=A$, and $u_t\sim0.90N(0,s^2)+0.05N(A,s^2)+0.05N(-A,s^2)$. As $s/A\to0$, a negative first jump of approximately $-A^2$ occurs on at least one side with probability tending to $1-0.95^2$, implying a likelihood ratio of at least approximately $A^2/(0.1A^2+s^2)\to10$. The nominal 95\% critical value is only 7.35, so the asymptotic coverage of $C_n(0.95)$ is at most $0.95^2\approx0.90$. Table~\ref{tab:mc3} illustrates this case.}

For a diminishing effect, invert the statistic using $c_p$. For a fixed effect with Gaussian errors, the same rule is asymptotically conservative. Because the criterion is a step function, the set is a union of accepted intervals $[q_{(k)},q_{(k+1)})\cap\Gamma$, possibly disconnected. Retain each interval in full. Keeping only accepted order statistics omits part of the final interval and can reduce coverage.

In computation, distinguish the full-domain set $C_n(p)$ from
\[
C_n^F(p)=\{\gamma\in\hat\Gamma:LR_n(\gamma)\le c_p\},\qquad
C_n^U(p)=(\Gamma\setminus\hat\Gamma)\cup C_n^F(p).
\]
Set $C_n^U(p)=\Gamma$ when $\hat\Gamma$ is empty. $C_n^F(p)$ reports inversion over feasible splits. $C_n^U(p)$ retains infeasible candidates rather than rejecting them. All three constructions agree on the event $\hat\Gamma=\Gamma$, whose probability tends to one.\footnote{Section~\ref{sec:mc} distinguishes their finite-sample summaries.}

The convex hull has at least the coverage of the original set when the same critical value is used. For a diminishing effect, a hull calibrated directly at
\[
c_p^{\mathrm{hull}}=-2\log(1-p)
\]
has asymptotic coverage $p$, as in \citet{Yu2026}. For example, $c_{0.95}^{\mathrm{hull}}\approx5.99$. This differs from the value $c_p\approx7.35$ at $p=0.95$ used for the original union set. The proof of Theorem~\ref{thm:LR} establishes this calibration using the two one-sided criterion minima. Under the conditional heteroskedasticity of Remark~\ref{rem:het} below, use $\hat\vartheta^2c_p^{\mathrm{hull}}$. For a fixed effect, Theorem~\ref{thm:LR}(b) continues to justify the conservative value $c_p$ for either the union set or its hull under Gaussian errors.

\begin{remark}[Conditional heteroskedasticity]\label{rem:het}
Suppose that $u_t=\sigma(x_t)\epsilon_t$, where $\{\epsilon_t\}$ is \textit{i.i.d.}\ with mean zero, unit variance and $\E|\epsilon_t|^r<\infty$, the sequences $\{\epsilon_t\}$, $\{x_t\}$ and $\{\eta_t\}$ are mutually independent, $\E[\sigma(x_t)^r(1+\|x_t\|)^r]<\infty$ for some $r>4$, and $\E\sigma(x_t)^2>0$. Let $\Omega=\E[x_tx_t'\sigma(x_t)^2]$ and assume $\delta_0'\Omega\delta_0>0$. Then Theorems~\ref{thm:cons}(a) and \ref{thm:dist}(b) hold with $\sigma^2=\E\sigma(x_t)^2$ as the limit of $\hat\sigma^2$ and $\lambda_\ell=\delta_0'\Omega\delta_0/(\lt\kappa_0^2)$, and $LR_n(\gamma_0)\convd\vartheta^2\xi$ with $\vartheta^2=\delta_0'\Omega\delta_0/(\sigma^2\kappa_0)$. In the stationary case $\vartheta^2$ involves moments conditional on $q_t=\gamma_0$, which must be estimated nonparametrically \citep{Hansen2000}. 

Here the variance ratio involves unconditional moments and is estimated by $\hat\vartheta^2=\sum_t(\hat\delta'x_t)^2\hat u_t^2/\{\hat\sigma^2\sum_t(\hat\delta'x_t)^2\}$, where $\hat\delta=\hat\beta_1-\hat\beta_2$ and $\hat u_t$ are the residuals from the regression of $y_t$ on $X_t(\hat\gamma)$. For the diminishing-effect confidence set, use the critical value $\hat\vartheta^2c_p$.
\end{remark}

\section{Testing for a threshold effect}\label{sec:test}

To test for the nonlinearity, we consider $H_0:\delta_n=0$. Let $S_0$ be the sum of squared residuals from the linear regression of $y_t$ on $x_t$, and define $\Wald_n(\gamma)=n[S_0-S_n(\gamma)]/S_n(\gamma)$. On a zero-denominator event, set $\Wald_n(\gamma)=0$ for zero improvement and $+\infty$ for positive improvement. Under the null the denominator divided by $n$ converges to $\sigma^2>0$. The threshold is not identified under the null. Following \citet{Andrews1993} and \citet{Hansen1996}, we therefore take the supremum over a trimmed set. For this test, trimming is based on sample quantiles,
\[
\Gamma_n^\pi=\Big\{\gamma\in\R:\ \pi\le\frac1n\sum_{t=1}^n\1\{q_t\le\gamma\}\le1-\pi\Big\},\qquad0<\pi<1/2 .
\]

\begin{proposition}[Sup-Wald test]\label{prop:test}
Let Assumptions~\ref{ass:q} and \ref{ass:x} hold.
\begin{enumerate}[label=(\alph*),leftmargin=*]
\item Under $H_0$, $\sup_{\gamma\in\Gamma_n^\pi}\Wald_n(\gamma)\convd\sup_{\pi\le r\le1-\pi}BB_d(r)'BB_d(r)/\{r(1-r)\}$, where $BB_d$ is a $d$-dimensional standard Brownian bridge.
\item If $\delta_n=n^{-\alpha}\delta_0$ with $0\le\alpha<1/2$ and $\gamma_0$ is fixed, then $\sup_{\gamma\in\Gamma_n^\pi}\Wald_n(\gamma)\convp\infty$.
\end{enumerate}
\end{proposition}

Proposition~\ref{prop:test} extends the sup-Wald approach of \citet{Andrews1993} and \citet{Hansen1996} to a nearly integrated threshold variable. Under whole-path independence, the null limit is the familiar nuisance-free Brownian-bridge distribution tabulated by \citet{Andrews1993}, and in particular does not depend on the local-to-unity parameter $c$.\footnote{Conditioning on the threshold path and ordering observations by $q_t$ turns the regime fraction into the index $r$. Independence then yields the usual Brownian partial-score limit, while estimating the common slope produces the Brownian bridge. Quantile trimming restricts $r$ to $[\pi,1-\pi]$. If the supremum were instead taken over the fixed parameter space $\Gamma$, every fixed threshold would have the same limiting regime proportion $\mu$, and the Wald process would be asymptotically constant across $\Gamma$, yielding a $\chi_d^2$ rather than the Andrews sup-Wald limit on $\{\pi<\mu<1-\pi\}$.}

This result is also closely related to Proposition~1 of \citet{GonzaloPitarakis2012}, who obtain the same Brownian-bridge form in a predictive threshold model with a nearly integrated slope regressor and a stationary threshold variable. \citet{Chen2015} and Corollary~3 of \citet{ChenStengos2022} consider related tests with integrated or more general nonstationary slope regressors; their null limits generally retain features of the nonstationary slope process. Our setting reverses the source of nonstationarity: the slope regressors are stationary, while the threshold variable is nearly integrated. Despite this difference, whole-path independence and quantile trimming recover the standard nuisance-free limit. This also contrasts with the more general stationary framework of \citet{Hansen1996}, where the null distribution can depend on unknown population moments and bootstrap inference is typically required. \citet{Hansen2000}, by contrast, focuses on estimation and likelihood ratio inference for the threshold location once a threshold effect is present.

Part~(b) highlights the distinction between detecting a threshold effect and locating its threshold. The sup-Wald test uses information from the two regimes and is consistent for $\alpha<1/2$, whereas consistent estimation of $\gamma_0$ relies on the much smaller number of observations near the threshold and requires $\alpha<1/4$. Thus, when $1/4\leq\alpha<1/2$, a threshold effect can be detected even though its location cannot be consistently estimated.\footnote{See Theorem~\ref{thm:cons}(b).}

Whole-path independence in Assumption~\ref{ass:x}(iii) is important for the null result. A martingale difference condition alone may still allow the regression error to depend on future innovations of the threshold variable, in which case the score and threshold path can share the same limiting Brownian motion and the Brownian-bridge argument fails.\footnote{For example, with $x_t=1$, $q_t=q_{t-1}+\eta_t$, and $u_t=\eta_{t+1}$, $u_t$ is a martingale difference under a natural predetermined filtration, but the threshold path and score are driven by the same Brownian motion. Section~\ref{sec:mc} illustrates the resulting size distortion. Contemporaneous error--innovation dependence considered in Section~\ref{sec:endog} also invalidates Proposition~\ref{prop:test}.}

\section{Endogenous threshold variable}\label{sec:endog}

It is well documented that endogeneity of a stationary threshold variable can lead to inconsistent threshold estimation \citep{Yu2013}. Two strands of the literature address this problem. One uses generalised method of moments procedures \citep{SeoShin2016,ChenStengos2024}, while another adopts control-function approaches \citep{KourtellosStengosTan2016,KourtellosStengosSun2022,YuLiaoPhillips2024}. Both rely on valid instruments, and two-stage least squares alone need not resolve the inconsistency \citep{Yu2013}. \citet{YuPhillipsEndogeneity2018} instead propose an integrated difference kernel estimator (IDKE) and show that consistent threshold estimation can be achieved without instrumental variables.

In this section, we extend model~\eqref{eq:model} to allow for an endogenous threshold variable. We consider the \textit{contemporaneous} endogeneity $u_t=\varrho\eta_t+e_t$ with exogenous slope regressors. Similar to \citet{YuPhillipsEndogeneity2018}, we show that, under this form of endogeneity, ordinary least squares retains the diminishing-effect limiting distribution of the threshold estimator and the corresponding likelihood ratio inference, without requiring instrumental variables or a control-function correction. For simplicity, we consider \textit{i.i.d.}\ exogenous slope regressors. We establish the distributional results under diminishing threshold effects below and examine the fixed-effect case numerically in Section~\ref{sec:mc}.

\begin{assumption}[Endogenous threshold variable]\label{ass:endog}
Assumption~\ref{ass:q} holds with $p>5$, and
\begin{enumerate}[label=(\roman*),leftmargin=*]
\item $u_t=\varrho\,\eta_t+e_t$, where $\varrho\in\R$ and $\{e_t\}$ is \textit{i.i.d.}\ with $\E e_t=0$, $\E e_t^2=\sigma_e^2$ and $\E|e_t|^r<\infty$ for some $r>4$, and $\sigma^2=\varrho^2\sigma_\eta^2+\sigma_e^2>0$.
\item $\{x_t\}$ is \textit{i.i.d.}\ with $\E\|x_t\|^r<\infty$ and $D=\E(x_tx_t')$ positive definite.
\item the sequences $\{x_t\}$, $\{e_t\}$ and $\{\eta_t\}$ are mutually independent.
\end{enumerate}
\end{assumption}

Under Assumption~\ref{ass:endog}, $\sigma^2=\E u_t^2$ and $\E(u_tq_t)=\E(u_t\eta_t)=\varrho\sigma_\eta^2$. Thus $\varrho\neq0$ makes the threshold variable endogenous, with $\E(u_t\mid q_t)\neq0$ in general, while the slope regressors remain exogenous. This specification includes jointly Gaussian errors and threshold innovations. The stronger moment condition on $\eta_t$ controls the local second moments needed for the variance limit; we do not expect it to be sharp.

\begin{proposition}[Contemporaneous error--innovation dependence]\label{prop:endog}
Under Assumptions~\ref{ass:par} and \ref{ass:endog}, the conclusions of Theorem~\ref{thm:cons}(a) hold. If $0<\alpha<1/4$, the conclusions of Theorems~\ref{thm:dist}(b) and \ref{thm:LR}(a) also hold, with the same $\lt$, $D$ and $\sigma^2$.
\end{proposition}

Proposition~\ref{prop:endog} shows that contemporaneous endogeneity does not affect the consistency of the threshold estimator or its limiting distribution under diminishing threshold effects. This robustness arises from a local cancellation around the threshold. To see this point clear, let $q_t=z_t+\eta_t$, where $z_t$ is known at time $t-1$, and let $A_h=(\gamma_0,\gamma_0+h]$. Then
\[
\E\bigl(u_t\1\{q_t\in A_h\}\mid\F_{t-1},\{x_s\}\bigr)
=
\varrho\,\Upsilon_h(z_t),
\qquad
\Upsilon_h(z)
=
\int_{\gamma_0-z}^{\gamma_0+h-z} e\,f_\eta(e)\,de .
\]
Since $\int_{\R}\Upsilon_h(z)\,dz=h\E\eta_t=0$, this predictable component has no leading local-time occupation term. Intuitively, the locally flat density of the nearly integrated state means that innovations observed near the threshold approximately follow their unconditional distribution. Its mean-zero endogenous contributions then cancel when visits are aggregated. The remaining predictable component is of smaller order than the local score, whose variance remains $\sigma^2D$.

The intuition is related to the IDKE of \citet{YuPhillipsEndogeneity2018}, which separates a threshold jump from a smooth endogenous conditional mean under \textit{i.i.d.}\ sampling. Here, no kernel correction is required: the local averaging generated by visits of the nearly integrated threshold variable produces an analogous cancellation for ordinary least squares under Assumption~\ref{ass:endog}. The closer technical connection is to zero-energy functionals in structural nonparametric cointegrating regression \citep{WangPhillips2009b,WangPhillips2011}. With a stationary threshold variable, this local cancellation generally need not occur, as illustrated in Table~\ref{tab:mc4}.\footnote{Section~\ref{sec:Sendog} of the Online Appendix illustrates the contrast in a simple step regression.}

\begin{remark}
Proposition~\ref{prop:endog} establishes consistency of the threshold estimator under contemporaneous endogeneity and justifies likelihood ratio inference under diminishing threshold effects. It does not establish the validity of conventional coefficient inference or the sup-Wald test, since the local cancellation need not hold for scores accumulated over entire regimes. The simulations in Section~\ref{sec:mc} illustrate size distortion in the uncorrected sup-Wald test.
\end{remark}

\section{Nonstationary slope regressors}\label{sec:nonstat}

We now allow both the slope regressors and the threshold variable to be
nonstationary. Note that in the threshold models of
\citet{GonzaloPitarakis2006}, \citet{Chen2015} and \citet{ChenStengos2022}, only the slope regressors
are nonstationary. Consider model \eqref{eq:model}--\eqref{eq:q} with
\begin{equation}\label{eq:deltans}
\delta_n=n^{-1/2-\tau}\delta_0,\qquad\delta_0\neq0,\qquad0\le\tau<1/4,
\end{equation}
as in \citet{Chen2015}. The factor $n^{-1/2}$ offsets the order $\sqrt n$ of $x_t$, so that $\delta_n'x_t=O_p(n^{-\tau})$. Thus, $\tau=0$ corresponds to a fixed effect after normalising the regressors, while $\tau>0$ gives a diminishing effect.

\begin{assumption}[Nonstationary slope regressors]\label{ass:ns}\leavevmode
\begin{enumerate}[label=(\roman*),leftmargin=*]
\item $n^{-1/2}x_{[ns]}\weak G(s)$ on $\mathbb D[0,1]^d$, where $G$ has continuous sample paths, $P\{\delta_0'G(s)=0\}=0$ for almost every $s\in[0,1]$, and the matrices
\[
M_\mu=\int_0^1G(s)G(s)'\1\{J_c(s)\le0\}\,ds\quad\text{and}\quad M-M_\mu,\qquad M=\int_0^1G(s)G(s)'\,ds,
\]
are positive definite almost surely.
\item $\{u_t\}$ satisfies Assumption~\ref{ass:x}(ii), and the sequences $\{x_t\}$, $\{u_t\}$ and $\{\eta_t\}$ are mutually independent.
\item $\gamma_0$ lies in the interior of $\Gamma$, $\beta$ and $\delta_0\neq0$ are fixed, and $\delta_n$ satisfies \eqref{eq:deltans}.
\end{enumerate}
\end{assumption}

Assumption~\ref{ass:ns} allows the slope regressors to be integrated or nearly integrated, with each column having the common order $\sqrt n$. This common normalisation excludes a constant or mixtures of stationary and integrated regressors, which would require separate normalisations and are not covered by the limiting results below. The assumption includes local-to-unity regressors of the form $x_t=\exp(\Lambda/n)x_{t-1}+v_t$, where $\Lambda$ is diagonal and $v_t$ is a linear process with a nonsingular long-run covariance matrix, so that $G$ is a multivariate Ornstein--Uhlenbeck process. It also accommodates the hybrid stochastic local unit root regressors of \citet{LiebermanPhillips2020} and \citet{ChenStengos2022}. In these examples, independence of $G$ and $J_c$, together with nondegeneracy of $G$ and $0<\mu<1$, ensures that the regime-specific information matrices are positive definite.

The information relevant for estimating the threshold has a different form. Define
\begin{equation}\label{eq:QL}
Q_\ell=\int_0^1G(s)G(s)'\,d\Lt(s,0),
\end{equation}
where the integration is with respect to the local-time measure of $J_c$ at the origin. Unlike the regime information matrices, which use observations throughout each regime, $Q_\ell$ samples the slope process only at times when the threshold variable visits the threshold. Assumption~\ref{ass:ns} ensures that $(\delta_0'G(s))^2$ is positive for Lebesgue-almost every $s$, almost surely. By independence of $G$ and $J_c$, its zero set receives no local-time mass.\footnote{This follows by conditioning on $G$ and using absolute continuity of $\E\Lt(ds,0)$.} Since $\lt>0$ almost surely, the resulting directional information satisfies
\[
\delta_0'Q_\ell\delta_0
=
\int_0^1(\delta_0'G(s))^2\,d\Lt(s,0)
>0
\qquad \text{almost surely}.
\]

Let $\mathcal T$ and $\xi$ be as in Section~\ref{sec:results}, independent of $(G,J_c)$. For $\tau=0$, define $\mathcal D_G$ as $\mathcal D$ in \eqref{eq:Dcal}, with marks
\[
z_{1i}=(\delta_0'G(s_{1i}))^2-2\delta_0'G(s_{1i})u_{1i},\qquad z_{2i}=(\delta_0'G(s_{2i}))^2+2\delta_0'G(s_{2i})u_{2i},
\]
where, conditionally on $(G,J_c)$, the times $s_{ki}$ are \textit{i.i.d.}\ draws from the probability measure $\Lt(ds,0)/\lt$ on $[0,1]$ and the $u_{ki}$ are \textit{i.i.d.}\ draws from $P_u$, all independent of $(N_1,N_2)$. Let $\mathcal V_G$ be the left end-point of the interval on which $\mathcal D_G$ attains its minimum.

\begin{theorem}[Nonstationary slope regressors]\label{thm:ns}
Let Assumptions~\ref{ass:q} and \ref{ass:ns} hold.
\begin{enumerate}[label=(\alph*),leftmargin=*]
\item Consistency. Uniformly in $\gamma\in\Gamma$, $n^{2\tau-1/2}[S_n(\gamma)-S_n(\gamma_0)]\convp\delta_0'Q_\ell\delta_0\,|\gamma-\gamma_0|$, and $\hat\gamma\convp\gamma_0$.
\item Diminishing threshold effect. If $0<\tau<1/4$, then $n^{1/2-2\tau}(\hat\gamma-\gamma_0)\convd\lambda_Q\mathcal T$ with $\lambda_Q=\sigma^2/(\delta_0'Q_\ell\delta_0)$, and $LR_n(\gamma_0)\convd\xi$.
\item Fixed threshold effect. If $\tau=0$, then $\sqrt n(\hat\gamma-\gamma_0)=O_p(1)$ and $LR_n(\gamma_0)\convd-\sigma^{-2}\min_v\mathcal D_G(v)$. If, in addition, $u_t$ has a continuous distribution, then $\sqrt n(\hat\gamma-\gamma_0)\convd\mathcal V_G$. If $u_t\sim N(0,\sigma^2)$, then $\limsup_nP\{LR_n(\gamma_0)>z\}\le1-(1-e^{-z/2})^2$ for every $z>0$.
\end{enumerate}
\end{theorem}

Theorem~\ref{thm:ns} shows that allowing the slope regressors to be nonstationary does not change the main rate results once the normalisation in \eqref{eq:deltans} is taken into account. As in the stationary-slope case, the threshold estimator converges at rate $n^{1/2-2\tau}$ under a diminishing effect, at rate $\sqrt n$ under a fixed effect, and the diminishing-effect likelihood ratio statistic retains the same pivotal limit. What changes is the information available for locating the threshold. With stationary slope regressors, averaging over boundary visits replaces $x_tx_t'$ by its expectation $D$, yielding the information matrix $\lt D$. With integrated slopes, the realised magnitude of the regressors varies over time, so the corresponding information is the path-dependent matrix $Q_\ell$ in \eqref{eq:QL}. The fixed-effect limit changes in the same way: the compound Cox marks now depend on the values of $G$ at the times selected by the local-time measure.

This result also extends \citet{Chen2015} and \citet{ChenStengos2022}, who allow nonstationary slope regressors but maintain a stationary threshold variable. In their setting, observations near $\gamma_0$ occur throughout the sample at frequency $f(\gamma_0)$, so the relevant information takes the form
$f(\gamma_0)\int_0^1G(s)G(s)'ds$. Here the threshold variable is itself nearly integrated, and boundary visits are instead distributed according to its local time. Ordinary time integration is therefore replaced by the local-time-weighted matrix $Q_\ell$. Thus, even with the same nonstationary slope process, information about the threshold depends not only on the path of $G$, but also on when the nearly integrated threshold variable visits $\gamma_0$.

The independence condition in Assumption~\ref{ass:ns}(ii) allows us to condition on the slope path and separate these two sources of nonstationarity. With correlated innovations, a joint local limit theory would be required. The extreme case $x_t=q_t$ illustrates the issue: then $G=J_c$ and $Q_\ell=\int_0^1J_c(s)^2\,d\Lt(s,0)=0$, because local time is supported on the zero set of $J_c$. A nonsingular joint long-run innovation covariance may avoid this degeneracy, but establishing the corresponding joint local limits is beyond the present analysis. The correlated design in Section~\ref{sec:mc} provides numerical evidence for this possible extension.

\section{Monte Carlo evidence}\label{sec:mc}

We now examine the finite-sample performance of threshold estimation, likelihood ratio inference and the sup-Wald test. The main results illustrate the convergence rates, the consistency boundary and the effect of contemporaneous endogeneity. Additional experiments on confidence-set construction and nonstationary slope regressors are reported in Section~\ref{sec:Smc} of the Online Appendix.

\subsection{Design}

We generate the data from
\[
y_t=x_t'\beta+n^{-\alpha}\delta_0'x_t\1\{q_t\le\gamma_0\}+u_t,
\qquad x_t=(1,x_{2t})',\qquad
q_t=e^{c/n}q_{t-1}+w_t,\quad q_0=0,
\]
where $\beta=(0,1)'$, $\delta_0=(1,1)'$, $\gamma_0=0.5$, and $x_{2t}$, $w_t$ and $u_t$ are mutually independent \textit{i.i.d.}\ $N(0,1)$ sequences. We consider $c\in\{-10,0,5\}$, $\alpha\in\{0,0.1,0.2\}$ and $n\in\{250,500,1000,2000\}$. A fourth design sets $c=0$, $w_t=0.5w_{t-1}+\eta_t$ and $x_{2t}=0.5x_{2,t-1}+\sqrt{0.75}\nu_t$, with mutually independent standard Gaussian innovations, also independent of $u_t$. These processes satisfy Assumptions~\ref{ass:q} and \ref{ass:x}.\footnote{The filtered threshold innovation starts at zero, and the stationary slope process uses a 100-observation burn-in.}

We search over $\Gamma=[-1.5,2.5]$ and require at least ten observations in each regime. We report the median absolute error (MdAE), the coverage (Cov) of $C_n^F(0.95)$ and its average Lebesgue measure (Len), computed from the exact union of accepted intervals. MdAE and Len condition on a feasible fit, whereas Cov conditions on a feasible true split. Median errors are less sensitive than root mean squared errors to samples with little local time. Each experiment uses 2{,}000 replications unless otherwise stated. The Monte Carlo standard error of a coverage estimate near 0.95 is about 0.005.

\subsection{Estimation and likelihood ratio inference}

Table~\ref{tab:mc1} reports the baseline estimation and confidence-set results. Median errors decline with $n$, but precision weakens as $\alpha$ approaches $1/4$. Mean reversion improves precision by increasing visits near the threshold. The dependent design has larger errors, consistent with its doubled long-run innovation standard deviation and smaller local time.

\begin{table}[!ht]
\centering
\caption{Threshold estimator and likelihood ratio confidence set: stationary slope regressors}\label{tab:mc1}
\footnotesize
\setlength{\tabcolsep}{4pt}
\begin{tabular}{r ccc ccc ccc ccc}
\toprule
 & \multicolumn{3}{c}{$c=-10$} & \multicolumn{3}{c}{$c=0$} & \multicolumn{3}{c}{$c=5$} & \multicolumn{3}{c}{$c=0$, dependent} \\
\cmidrule(lr){2-4}\cmidrule(lr){5-7}\cmidrule(lr){8-10}\cmidrule(lr){11-13}
$n$ & MdAE & Cov & Len & MdAE & Cov & Len & MdAE & Cov & Len & MdAE & Cov & Len \\
\midrule

% Begin output/tables/table1.tex
\multicolumn{13}{l}{Panel A: $\alpha=0.0$} \\
250 & 0.055 & 0.973 & 0.417 & 0.174 & 0.979 & 0.968 & 0.434 & 0.976 & 1.179 & 0.277 & 0.976 & 1.572 \\
500 & 0.042 & 0.980 & 0.282 & 0.110 & 0.985 & 0.781 & 0.263 & 0.979 & 1.085 & 0.188 & 0.981 & 1.321 \\
1000 & 0.028 & 0.988 & 0.207 & 0.076 & 0.983 & 0.656 & 0.181 & 0.984 & 0.996 & 0.143 & 0.986 & 1.112 \\
2000 & 0.019 & 0.977 & 0.146 & 0.052 & 0.983 & 0.523 & 0.129 & 0.984 & 0.852 & 0.101 & 0.976 & 0.891 \\
\midrule
\multicolumn{13}{l}{Panel B: $\alpha=0.1$} \\
250 & 0.116 & 0.962 & 1.080 & 0.334 & 0.970 & 1.773 & 0.736 & 0.974 & 1.810 & 0.493 & 0.972 & 2.517 \\
500 & 0.086 & 0.961 & 0.805 & 0.245 & 0.972 & 1.667 & 0.541 & 0.978 & 1.886 & 0.399 & 0.980 & 2.377 \\
1000 & 0.068 & 0.973 & 0.627 & 0.193 & 0.967 & 1.541 & 0.422 & 0.968 & 1.891 & 0.337 & 0.979 & 2.208 \\
2000 & 0.059 & 0.965 & 0.490 & 0.158 & 0.973 & 1.342 & 0.333 & 0.973 & 1.873 & 0.281 & 0.975 & 2.039 \\
\midrule
\multicolumn{13}{l}{Panel C: $\alpha=0.2$} \\
250 & 0.318 & 0.946 & 2.408 & 0.699 & 0.974 & 2.616 & 1.046 & 0.974 & 2.287 & 0.912 & 0.984 & 3.228 \\
500 & 0.308 & 0.957 & 2.208 & 0.588 & 0.970 & 2.781 & 0.926 & 0.977 & 2.530 & 0.747 & 0.988 & 3.350 \\
1000 & 0.244 & 0.961 & 2.007 & 0.555 & 0.967 & 2.787 & 0.807 & 0.980 & 2.751 & 0.700 & 0.980 & 3.314 \\
2000 & 0.214 & 0.964 & 1.826 & 0.507 & 0.967 & 2.786 & 0.753 & 0.983 & 2.941 & 0.683 & 0.987 & 3.370 \\

% End output/tables/table1.tex

\midrule
\multicolumn{13}{l}{Panel D: frequency of infeasible samples} \\
 & \multicolumn{4}{c}{Fail} & \multicolumn{4}{c}{Inf} & & & & \\
\cmidrule(lr){2-5}\cmidrule(lr){6-9}
$n$ & $c=-10$ & $c=0$ & $c=5$ & dep. & $c=-10$ & $c=0$ & $c=5$ & dep. & & & & \\

% Begin output/tables/table1_panelD.tex
250 & 0.000 & 0.117 & 0.344 & 0.172 & 0.003 & 0.164 & 0.366 & 0.102 & & & & \\
500 & 0.000 & 0.070 & 0.222 & 0.122 & 0.000 & 0.111 & 0.283 & 0.064 & & & & \\
1000 & 0.000 & 0.053 & 0.179 & 0.075 & 0.000 & 0.077 & 0.218 & 0.061 & & & & \\
2000 & 0.000 & 0.042 & 0.106 & 0.058 & 0.000 & 0.045 & 0.148 & 0.029 & & & & \\

% End output/tables/table1_panelD.tex

\bottomrule
\end{tabular}

\vspace{0.5ex}
\parbox{0.95\textwidth}{\footnotesize Notes: MdAE is the median of $|\hat\gamma-\gamma_0|$ over samples with a feasible split. Cov is the coverage of $C_n^F(0.95)$ over samples with a feasible true split. Len is its average Lebesgue measure over samples with any feasible fit. ``Dependent'': $w_t$ and $x_{2t}$ are AR(1) with coefficient 0.5. Panel D is common to Panels A--C: Fail is the frequency of samples without a feasible split, and Inf the frequency, among samples with a feasible split, of an infeasible split at $\gamma_0$. 2{,}000 replications.}
\end{table}

Conditional coverage is close to or above 0.95, although the confidence sets can remain wide. At $\alpha=0.2$, average length does not decline with $n$ in the unit-root and locally explosive designs over the sample sizes considered.

Persistent threshold paths can also leave too few observations in one regime for estimation. Panel D of Table~\ref{tab:mc1} reports the frequency of samples without a feasible split and, among the remaining samples, the frequency of an infeasible true split. For $c=5$ and $n=250$, these frequencies are 34.4\% and 36.6\%, respectively. Consequently, about 42\% of replications enter the coverage calculation, compared with 66\% for MdAE and Len. These frequencies decline with $n$ and are negligible for $c=-10$. Table~\ref{tab:mcU} of the Online Appendix reports unconditional results for $C_n^U(0.95)$, which retains infeasible candidates and equals $\Gamma$ when no split is feasible. Coverage remains high, but the sets can occupy much of the parameter space. High coverage therefore need not imply precise threshold localisation.

Table~\ref{tab:mc2} examines the convergence rates and the consistency boundary using larger samples. The normalised median errors in Panel A remain broadly stable, in line with Theorem~\ref{thm:dist}. Panel B instead shows persistent unnormalised errors at and beyond $\alpha=1/4$. High coverage in these designs reflects weak information about the threshold. For $\alpha>1/4$, the LR statistic tends to zero uniformly on $\Gamma$, so the confidence set approaches the entire parameter space.\footnote{See Theorem~\ref{thm:cons}(b).}

\begin{table}[!ht]
\centering
\caption{Convergence rates and the consistency boundary: medians of $|\hat\gamma-\gamma_0|$ ($c=0$)}\label{tab:mc2}
\small
\begin{tabular}{llcccc}
\toprule
 & & $n=500$ & $n=2{,}000$ & $n=8{,}000$ & $n=32{,}000$ \\
\midrule

\multicolumn{6}{l}{Panel A: stationary slope regressors, $\delta_n=n^{-\alpha}\delta_0$} \\
$\alpha=0$ & $n^{0.5}\,\mathrm{med}|\hat\gamma-\gamma_0|$ & 2.71 & 2.39 & 2.34 & 2.46 \\
 & Coverage & 0.988 & 0.982 & 0.979 & 0.988 \\
$\alpha=0.1$ & $n^{0.3}\,\mathrm{med}|\hat\gamma-\gamma_0|$ & 1.73 & 1.66 & 1.26 & 1.53 \\
 & Coverage & 0.972 & 0.975 & 0.967 & 0.973 \\
$\alpha=0.2$ & $n^{0.1}\,\mathrm{med}|\hat\gamma-\gamma_0|$ & 1.08 & 1.24 & 1.09 & 1.20 \\
 & Coverage & 0.962 & 0.974 & 0.970 & 0.966 \\
\midrule
\multicolumn{6}{l}{Panel B: stationary slope regressors, $\alpha\ge1/4$ (no normalisation)} \\
$\alpha=0.25$ & $\mathrm{med}|\hat\gamma-\gamma_0|$ & 0.81 & 0.70 & 0.68 & 0.68 \\
 & Coverage & 0.973 & 0.973 & 0.988 & 0.990 \\
$\alpha=0.35$ & $\mathrm{med}|\hat\gamma-\gamma_0|$ & 1.03 & 1.13 & 1.12 & 1.18 \\
 & Coverage & 0.979 & 0.991 & 0.996 & 0.999 \\
\bottomrule
\end{tabular}

\vspace{0.5ex}
\parbox{0.95\textwidth}{\footnotesize Notes: Both panels use the design of Table~\ref{tab:mc1} with $c=0$. Panel A reports median absolute errors multiplied by $n^{1/2-2\alpha}$. Panel B reports unnormalised errors at and beyond $\alpha=1/4$. Coverage refers to $C_n^F(0.95)$ over samples with a feasible true split. There are 1{,}000 replications.}
\end{table}

\subsection{Error distributions and alternative inference}

Table~\ref{tab:mc3} isolates the role of the error distribution under a fixed intercept shift. All three distributions have the same variance and produce similar median errors, but their LR coverage differs substantially. Gaussian and Laplace errors yield coverage above the nominal level. The mixture yields coverage of about 0.90 throughout, consistent with the asymptotic upper bound of $0.95^2$ in footnote~\ref{rem:nongauss}. Thus, the Gaussian guarantee of Theorem~\ref{thm:LR}(b) cannot be extended on the basis of the error variance alone.

\begin{table}[!ht]
\centering
\caption{Fixed threshold effect: coverage under different error distributions}\label{tab:mc3}
\small
\begin{tabular}{r cc cc cc}
\toprule
 & \multicolumn{2}{c}{Gaussian} & \multicolumn{2}{c}{Laplace} & \multicolumn{2}{c}{Mixture} \\
\cmidrule(lr){2-3}\cmidrule(lr){4-5}\cmidrule(lr){6-7}
$n$ & MdAE & Cov & MdAE & Cov & MdAE & Cov \\
\midrule

% Begin output/tables/table5.tex
250 & 0.077 & 0.991 & 0.075 & 0.972 & 0.080 & 0.903 \\
500 & 0.052 & 0.990 & 0.050 & 0.972 & 0.053 & 0.905 \\
1000 & 0.036 & 0.992 & 0.036 & 0.972 & 0.037 & 0.904 \\
2000 & 0.025 & 0.991 & 0.026 & 0.972 & 0.027 & 0.894 \\

% End output/tables/table5.tex

\bottomrule
\end{tabular}

\vspace{0.5ex}
\parbox{0.95\textwidth}{\footnotesize Notes: $y_t=\1\{q_t\le0.5\}+u_t$ with $q_t$ a Gaussian random walk. The three error distributions have variance $0.1025$: Gaussian, Laplace, and the mixture $0.90\,N(0,0.05^2)+0.05\,N(1,0.05^2)+0.05\,N(-1,0.05^2)$ of footnote~\ref{rem:nongauss}. MdAE uses feasible fits and Cov uses feasible true splits. 20{,}000 replications.}
\end{table}

Table~\ref{tab:mc6} of the Online Appendix reports alternative confidence-set calibrations and a fixed-regressor residual bootstrap. Under a diminishing effect, hull calibration brings coverage closer to the nominal level, and the variance correction of Remark~\ref{rem:het} substantially improves coverage under heteroskedasticity. The fixed-effect bootstrap gives encouraging results in the designs considered, but its asymptotic validity remains unproved.

\subsection{Testing for a threshold effect}

Table~\ref{tab:mc7} reports rejection frequencies using quantile trimming $\pi=0.15$ and simulated Andrews critical values with $d=2$. The 10\%, 5\% and 1\% critical values are 10.01, 11.72 and 15.68, respectively. Under independence, size is close to the nominal 5\% level. Power increases with $n$ for $\alpha=0.25$, particularly under mean reversion, even though least squares cannot consistently locate the threshold. The weaker alternatives with $\alpha=0.4$ have low power at these sample sizes. Panel B shows size distortion when errors depend on threshold innovations, whether contemporaneously or through $u_t=w_{t+1}$, a specification permitted by a martingale difference restriction.

\begin{table}[!ht]
\centering
\caption{Sup-Wald test: rejection frequencies at the 5\% level}\label{tab:mc7}
\small
\begin{tabular}{r ccc ccc ccc}
\toprule
 & \multicolumn{3}{c}{size} & \multicolumn{3}{c}{power, $\alpha=0.25$} & \multicolumn{3}{c}{power, $\alpha=0.4$} \\
\cmidrule(lr){2-4}\cmidrule(lr){5-7}\cmidrule(lr){8-10}
$n$ & $c=-10$ & $c=0$ & $c=5$ & $c=-10$ & $c=0$ & $c=5$ & $c=-10$ & $c=0$ & $c=5$\\
\midrule
\multicolumn{10}{l}{Panel A: independent errors and threshold innovations} \\

% Begin output/tables/table7_panelA.tex
250 & 0.050 & 0.049 & 0.043 & 0.442 & 0.254 & 0.120 & 0.108 & 0.076 & 0.049 \\
500 & 0.041 & 0.052 & 0.054 & 0.628 & 0.353 & 0.138 & 0.113 & 0.080 & 0.054 \\
1000 & 0.044 & 0.039 & 0.051 & 0.802 & 0.464 & 0.168 & 0.140 & 0.095 & 0.059 \\
2000 & 0.048 & 0.043 & 0.048 & 0.917 & 0.534 & 0.215 & 0.160 & 0.091 & 0.060 \\

% End output/tables/table7_panelA.tex

\midrule
\multicolumn{10}{l}{Panel B: size when $u_t$ is correlated with the innovations of $q_t$ ($c=0$)} \\
 & \multicolumn{3}{c}{$\rho_{uw}=0.5$} & \multicolumn{3}{c}{$\rho_{uw}=0.9$} & \multicolumn{3}{c}{$u_t=w_{t+1}$} \\
\cmidrule(lr){2-4}\cmidrule(lr){5-7}\cmidrule(lr){8-10}

% Begin output/tables/table7_panelB.tex
500 & \multicolumn{3}{c}{0.077} & \multicolumn{3}{c}{0.155} & \multicolumn{3}{c}{0.185} \\
2000 & \multicolumn{3}{c}{0.081} & \multicolumn{3}{c}{0.146} & \multicolumn{3}{c}{0.172} \\

% End output/tables/table7_panelB.tex

\bottomrule
\end{tabular}

\vspace{0.5ex}
\parbox{0.95\textwidth}{\footnotesize Notes: design of Table~\ref{tab:mc1} with $\delta_n=0$ (size) or $\delta_n=n^{-\alpha}(1,1)'$ (power). Trimming is $\pi=0.15$. Critical values use the simulated Andrews distribution with $d=2$. In Panel B, $u_t=\rho_{uw}w_t+(1-\rho_{uw}^2)^{1/2}\nu_t$ or $u_t=w_{t+1}$. 2{,}000 replications.}
\end{table}

\subsection{Endogenous threshold variable}

Table~\ref{tab:mc4} compares integrated and stationary threshold variables under the same diminishing effect, $\alpha=0.1$. With a unit-root threshold variable, median errors and LR coverage change relatively little as contemporaneous error--innovation correlation increases. These results support Proposition~\ref{prop:endog}. The small negative median bias decreases with $n$ and is consistent with the left end-point convention in \eqref{eq:gammahat}. With a stationary threshold variable, bias persists and coverage deteriorates sharply, reaching zero under strong correlation. This contrast illustrates the cancellation mechanism under the particular endogenous specification considered here.

\begin{table}[!ht]
\centering
\caption{Endogenous threshold variable under a diminishing effect: $\mathrm{corr}(u_t,w_t)=\rho_{uw}$}\label{tab:mc4}
\small
\begin{tabular}{r ccc ccc ccc}
\toprule
 & \multicolumn{3}{c}{$\rho_{uw}=0$} & \multicolumn{3}{c}{$\rho_{uw}=0.5$} & \multicolumn{3}{c}{$\rho_{uw}=0.9$} \\
\cmidrule(lr){2-4}\cmidrule(lr){5-7}\cmidrule(lr){8-10}
$n$ & MdB & MdAE & Cov & MdB & MdAE & Cov & MdB & MdAE & Cov \\
\midrule

\multicolumn{10}{l}{Panel A: unit-root $q_t$, $\alpha=0.1$} \\
250 & -0.058 & 0.349 & 0.968 & -0.061 & 0.358 & 0.973 & -0.048 & 0.385 & 0.985 \\
500 & -0.032 & 0.234 & 0.970 & -0.024 & 0.258 & 0.983 & -0.020 & 0.277 & 0.984 \\
1000 & -0.026 & 0.185 & 0.976 & -0.021 & 0.188 & 0.980 & -0.021 & 0.195 & 0.983 \\
2000 & -0.008 & 0.157 & 0.969 & -0.006 & 0.158 & 0.972 & -0.008 & 0.162 & 0.978 \\
\midrule
\multicolumn{10}{l}{Panel B: stationary $q_t$ (AR(1), $\phi=0.5$), $\alpha=0.1$} \\
250 & -0.006 & 0.041 & 0.967 & -0.188 & 0.372 & 0.631 & -0.932 & 0.932 & 0.025 \\
500 & -0.004 & 0.024 & 0.963 & -0.270 & 0.399 & 0.529 & -0.901 & 0.901 & 0.004 \\
1000 & -0.002 & 0.011 & 0.978 & -0.517 & 0.542 & 0.381 & -0.886 & 0.886 & 0.000 \\
2000 & -0.001 & 0.007 & 0.970 & -0.770 & 0.774 & 0.195 & -0.871 & 0.871 & 0.000 \\
\bottomrule
\end{tabular}

\vspace{0.5ex}
\parbox{0.95\textwidth}{\footnotesize Notes: The design of Table~\ref{tab:mc1} is used with $\alpha=0.1$ and $u_t=\rho_{uw}w_t+(1-\rho_{uw}^2)^{1/2}\nu_t$, where $\nu_t$ is an independent \textit{i.i.d.}\ $N(0,1)$ sequence. In Panel A, $q_t$ is a Gaussian random walk. In Panel B, $q_t=0.5q_{t-1}+w_t$. MdB and MdAE are the median signed and absolute errors over samples with a feasible fit. Cov conditions on a feasible true split. There are 2{,}000 replications. The complete fixed- and diminishing-effect results are reported in Table~\ref{tab:mc4full} of the Online Appendix.}
\end{table}

The complete comparison, including fixed effects, appears in Table~\ref{tab:mc4full} of the Online Appendix. The unit-root fixed-effect design shows a similar pattern, although its limiting distribution is not covered by Proposition~\ref{prop:endog}.

\subsection{Nonstationary slope regressors}

Tables~\ref{tab:mc5}--\ref{tab:mcQ} of the Online Appendix examine the nonstationary-slope extension. The independent designs satisfy Assumption~\ref{ass:ns}, whereas the correlated-innovation design explores a possible extension. Median errors decline slowly and confidence sets remain wide. Comparison with the simulated limiting distribution shows particularly slow convergence of the upper quantiles, reflecting small values of $Q_\ell$ and the bounds imposed by the compact search space. The asymptotic approximation can therefore be poor in the tails at the sample sizes considered.

\section{Empirical illustration: public debt and growth}\label{sec:app}

We revisit the debt--growth nexus using a threshold model that allows the debt-to-GDP ratio to be integrated or nearly integrated. Conventional threshold theory assumes a stationary threshold variable, whereas debt ratios may exhibit unit-root or near-unit-root behaviour. We use the US debt and growth data distributed by \citet{Hansen2017}, originally compiled by \citet{ReinhartRogoff2010}, to illustrate threshold estimation in short-run growth dynamics with a persistent debt ratio. The full annual sample covers 1791--2009, and we also examine the post-war subsample, 1946--2009. This subsample is chosen independently of the persistence diagnostics. The historical distinction is relevant because \citet{Yoon2012} relates changes in debt persistence to the wartime debt surge, while \citet{Martins2021} compares debt--growth relationships before and after World War II.

Let $d_t$ denote gross federal debt/GDP in percentage points and $g_t$ annual real GDP growth. For each sample, we estimate
\[
\begin{aligned}
g_t&=(a_1+r_1g_{t-1})\1\{d_{t-1}\leq\gamma\}\\
&\quad +(a_2+r_2g_{t-1})\1\{d_{t-1}>\gamma\}+u_t.
\end{aligned}
\]
The slope vector is $x_t=(1,g_{t-1})'$. Hence, debt enters only as the threshold variable and no integrated regressor appears in the growth equation. The specification allows both the growth intercept and its autoregressive coefficient to differ across debt regimes. The first observation supplies the lags, leaving 218 and 63 regression observations. We search over $\Gamma=[30,90]$, require at least ten observations in each regime, and choose the smallest minimizer.

Table~\ref{tab:app} reports coefficient estimates and heteroskedasticity- and autocorrelation-consistent (HAC) standard errors conditional on the estimated sample splitting. The covariance matrix uses the four-column interacted design in calendar order, Bartlett weights and bandwidth $L=\lfloor4(n/100)^{2/9}\rfloor$, giving $L=4$ and $L=3$ in the two samples. Treating growth as stationary does not establish the whole-path independence required by Corollary~\ref{cor:slope}, because $g_{t-1}$ is a lagged dependent variable. The standard errors are therefore descriptive regression summaries rather than an application of that corollary. We do not report threshold tests or likelihood ratio confidence sets for this dynamic illustration.

\begin{table}[!ht]
\centering
\caption{US debt persistence and growth dynamics across debt regimes}\label{tab:app}
\small
\begin{tabular}{l rr rr}
\toprule
 & \multicolumn{2}{c}{1791--2009} & \multicolumn{2}{c}{1946--2009} \\
\cmidrule(lr){2-3}\cmidrule(lr){4-5}
 & Statistic & Ref. $p$ & Statistic & Ref. $p$ \\
\midrule

% Begin output/tables/table10.tex
\multicolumn{5}{l}{Panel A: debt/GDP persistence} \\
ADF & $-1.282\,[2]$ & 0.656 & $-1.099\,[1]$ & 0.710 \\
DF-GLS & $-1.425\,[2]$ & 0.171 & $-0.678\,[1]$ & 0.605 \\
Phillips--Perron & $-1.177\,[14]$ & 0.713 & $-3.272\,[10]$ & 0.023 \\
KPSS & $0.976\,[14]$ & $<0.001$ & $0.196\,[10]$ & 0.359 \\
AR(1) slope & $0.996$ & & $0.896$ & \\
\midrule
\multicolumn{5}{l}{Panel B: dynamic threshold fits} \\
Regression observations & \multicolumn{2}{c}{218} & \multicolumn{2}{c}{63} \\
Threshold estimate (\%) & \multicolumn{2}{c}{$30.85$} & \multicolumn{2}{c}{$34.51$} \\
Low/high regime observations & \multicolumn{2}{c}{130/88} & \multicolumn{2}{c}{10/53} \\
SSR reduction (\%) & \multicolumn{2}{c}{$3.08$} & \multicolumn{2}{c}{$2.22$} \\
Observations within 2.5 pp & \multicolumn{2}{c}{17} & \multicolumn{2}{c}{12} \\
\addlinespace
 & Low debt & High debt & Low debt & High debt \\
\cmidrule(lr){2-3}\cmidrule(lr){4-5}
Intercept ($\hat a_j$) & $2.915$ & $2.404$ & $1.757$ & $2.918$ \\
 & $(0.733)$ & $(0.539)$ & $(0.933)$ & $(0.480)$ \\
Lagged growth ($\hat r_j$) & $0.172$ & $0.477$ & $0.206$ & $0.147$ \\
 & $(0.135)$ & $(0.119)$ & $(0.158)$ & $(0.101)$ \\

% End output/tables/table10.tex

\bottomrule
\end{tabular}
\vspace{0.5ex}
\parbox{0.95\textwidth}{\footnotesize
\textit{Notes:} Panel A uses 219 and 64 debt observations. Brackets report the selected lag order or Bartlett truncation lag. ADF uses BIC and DF-GLS modified AIC. Reference $p$-values use 10{,}000 Gaussian random walks for the unit-root null and Gaussian white noise for KPSS, with lag selection repeated in every draw. These references do not account for historical breaks. Panel B uses an intercept and lagged growth, with debt entering only as the threshold. The search is over $[30,90]$ with at least ten observations per regime. Parentheses contain calendar-order HAC standard errors with Bartlett bandwidths 4 and 3. Their interpretation is discussed in the text. SSR reductions are relative to the corresponding linear AR(1) model. Visit counts use a 2.5 percentage-point neighbourhood.}
\end{table}

Panel A shows substantial persistence in the full debt series. Its AR(1) coefficient estimate is 0.996, the unit-root diagnostics do not reject their reference nulls, and KPSS rejects its Gaussian white-noise reference. The post-war evidence is mixed. ADF and DF-GLS do not reject, whereas Phillips--Perron rejects and KPSS does not reject its reference null. These diagnostics motivate allowing for persistence without establishing a unit root or excluding historical changes.

The full-sample fit places the threshold at 30.85 percent, close to the lower search bound. Growth persistence is greater in the high-debt regime, with estimated autoregressive coefficients of 0.477 and 0.172. The threshold specification reduces SSR by only 3.08 percent relative to a linear AR(1). After the war, the corresponding coefficient estimates are 0.147 and 0.206, and the SSR reduction is 2.22 percent. Thus the direction of the persistence contrast changes across samples and the improvement in fit is modest.

The post-war threshold remains the smallest feasible split under the ten-observation rule. Its low-debt regime comprises 1973--1982, so its coefficients describe a short historical episode and its standard errors are sensitive to the covariance estimator. The full-sample threshold is also sensitive to the search convention. Lowering the search bound from 30 to 20 percent moves its estimate to 20.27 percent, while the post-war estimate is unchanged. Only seventeen full-sample observations and twelve post-war observations lie within 2.5 percentage points of the reported threshold. These counts are not local-time estimates, but they illustrate why a large total sample need not provide much information about a boundary. The fits describe sample-specific growth dynamics and provide limited evidence for a stable structural debt threshold.

\section{Conclusion}\label{sec:conclusion}

This paper develops least squares estimation and inference for threshold regression with an integrated or local-to-unity threshold variable and a threshold fixed in levels. Information about the threshold comes from only order $\sqrt n$ visits to its neighbourhood, governed by diffusion local time. As a result, the threshold estimator converges at rate $n^{1/2-2\alpha}$ and is consistent only for $\alpha<1/4$. A fixed effect yields a compound Cox argmin limit, while a diminishing effect gives a two-sided Brownian criterion with a random local-time scale. Despite this random scale, the likelihood ratio statistic remains pivotal under a diminishing effect and the critical values of \citet{Hansen2000} continue to apply; under a fixed effect, they are conservative with Gaussian errors.

The results also separate detection of a threshold effect from estimation of its location. Under whole-path independence, the sup-Wald test remains consistent for $\alpha<1/2$, including cases in which the threshold itself cannot be consistently located, and the slope coefficients retain conventional $t$-ratios because they are estimated from whole regimes. The extensions show that the diminishing-effect results are robust to the specified form of contemporaneous endogeneity, while with independent nonstationary slope regressors the convergence rates are unchanged after normalisation but threshold information becomes path dependent.

The Monte Carlo results support the asymptotic theory and show that high coverage can coexist with wide confidence sets when visits near the threshold are sparse. The US debt--growth illustration reinforces the same point: information about a threshold depends on observations near the boundary rather than on total sample size alone.

Several questions remain open. Regressors dependent on the threshold innovations and lagged dependent variables require additional joint occupation and score limits. A theory for a slowly expanding parameter space would also be useful. The fixed-regressor residual bootstrap has encouraging coverage in the designs examined, but its validity for the fixed-effect likelihood ratio remains unproved. Finally, nonstationary slope vectors with an intercept or mixed orders of integration require separate normalisations, and the approximation to the upper quantiles of the existing nonstationary-slope limit is slow.

\bibliographystyle{elsarticle-harv}
\bibliography{references_threshold_model_unit_root}

\newpage
\appendix
\setcounter{lemma}{0}
\renewcommand{\thelemma}{\Alph{section}.\arabic{lemma}}
\renewcommand{\theHlemma}{\Alph{section}.\arabic{lemma}}
\section{Lemmas}\label{app:A}

We present the lemmas used in the proof of the main theorems in Appendix~\ref{app:B}. Section~\ref{sec:Sproofs} of the Online Appendix present the proofs of the lemmas.

\paragraph{Notation.}\label{app:notation}

$C$ is a generic finite constant that does not depend on $n$. The notation $\|\cdot\|$ is the Euclidean norm for vectors and the spectral norm for matrices. $f$ is the density of $\eta_t$ (written $f_\eta$ in Sections~\ref{sec:results} and \ref{sec:endog}), $f^\sharp(y)=\sup_{|s|\le1}f(y+s)$, and $\|g\|_*=\|g\|_1+\|g\|_\infty$ for measurable $g:\R\to\R$.

\emph{Threshold variable.} $\F_t=\sigma(\eta_s:s\le t)$, and $z_t=q_t-\eta_t=\rho_nq_{t-1}+\sum_{j\ge1}\varphi_j\eta_{t-j}$ is the $\F_{t-1}$-measurable part of $q_t$. $\mathcal Q=\sigma(\eta_s:s\in\mathbb Z)$ is generated by the complete innovation sequence and contains the threshold path, and $\mathcal G=\mathcal Q\vee\sigma(x_s:s\ge1)$. For an interval $A$, $\1_t(A)=\1\{q_t\in A\}$ and $N(A)=\sum_{t=1}^n\1_t(A)$. For $g\ge0$, $\Lambda_n(g)=n^{-1/2}\sum_{t=1}^ng(\gamma_0-z_t)$. For $v\in\R$ and $h>0$, $I(v,h)=(\gamma_0,\gamma_0+vh]$ if $v\ge0$ and $I(v,h)=(\gamma_0+vh,\gamma_0]$ if $v<0$. As in \citet{ParkPhillips2001}, convergence to functionals of $J_c$ is joint with \eqref{eq:FCLT} and is stated in a Skorokhod representation that also carries $\{x_t\}$ and $\{u_t\}$.

\emph{Criterion.} $A_\gamma=X_\gamma'X_\gamma$ and $C_\gamma=(X-X_\gamma)'(X-X_\gamma)$ are the moment matrices of the two regimes. $W(\gamma)$ is the window between $\gamma_0$ and $\gamma$: $W(\gamma)=(\gamma_0,\gamma]$ if $\gamma>\gamma_0$ and $W(\gamma)=(\gamma,\gamma_0]$ if $\gamma<\gamma_0$. The observations with $q_t\in W(\gamma)$ are misclassified when the threshold is placed at $\gamma$. With $X_W$ the matrix with rows $x_t'\1_t(W(\gamma))$,
\[
B(\gamma)=X_W'X_W=\sum_tx_tx_t'\1_t(W(\gamma)),\qquad b(\gamma)=X_W'u=\sum_tx_tu_t\1_t(W(\gamma)).
\]
The size of the regime-wise inverse moment matrices and scores is measured by
\[
\varpi_n=\sup_{\gamma\in\Gamma}\big(\|A_\gamma^{-1}\|+\|C_\gamma^{-1}\|\big),\qquad\varkappa_n=\sup_{\gamma\in\Gamma}\big(\|X_\gamma'u\|+\|(X-X_\gamma)'u\|\big),
\]
with $\varpi_n=\infty$ if a regime design is singular for some $\gamma\in\Gamma$. Finally, for window widths $h_n>0$, specified where the lemmas are applied, the local score is
\begin{equation}\label{eq:Mn}
M_n(v)=\frac{\operatorname{sgn}(v)}{(\sqrt n\,h_n)^{1/2}}\sum_{t=1}^nx_tu_t\1\{q_t\in I(v,h_n)\}.
\end{equation}

\begin{lemma}[Criterion decomposition and remainder bounds]\label{lem:decomp}\label{lem:orders}
Suppose that $\gamma_0\in\Gamma$.
\begin{enumerate}[label=(\alph*),leftmargin=*]
\item For every $\gamma\neq\gamma_0$ such that $A_\gamma$ and $C_\gamma$ are nonsingular,
\begin{equation}\label{eq:decomp}
S_n(\gamma)-S_n(\gamma_0)=R_{1n}(\gamma)-R_{2n}(\gamma)-R_{3n}(\gamma)+R_{4n}(\gamma)-R_{5n}(\gamma),
\end{equation}
where, with $B=B(\gamma)$ and $b=b(\gamma)$, $R_{1n}(\gamma)=\delta_n'B\delta_n$, $R_{5n}(\gamma)=u'(P_\gamma-P_{\gamma_0})u$, and
\[
\begin{aligned}
R_{2n}(\gamma)&=\delta_n'BA_\gamma^{-1}B\delta_n, & R_{3n}(\gamma)&=2\delta_n'b, & R_{4n}(\gamma)&=2\delta_n'BA_\gamma^{-1}X_\gamma'u & &\text{if }\gamma>\gamma_0,\\
R_{2n}(\gamma)&=\delta_n'BC_\gamma^{-1}B\delta_n, & R_{3n}(\gamma)&=-2\delta_n'b, & R_{4n}(\gamma)&=-2\delta_n'BC_\gamma^{-1}(X-X_\gamma)'u & &\text{if }\gamma<\gamma_0.
\end{aligned}
\]
\item If $\varpi_n<\infty$, then for every $\gamma\in\Gamma\setminus\{\gamma_0\}$, with $B=B(\gamma)$ and $b=b(\gamma)$,
\begin{align}
0\le R_{2n}(\gamma)&\le\varpi_n\|B\|\,R_{1n}(\gamma),\label{eq:R2}\\
|R_{4n}(\gamma)|&\le2\varpi_n\varkappa_n\|B\|^{1/2}R_{1n}(\gamma)^{1/2},\label{eq:R4}\\
|R_{5n}(\gamma)|&\le4\varpi_n\varkappa_n\|b\|+2\varpi_n\|b\|^2+2\varpi_n^2\varkappa_n^2\|B\|.\label{eq:R5}
\end{align}
\item Under Assumptions~\ref{ass:q} and \ref{ass:x}, $P(\hat\Gamma=\Gamma)\to1$, $\varpi_n=O_p(n^{-1})$, $\varkappa_n=O_p(n^{1/2})$, $\sup_{\gamma\in\Gamma}\|B(\gamma)\|=O_p(n^{1/2})$ and $\sup_{\gamma\in\Gamma}\|b(\gamma)\|=O_p(n^{1/4})$. Consequently
\begin{equation}\label{eq:rho}
\begin{gathered}
\sup_{\gamma\in\Gamma}\varpi_n\|B(\gamma)\|=O_p(n^{-1/2}),\qquad\sup_{\gamma\in\Gamma}\varpi_n\varkappa_n\|B(\gamma)\|^{1/2}=O_p(n^{-1/4}),\\
\sup_{\gamma\in\Gamma}|R_{5n}(\gamma)|=O_p(n^{-1/4}).
\end{gathered}
\end{equation}

\end{enumerate}
\end{lemma}

The decomposition separates misclassification from coefficient estimation. The term $R_{1n}$ measures the squared difference in the regression signal for misclassified observations, and $R_{3n}$ captures its interaction with the regression error. The terms $R_{2n}$, $R_{4n}$ and $R_{5n}$ account for refitting the slope coefficients. Persistence changes their stochastic orders: only $O_p(\sqrt n)$ observations contribute to $B$, and their centred score has order $O_p(n^{1/4})$. The much larger regime samples make refitting the slopes negligible for threshold localisation, leaving $R_{1n}-R_{3n}$ as the leading local criterion.

Let $0\le\alpha<1/4$ and $a_n=n^{1/2-2\alpha}$, and fix $\bar h>0$ with $[\gamma_0-\bar h,\gamma_0+\bar h]\subset\Gamma$. In the proofs for nonstationary slope regressors, $\alpha$ is read as $\tau$. For $K>0$ let $\Gamma_K=\{\gamma:K/a_n\le|\gamma-\gamma_0|\le\bar h\}$.

\begin{lemma}[Rate of convergence]\label{lem:rate}\leavevmode
\begin{enumerate}[label=(\alph*),leftmargin=*]
\item Suppose that
\begin{enumerate}[label=(\roman*),leftmargin=*]
\item $\hat\gamma\convp\gamma_0$ and $P(\hat\Gamma=\Gamma)\to1$.
\item The three quantities on the left-hand sides of \eqref{eq:rho} are $o_p(1)$.
\item \emph{(signal)} For every $\epsilon>0$ there are $\epsilon_0>0$ and $K_0$ such that $\limsup_nP\{R_{1n}(\gamma_0\pm K/a_n)<\epsilon_0K\}<\epsilon$ for all $K\ge K_0$.
\item \emph{(error)} For every $\epsilon>0$ there is $K_1$ such that $\limsup_nP(\mathcal E_K)<\epsilon$ for all $K\ge K_1$, where $\mathcal E_K$ is the event that $|R_{3n}(\gamma)|\ge\frac12R_{1n}(\gamma)$ for some $\gamma\in\Gamma_K$.
\end{enumerate}
Then
\begin{equation}\label{eq:outward}
\lim_{K\to\infty}\limsup_{n\to\infty}
P\left\{\inf_{\gamma\in\Gamma_K}[S_n(\gamma)-S_n(\gamma_0)]\le0\right\}=0.
\end{equation}
Consequently $a_n(\hat\gamma-\gamma_0)=O_p(1)$.
\item Suppose that $\delta_n'x_t=n^{-\alpha}\delta_0'\tilde x_t$ with $\tilde x_t$ $\mathcal G$-measurable, and that, conditionally on $\mathcal G$, $\{u_t\}$ is \textit{i.i.d.}\ with mean zero and variance $\sigma^2$. Then (iii) implies (iv).
\end{enumerate}
\end{lemma}

The uniform bound \eqref{eq:outward} states that, within a fixed neighbourhood of $\gamma_0$ but beyond the local scale $K/a_n$, the criterion exceeds its value at $\gamma_0$, with probability approaching one as $K\to\infty$. Together with global separation, it localises the minimising plateaux and the two one-sided minima used for LR inference.
\begin{lemma}[Window occupation and local scores]\label{lem:local}\label{lem:score}
Let Assumptions~\ref{ass:q} and \ref{ass:x} hold. For parts (a)--(c), let $\psi:\R^d\to\R^k$ satisfy $\E\|\psi(x_t)\|^{r/2}<\infty$.
\begin{enumerate}[label=(\alph*),leftmargin=*]
\item $\sup_{\gamma\in\Gamma}\|n^{-1}\sum_t\psi(x_t)\1\{q_t\le\gamma\}-\mu\,\E\psi(x_t)\|\convp0$.
\item $\sup_{\gamma_1<\gamma_2,\ \gamma_1,\gamma_2\in\Gamma}\|n^{-1/2}\sum_t\psi(x_t)\1\{\gamma_1<q_t\le\gamma_2\}-(\gamma_2-\gamma_1)\lt\,\E\psi(x_t)\|\convp0$.
\item If $h_n\to0$ and $\sqrt nh_n\to\infty$, then for every $\bar h>0$, $\sup_{h_n\le h\le\bar h}\|(\sqrt nh)^{-1}\sum_t\psi(x_t)\1\{\gamma_0<q_t\le\gamma_0+h\}-\lt\,\E\psi(x_t)\|\convp0$, and the same holds for windows $(\gamma_0-h,\gamma_0]$.

\item If $h_n\to0$ and $\sqrt nh_n\to\infty$, then, for every $\bar v<\infty$, $M_n\weak\sigma\lt^{1/2}D^{1/2}\mathcal W$ on $\mathbb D[-\bar v,\bar v]^d$, where $\mathcal W$ is a two-sided $d$-dimensional standard Brownian motion independent of $J_c$. The same holds with $h_n\equiv1$ on $\mathbb D[\underline\gamma-\gamma_0,\bar\gamma-\gamma_0]^d$.
\end{enumerate}
\end{lemma}

Parts (b)--(d) replace the stationary threshold density by random local time. Shrinking windows contain a diverging number of visits when $\sqrt nh_n\to\infty$, which permits a Gaussian score limit even though their fraction of the sample vanishes.

\begin{lemma}[Rare visits]\label{lem:cox}
Let Assumptions~\ref{ass:q} and \ref{ass:x} hold, and define the point process
\[
\mathcal N_n=\sum_{t=1}^n\delta_{(t/n,\,\sqrt n(q_t-\gamma_0),\,x_t,\,u_t)},
\]
where $\delta_y$ denotes the Dirac measure at $y$.
\begin{enumerate}[label=(\alph*),leftmargin=*]
\item For every $K<\infty$, the restriction of $\mathcal N_n$ to $[0,1]\times[-K,K]\times\R^{d+1}$ converges jointly with \eqref{eq:FCLT} to a Cox process directed by $\Lt(ds,0)\,dv\,P_x(dx)P_u(du)$.
\item For each fixed $K>0$, a constant $C$ independent of $n$ and $M$ satisfies
\[
P\{\exists t:\ |\sqrt n(q_t-\gamma_0)|\le K,\ \|x_t\|+|u_t|>M\}
\le CK\,P(\|x_t\|+|u_t|>M)\longrightarrow0
\]
as $M\to\infty$, uniformly in $n$.
\end{enumerate}
\end{lemma}

At the fixed-effect scale, only finitely many observations remain near the boundary. Their individual regression losses therefore survive as jumps. Part (b) permits the marked limit to be applied to those quadratic losses by truncation.

\begin{lemma}[Endogenous error sums]\label{lem:endogsums}
Let Assumptions~\ref{ass:par} and \ref{ass:endog} hold, and let $b(\gamma)$, $X_\gamma'u$ and $M_n$ be formed with $u_t=\varrho\eta_t+e_t$.
\begin{enumerate}[label=(\alph*),leftmargin=*]
\item The feasibility, inverse-matrix, full-regime-score and window-design orders of Lemma~\ref{lem:orders}(c) hold. The window score and profiling remainder satisfy $\sup_{\gamma\in\Gamma}\|b(\gamma)\|=O_p(n^{1/4}\sqrt{\log n})$ and $\sup_{\gamma\in\Gamma}|R_{5n}(\gamma)|=O_p(n^{-1/4}\sqrt{\log n})$. The first two orders in \eqref{eq:rho} are unchanged, and all three quantities are $o_p(1)$.
\item If $h_n\to0$, $\sqrt nh_n\to\infty$ and $h_n\log n\to0$, then $M_n\weak\sigma\lt^{1/2}D^{1/2}\mathcal W$ on $\mathbb D[-\bar v,\bar v]^d$, with $\sigma^2=\varrho^2\sigma_\eta^2+\sigma_e^2$ and $\mathcal W$ as in Lemma~\ref{lem:score}(d).
\item If $0<\alpha<1/4$, condition (iv) of Lemma~\ref{lem:rate} holds.
\end{enumerate}
\end{lemma}

Endogeneity affects the local score through both a martingale and a predictable term. The latter has zero integral and becomes negligible over diminishing windows. The lemma also verifies the variance and outward error bounds needed to retain the baseline estimator and LR limits.

\begin{lemma}[Nonstationary slope regressors]\label{lem:localns}
Let Assumptions~\ref{ass:q} and \ref{ass:ns}(i)--(ii) hold. Jointly with \eqref{eq:FCLT} and $n^{-1/2}x_{[n\cdot]}\weak G$:
\begin{enumerate}[label=(\alph*),leftmargin=*]
\item $\sup_{\gamma\in\Gamma}\|n^{-2}\sum_tx_tx_t'\1\{q_t\le\gamma\}-M_\mu\|\convp0$.
\item $\sup_{\gamma_1<\gamma_2,\ \gamma_1,\gamma_2\in\Gamma}\|n^{-3/2}\sum_tx_tx_t'\1\{\gamma_1<q_t\le\gamma_2\}-(\gamma_2-\gamma_1)Q_\ell\|\convp0$.
\item If $h_n\to0$ and $\sqrt nh_n\to\infty$, then $\sup_{h_n\le h\le\bar h}\|(n\sqrt nh)^{-1}\sum_tx_tx_t'\1\{\gamma_0<q_t\le\gamma_0+h\}-Q_\ell\|\convp0$ for every $\bar h>0$, and the same holds for windows to the left of $\gamma_0$.
\item Under the conditions of (c), $\operatorname{sgn}(v)(\sqrt nh_n)^{-1/2}n^{-1/2}\sum_tx_tu_t\1\{q_t\in I(v,h_n)\}\weak\sigma Q_\ell^{1/2}\mathcal W(v)$ on $\mathbb D[-\bar v,\bar v]^d$, where $\mathcal W$ is a two-sided $d$-dimensional standard Brownian motion independent of $(G,J_c)$.
\item For every $K<\infty$, the point process $\sum_t\delta_{(t/n,\,\sqrt n(q_t-\gamma_0),\,n^{-1/2}x_t,\,u_t)}$ restricted to $[0,1]\times[-K,K]\times\R^{d+1}$ converges in distribution to a point process that, conditionally on $(G,J_c)$, is the image of a Poisson process on $[0,1]\times[-K,K]\times\R$ with mean measure $\Lt(ds,0)\,dv\,P_u(du)$ under $(s,v,u)\mapsto(s,v,G(s),u)$.
\item $P(\hat\Gamma=\Gamma)\to1$, $\varpi_n=O_p(n^{-2})$, $\varkappa_n=O_p(n)$, $\sup_{\gamma\in\Gamma}\|B(\gamma)\|=O_p(n^{3/2})$ and $\sup_{\gamma\in\Gamma}\|b(\gamma)\|=O_p(n^{3/4})$. Consequently \eqref{eq:rho} holds.
\end{enumerate}
\end{lemma}

Information now weights the regressor path at the calendar times of threshold visits. This produces $Q_\ell=\int GG'\,d\Lt(\cdot,0)$, rather than $\lt\int GG'\,ds$. Once these weighted limits are established, the baseline profiling and argmin arguments apply.

\section{Proofs of the main results}\label{app:B}

We follow the projection, localisation and argmin arguments of \citet{Hansen2000} and \citet{Chen2015}, using Lemmas~\ref{lem:decomp}--\ref{lem:localns} to verify the changed occupation orders and random information. Fixed effects additionally require a marked jump limit and the smallest-minimiser mapping. We work on $\{\hat\Gamma=\Gamma\}$, whose probability tends to one.

\subsection{Proof of Theorem~\ref{thm:cons}}

(a) We apply the compact argmin argument of \citet[Theorem 2.1]{NeweyMcFadden1994} on the event that the random information $\lt\kappa_0$ is bounded away from zero. Put $\mathcal A_{jn}=n^{2\alpha-1/2}R_{jn}$, $j=1,\ldots,5$. Lemmas~\ref{lem:orders} and \ref{lem:local}(b) give, uniformly in $\gamma\in\Gamma$,
\begin{equation}\label{eq:T1}
\begin{aligned}
&(1)\quad\mathcal A_{1n}=n^{-1/2}\delta_0'B(\gamma)\delta_0\convp\lt\kappa_0|\gamma-\gamma_0|,\\
&(2)\quad\mathcal A_{2n}=O_p(n^{-1/2}),\qquad
(3)\quad\mathcal A_{3n}=O_p(n^{\alpha-1/4}),\\
&(4)\quad\mathcal A_{4n}=O_p(n^{\alpha-1/2}),\qquad
(5)\quad\mathcal A_{5n}=O_p(n^{2\alpha-3/4}).
\end{aligned}
\end{equation}
The dominant remainder vanishes because $\alpha<1/4$, proving \eqref{eq:globalSn}. For any $\epsilon,\zeta>0$, on $\{\lt\kappa_0>\zeta\}$ the normalised criterion is eventually positive outside $|\gamma-\gamma_0|<\epsilon$, with probability tending to one. Hence $\limsup_nP(|\hat\gamma-\gamma_0|\ge\epsilon)\le P(\lt\kappa_0\le\zeta)\to0$ as $\zeta\downarrow0$. This gives consistency and the global separation used below.

For the variance estimator, $R_{1n}-R_{2n}\ge0$ and the same bounds imply $S_n(\hat\gamma)-S_n(\gamma_0)=O_p(n^{1/4-\alpha})$. Also $S_n(\gamma_0)=u'u-u'P_{\gamma_0}u$, with $u'P_{\gamma_0}u\le\varpi_n\varkappa_n^2=O_p(1)$. Hence $\hat\sigma^2=n^{-1}u'u+o_p(1)\convp\sigma^2$.

(b) For $1/4\le\alpha<1/2$ the error term is at least of the order of the signal, so we normalise by $n^{\alpha-1/4}$ instead. Uniformly in $\gamma\in\Gamma$, by \eqref{eq:T1}(1), Lemma~\ref{lem:score}(d) with $h_n\equiv1$ and Lemma~\ref{lem:orders},
\[
\begin{aligned}
&n^{\alpha-1/4}R_{1n}=n^{1/4-\alpha}\mathcal A_{1n}\convp\1\{\alpha=1/4\}\lt\kappa_0|\gamma-\gamma_0|,\\
&n^{\alpha-1/4}R_{3n}=2\delta_0'M_n(\gamma-\gamma_0)\weak2\sigma(\lt\kappa_0)^{1/2}\mathcal B(\gamma-\gamma_0),\\
&n^{\alpha-1/4}\{R_{2n}+|R_{4n}|+|R_{5n}|\}=O_p(n^{-1/4-\alpha}+n^{-1/4}+n^{\alpha-1/2})=o_p(1),
\end{aligned}
\]
where $\mathcal B=\kappa_0^{-1/2}\delta_0'D^{1/2}\mathcal W$ is a two-sided standard Brownian motion independent of $J_c$. Conditionally on $J_c$, the limit is a Gaussian process with continuous paths and nondegenerate increments, so it has a unique minimiser on $\Gamma-\gamma_0$ \citep[Lemma 2.6]{KimPollard1990}, and the argmax continuous mapping theorem applies. By the law of the iterated logarithm the limit takes negative values arbitrarily close to zero on both sides, so the minimiser is not zero. \hfill$\square$

\subsection{Proof of Theorem~\ref{thm:dist}}

(a) Apply Lemma~\ref{lem:rate} with $\alpha=0$. Consistency and the profiling bounds give (i)--(ii). By Lemma~\ref{lem:cox}(a)--(b) and the truncation argument of \citet[Theorem 4.2]{Billingsley1968}, the limiting signal over a window of scaled width $K$ has conditional mean $K\lt\kappa_0$ and variance $K\lt\E(\delta_0'x_t)^4$. On $\{\lt\kappa_0>2\epsilon_0\}$, conditional Chebyshev bounds its probability of falling below $K\epsilon_0$ by $4\E(\delta_0'x_t)^4/(K\epsilon_0\kappa_0)$. Let $K\to\infty$, then $\epsilon_0\downarrow0$, to verify (iii). Lemma~\ref{lem:rate}(b) gives (iv), so the rate and \eqref{eq:outward} follow without the continuity condition.

On $|v|\le K$, bounded rare-visit counts and independent marks give $\|B\|,\|b\|=O_p(1)$. The profiling bounds imply
\begin{equation}\label{eq:T2Dn}
\begin{aligned}
\mathcal S_n(v)&=S_n(\gamma_0+v/\sqrt n)-S_n(\gamma_0)=\mathcal D_n(v)+O_p(n^{-1/2}),\\
\mathcal D_n(v)&=\sum_t[(\delta_0'x_t)^2\mp2\delta_0'x_tu_t]\1_t(I(v,n^{-1/2})),
\end{aligned}
\end{equation}
with the upper sign for $v>0$ and the lower sign for $v<0$, uniformly on that compact set. Lemma~\ref{lem:cox}(a)--(b), first with bounded marks and then with the truncation removed, gives joint $J_1$ convergence of $\mathcal D_n$ and its jump-count process. Limiting locations are distinct and avoid $\pm K$. Under the continuity condition, jumps are nonzero and plateau levels are distinct almost surely. Negligible profiling preserves the selected plateau, verifying the conditions of the smallest-argmax theorem of \citet[Theorem 3.2]{SeijoSen2011b}. Equation \eqref{eq:outward} and global separation from Theorem~\ref{thm:cons}(a) localise both end-points. Since $\mathcal D$ tends to infinity at both ends, let $K\to\infty$ to obtain the selected-endpoint limit.

(b) We closely follow \citet[Theorem 1]{Hansen2000} and the proof of \citet[Theorem 3.1]{Chen2015}, verifying the different rate and random information. Put $h_n=a_n^{-1}=n^{2\alpha-1/2}$. Lemma~\ref{lem:local}(c) gives $R_{1n}(\gamma_0\pm Kh_n)\convp K\lt\kappa_0$. Positivity of $\lt\kappa_0$ verifies the signal condition, and Lemma~\ref{lem:rate}(b) gives the error condition. Hence $a_n(\hat\gamma-\gamma_0)=O_p(1)$.

For $\mathcal S_n(v)=S_n(\gamma_0+vh_n)-S_n(\gamma_0)$, Lemmas~\ref{lem:local} and \ref{lem:orders} give, uniformly on compacts,
\begin{equation}\label{eq:T2}
\begin{aligned}
R_{1n}&\convp\lt\kappa_0|v|, & R_{3n}&=2\delta_0'M_n(v)\weak2\sigma(\lt\kappa_0)^{1/2}\mathcal B(v),\\
R_{2n}&=O_p(n^{2\alpha-1}), & R_{4n},R_{5n}&=O_p(n^{\alpha-1/2}).
\end{aligned}
\end{equation}
Thus $\mathcal S_n\weak\mathcal S$, where $\mathcal S(v)=\lt\kappa_0|v|-2\sigma(\lt\kappa_0)^{1/2}\mathcal B(v)$. Given $J_c$, the limit is continuous, coercive and has a unique minimiser. The tight rate and the argmin theorem \citep[Theorem 2.7]{KimPollard1990} therefore give convergence to its minimiser. Brownian scaling, conditionally on $\lt$, yields
\begin{equation}\label{eq:scaling}
\{\mathcal S(\lambda_\ell r)\}_{r\in\R}\overset d=
\{\sigma^2[|r|-2\mathcal B(r)]\}_{r\in\R},\qquad
\lambda_\ell=\frac{\sigma^2}{\lt\kappa_0}.
\end{equation}
Its minimiser is $\lambda_\ell\mathcal T$, with $\mathcal T$ independent of $\lt$. \hfill$\square$

\subsection{Proof of Theorem~\ref{thm:LR}}

(a) The criterion convergence, outward localisation and variance consistency give
\[
LR_n(\gamma_0)=-\hat\sigma^{-2}\min_v\mathcal S_n(v)
\convd-\sigma^{-2}\min_v\mathcal S(v)=\xi.
\]
Equation \eqref{eq:scaling} removes the random information. The two half-minima, with reversed sign and divided by $\sigma^2$, are independent exponential variables with mean 2 \citep{Hansen2000}, giving the stated distribution.

For $z>0$ and $H_n(z)=\operatorname{conv}\{\gamma\in\Gamma:LR_n(\gamma)\le z\}$, let $m_n^\pm$ denote the minimum of $S_n(\gamma)-S_n(\gamma_0)$ over the two closed halves of $\Gamma$ split at $\gamma_0$. Both are nonpositive. The hull contains $\gamma_0$ precisely when each half has an accepted point, so
\[
\{\gamma_0\in H_n(z)\}=\{|m_n^--m_n^+|\le\hat\sigma^2z\}.
\]
Equation \eqref{eq:outward}, together with global separation in Theorem~\ref{thm:cons}(a), localises both half-minima because each is at most zero. Compact convergence therefore gives $(-m_n^-/\sigma^2,-m_n^+/\sigma^2)\convd(E_-,E_+)$ for independent exponential variables with mean 2. Hence hull coverage tends to $P(|E_--E_+|\le z)=1-e^{-z/2}$. Under Remark~\ref{rem:het}, both half-minima acquire the factor $\vartheta^2$.

(b) For Gaussian fixed effects, consistency permits restriction to $|\gamma-\gamma_0|\le\epsilon_n$, where $\epsilon_n\downarrow0$ and $\sqrt n\epsilon_n\to\infty$. On these windows $\|B\|=O_p(\sqrt n\epsilon_n)$ and $\|b\|=O_p((\sqrt n\epsilon_n)^{1/2})$, so all profiling terms are $o_p(1)$. Let $t(1),t(2),\ldots$ index the observations with $q_t>\gamma_0$ in increasing order of $q_t$, let $d_i=\delta_0'x_{t(i)}$, and let $Z_k^R=\sum_{i\le k}(d_i^2-2d_iu_{t(i)})$. Define $Z_k^L$ in the same way from the observations with $q_t\le\gamma_0$, taken in decreasing order of $q_t$ and with $+2d_iu_{t(i)}$ in place of $-2d_iu_{t(i)}$, as in \eqref{eq:Dcal}. Then
\begin{equation}\label{eq:T3b}
\hat\sigma^2LR_n(\gamma_0)\le\max\{\sup_{k\ge0}(-Z_k^R),\sup_{k\ge0}(-Z_k^L)\}+o_p(1).
\end{equation}
Given $\mathcal G$, $\exp\{-Z_k^R/(2\sigma^2)\}$ and its left counterpart are independent mean-one nonnegative martingales. Ville's inequality \citep{Ville1939} bounds each crossing probability at level $\sigma^2z$ by $e^{-z/2}$. Their joint noncrossing probability is therefore at least $(1-e^{-z/2})^2$. Variance consistency gives the conservative bound, following the maximal-inequality argument of \citet[Theorem 3]{Hansen2000}.

(c) Lemma~\ref{lem:cox}(a)--(b) and \eqref{eq:T2Dn} give compact $J_1$ convergence to $\mathcal D$, using the same mark truncation as in Theorem~\ref{thm:dist}(a). The minimum is continuous at the limiting step paths even when plateau levels tie. Equation \eqref{eq:outward}, global separation and coercivity allow the compact interval to grow. Variance consistency gives the fixed-effect LR limit. \hfill$\square$

\subsection{Proof of Corollary~\ref{cor:slope}}

Let $H_j$ and $s_j$ be the design matrix and score in true regime $j$, and let $\tilde\beta_j-\beta_j=H_j^{-1}s_j$ denote the oracle estimator. Lemma~\ref{lem:local}(a) gives $H_1/n\convp\mu D$ and $H_2/n\convp(1-\mu)D$. Given $\mathcal G$, the errors are independent across the two regimes. The conditional Lindeberg CLT applies because $\max_t\|x_t\|^2/n=O_p(n^{2/r-1})\convp0$. As in the proof of Lemma~\ref{lem:local}(d), the conditional characteristic function identifies the joint oracle mixed-normal limits and standard normal studentised limits, with Gaussian drivers independent of $J_c$.

It remains to show oracle equivalence. Suppose $\hat\gamma>\gamma_0$, since the other case is symmetric. With $B=B(\hat\gamma)$ and $b=b(\hat\gamma)$,
\[
\hat\beta_1-\beta_1=(H_1+B)^{-1}(s_1+b-B\delta_n),\qquad
\hat\beta_2-\beta_2=(H_2-B)^{-1}(s_2-b).
\]
The threshold rate and Lemma~\ref{lem:local} give $\|B\|=O_p(n^{2\alpha})$ and $\|b\|=O_p(n^\alpha)$. At $\alpha=0$ the same orders follow from the bounded rare-visit counts and independent finite-moment marks. Since the inverse regime matrices are $O_p(n^{-1})$, the inverse-matrix identity gives
\[
\sqrt n\,\|\hat\beta_j-\tilde\beta_j\|
=O_p(n^{2\alpha-1}+n^{\alpha-1/2})=o_p(1),\qquad j=1,2.
\]
Also $(\hat H_j-H_j)/n=o_p(1)$ and $\hat\sigma^2\convp\sigma^2$. Slutsky's theorem transfers the oracle and studentised limits. \hfill$\square$

\subsection{Proof of Proposition~\ref{prop:test}}

Under $H_0$ the threshold variable affects $\Wald_n(\gamma)$ only through the order of the observations. Order them by $q_t$, let $x_{(i)}$ and $u_{(i)}$ be the regressors and the error of the $i$-th observation, and let $\mathcal H_k=n^{-1}\sum_{i\le k}x_{(i)}x_{(i)}'$ and $U_k=n^{-1/2}\sum_{i\le k}x_{(i)}u_{(i)}$. For $\gamma$ with $k$ observations at or below it, standard algebra \citep{Andrews1993,Hansen1996} gives
\[
\Wald_n(\gamma)=\frac{(U_k-\mathcal H_k\mathcal H_n^{-1}U_n)'(\mathcal H_k-\mathcal H_k\mathcal H_n^{-1}\mathcal H_k)^{-1}(U_k-\mathcal H_k\mathcal H_n^{-1}U_n)}{S_n(\gamma)/n},
\]
and $\gamma\in\Gamma_n^\pi$ if and only if $\pi n\le k\le(1-\pi)n$. Conditionally on $\mathcal Q$, the maximal inequality \eqref{eq:reorderingbounds} in the Online Appendix gives $\sup_r\|\mathcal H_{[nr]}-rD\|\convp0$, and conditionally on $\mathcal G$, $U_{[nr]}$ is a partial sum of independent vectors with covariance $\sigma^2\mathcal H_{[nr]}$, so $U_{[n\cdot]}\weak\sigma D^{1/2}\mathcal W$ for a $d$-dimensional standard Brownian motion $\mathcal W$ independent of $J_c$. Also $S_n(\gamma)/n\convp\sigma^2$ uniformly on $\Gamma_n^\pi$. The continuous mapping theorem gives the limit in (a), as in \citet{Andrews1993}. Under the alternative, the true regime consists of the first $k_0$ observations in the $q$-order, with $k_0/n\convp\mu\in(0,1)$ (Lemma~\ref{lem:local}(a) with $\psi=1$), so for $k=[rn]$ the numerator contains the term $n^{1/2-\alpha}\{(r\wedge\mu-r\mu)D\delta_0+o_p(1)\}+O_p(1)$, and $\Wald_n(\gamma)$ diverges at the rate $n^{1-2\alpha}$. \hfill$\square$

\subsection{Proof of Proposition~\ref{prop:endog}}

The criterion decomposition and design bounds are unchanged. Lemma~\ref{lem:endogsums}(a) changes only the score remainders in \eqref{eq:T1}, adding $\sqrt{\log n}$ to (3) and (5). They still vanish for $\alpha<1/4$, so the consistency, global separation and variance arguments of Theorem~\ref{thm:cons}(a) apply.

The signal condition in Lemma~\ref{lem:rate} is unchanged, and Lemma~\ref{lem:endogsums}(c) supplies its outward error bound. This gives the rate and \eqref{eq:outward}. At $h_n=n^{2\alpha-1/2}$, $h_n\log n\to0$, so Lemma~\ref{lem:endogsums}(b) gives the baseline local score with variance $\varrho^2\sigma_\eta^2+\sigma_e^2$. Its order $\sup_{|v|\le\bar v}\|b\|=O_p(n^\alpha)$ also gives the profiling remainders in \eqref{eq:T2}. The argmin and LR arguments of Theorems~\ref{thm:dist}(b) and \ref{thm:LR}(a) therefore apply. \hfill$\square$

\subsection{Proof of Theorem~\ref{thm:ns}}

We closely follow the normalised-regressor argument of \citet[Theorem 3.1]{Chen2015} and \citet[Appendix B]{ChenStengos2022}. The change is the information measure. Put $\tilde x_t=n^{-1/2}x_t$ and $I_Q=\delta_0'Q_\ell\delta_0>0$, so $\delta_n'x_t=n^{-\tau}\delta_0'\tilde x_t$. Lemma~\ref{lem:localns} establishes the required weighted occupation and score limits, and its part (f) gives the profiling orders \eqref{eq:rho}.

(a) The consistency calculation \eqref{eq:T1} applies with $\alpha=\tau$ and $\lt\kappa_0=I_Q$: its leading term is $n^{-3/2}\delta_0'B(\gamma)\delta_0\convp I_Q|\gamma-\gamma_0|$, and its dominant remainder is $O_p(n^{\tau-1/4})$. The random separation argument of Theorem~\ref{thm:cons}(a) proves consistency.

(b) For $h_n=n^{2\tau-1/2}$, Lemma~\ref{lem:localns}(c)--(d) gives the signal condition in Lemma~\ref{lem:rate} and the local criterion
\[
I_Q|v|-2\sigma I_Q^{1/2}\mathcal B(v).
\]
The conditional error bound is still Lemma~\ref{lem:rate}(b). The profiling terms have the orders in \eqref{eq:T2}, with $\alpha=\tau$, and vanish. The rate, argmin and LR arguments of Theorems~\ref{thm:dist}(b) and \ref{thm:LR}(a) therefore apply with $\lt\kappa_0$ replaced by $I_Q$.

(c) At $\tau=0$, Lemma~\ref{lem:localns}(e) gives the marked criterion $\mathcal D_G$. For the signal condition, its sum over a window of scaled width $K$ has conditional mean $KI_Q$ and variance $KV_Q$, where $V_Q=\int_0^1(\delta_0'G)^4d\Lt(\cdot,0)<\infty$ almost surely. On $\{I_Q\ge\epsilon,V_Q\le M\}$, conditional Chebyshev bounds the probability of a signal below $K\epsilon/2$ by $4M/(K\epsilon^2)$. First let $K\to\infty$, then remove this localisation. This proves the rate without requiring an unconditional fourth moment for $G$. The fixed-effect mapping arguments in Theorems~\ref{thm:dist}(a) and \ref{thm:LR}(c) now apply. Distinct plateau levels follow from continuity of $u_t$ and nonzero $\delta_0'G$ at local-time-selected visits. The Gaussian bound follows from the same exponential martingale as Theorem~\ref{thm:LR}(b), with $d_i=\delta_0'\tilde x_{t(i)}$. \hfill$\square$

\clearpage
% ---------------------------------------------------------------------------
% Online Appendix (Part II): own numbering S1, S1.1, (S.1), Table S1, Figure S1
% ---------------------------------------------------------------------------
\begin{center}
{\Large\bfseries Online Appendix to\\[2pt] ``Threshold Regression with a Nearly Integrated Threshold Variable''}
\end{center}
\makeatletter
\@removefromreset{equation}{section}
\makeatother
\setcounter{section}{0}\setcounter{subsection}{0}\setcounter{subsubsection}{0}
\setcounter{equation}{0}\setcounter{figure}{0}\setcounter{table}{0}
\renewcommand{\thesection}{S\arabic{section}}
\renewcommand{\thesubsection}{S\arabic{section}.\arabic{subsection}}
\renewcommand{\thesubsubsection}{S\arabic{section}.\arabic{subsection}.\arabic{subsubsection}}
\renewcommand{\theequation}{S.\arabic{equation}}
\renewcommand{\thefigure}{S\arabic{figure}}
\renewcommand{\thetable}{S\arabic{table}}
% unique hyperref anchors, so that S-numbered items do not clash with the paper's
\renewcommand{\theHsection}{S\arabic{section}}
\renewcommand{\theHsubsection}{S\arabic{section}.\arabic{subsection}}
\renewcommand{\theHsubsubsection}{S\arabic{section}.\arabic{subsection}.\arabic{subsubsection}}
\renewcommand{\theHequation}{S.\arabic{equation}}
\renewcommand{\theHfigure}{S\arabic{figure}}
\renewcommand{\theHtable}{S\arabic{table}}

This Online Appendix proves the lemmas in Appendix~\ref{app:A} of the paper, explains the cancellation mechanism behind the endogeneity result of Proposition~\ref{prop:endog}, and present additional simulation results. Section~\ref{sec:Sproofs} contains the proofs of lemmas. Section~\ref{sec:Sendog} develops the intuition with a heuristic example and a figure.  Section~\ref{sec:Smc} presents the additional Monte Carlo simulation results. Section~\ref{sec:Snumerical} records the numerical approximation for the nonstationary-slope limit.

\section{Proofs of the lemmas in Appendix A}\label{sec:Sproofs}

The proofs follow the order of the six lemmas in the paper. Lemmas~\ref{lem:decomp} and~\ref{lem:rate} use the usual least squares decomposition and signal--error comparison. The main change is established in Lemma~\ref{lem:local}: visits to a threshold window are governed by local time rather than a stationary density. Lemma~\ref{lem:cox} treats the smaller windows in which individual regression losses remain visible. Lemmas~\ref{lem:endogsums} and~\ref{lem:localns} then adapt these arguments to endogeneity and nonstationary slope regressors. Technical bounds are developed where they enter these proofs.

\subsection*{Proof of Lemma~\texorpdfstring{\ref{lem:decomp}}{A.1}
(Criterion decomposition and remainder bounds)}

We follow the profiled least squares argument in Lemma A.5 of
\citet{Chen2015}. We first derive the decomposition separately for
$\gamma>\gamma_0$ and $\gamma<\gamma_0$, then bound the remainder terms
and establish their uniform orders. The algebraic decomposition does not
depend on persistence, which enters through the occupation and score
bounds used in part (c).

To prove part (a), note that the two regime designs have disjoint support,
so $X_\gamma'(X-X_\gamma)=0$. Whenever $A_\gamma$ and $C_\gamma$ are
nonsingular, the projection matrix therefore takes the form
\begin{equation}\label{eq:oa-decomp-projection}
P_\gamma
=X_\gamma A_\gamma^{-1}X_\gamma'
 +(X-X_\gamma)C_\gamma^{-1}(X-X_\gamma)'.
\end{equation}
Write $a_\gamma=X_\gamma'u$ and $c_\gamma=(X-X_\gamma)'u$. At the true
threshold, the regression signal $X\beta+X_{\gamma_0}\delta_n$ lies in
the range of $P_{\gamma_0}$. Hence
\[
S_n(\gamma_0)=u'(I-P_{\gamma_0})u,\qquad
u'(I-P_\gamma)u-S_n(\gamma_0)=-R_{5n}(\gamma).
\]
This isolates the change in the fitted error component. The remaining
terms arise from assigning observations in $W(\gamma)$ to the wrong regime.

Consider first $\gamma>\gamma_0$. Since
$X_{\gamma_0}=X_\gamma-X_W$, the true regression signal differs from
$X\beta+X_\gamma\delta_n$ by $-X_W\delta_n$. The matrix $I-P_\gamma$
annihilates both $X$ and $X_\gamma$, giving
\[
S_n(\gamma)
=(u-X_W\delta_n)'(I-P_\gamma)(u-X_W\delta_n).
\]
In this case, $X_\gamma'X_W=B$ and $(X-X_\gamma)'X_W=0$.
Substituting these relations into \eqref{eq:oa-decomp-projection} yields
\[
X_W'P_\gamma X_W=BA_\gamma^{-1}B,\qquad
X_W'P_\gamma u=BA_\gamma^{-1}a_\gamma.
\]
Expanding the residual sum of squares and combining these identities
with the expression for the fitted error component above, we obtain
\begin{equation}\label{eq:oa-decomp-right}
\begin{aligned}
S_n(\gamma)-S_n(\gamma_0)
={}&\underbrace{\delta_n'B\delta_n}_{R_{1n}}
 -\underbrace{\delta_n'BA_\gamma^{-1}B\delta_n}_{R_{2n}}
 -\underbrace{2\delta_n'b}_{R_{3n}}\\
 &+\underbrace{2\delta_n'BA_\gamma^{-1}a_\gamma}_{R_{4n}}
 -R_{5n}(\gamma).
\end{aligned}
\end{equation}

Now consider $\gamma<\gamma_0$. We have
$X_{\gamma_0}=X_\gamma+X_W$, so the corresponding residual sum of squares is
\[
S_n(\gamma)
=(u+X_W\delta_n)'(I-P_\gamma)(u+X_W\delta_n).
\]
The window observations now belong to the high regime at $\gamma$,
implying $X_\gamma'X_W=0$ and $(X-X_\gamma)'X_W=B$. Using these relations
in \eqref{eq:oa-decomp-projection} and expanding as before gives
\begin{equation}\label{eq:oa-decomp-left}
\begin{aligned}
S_n(\gamma)-S_n(\gamma_0)
={}&\delta_n'B\delta_n-\delta_n'BC_\gamma^{-1}B\delta_n
 +2\delta_n'b\\
 &-2\delta_n'BC_\gamma^{-1}c_\gamma-R_{5n}(\gamma).
\end{aligned}
\end{equation}
Comparing \eqref{eq:oa-decomp-right} and \eqref{eq:oa-decomp-left}
gives the stated decomposition, with $R_{3n}=-2\delta_n'b$ and
$R_{4n}=-2\delta_n'BC_\gamma^{-1}c_\gamma$ in the second case.
In both cases, $R_{1n}$ is the loss from misclassification, $R_{2n}$ is
the part absorbed by refitting the regime coefficients, $R_{3n}$ is the
local error contribution, and $R_{4n}$ is its interaction with slope
refitting. The term $R_{5n}$ records the change in the fitted part of the
regression error. This proves part (a).

For part (b), positive semidefiniteness of $B$ gives
\[
\|B\delta_n\|^2
=\delta_n'B^2\delta_n
\le\|B\|\delta_n'B\delta_n
=\|B\|R_{1n}(\gamma).
\]
Combining this inequality with the definitions of $\varpi_n$ and
$\varkappa_n$ yields, for either regime,
\begin{equation}\label{eq:oa-decomp-R24}
0\le R_{2n}\le\varpi_n\|B\|R_{1n},\qquad
|R_{4n}|
\le2\varpi_n\varkappa_n\|B\|^{1/2}R_{1n}^{1/2}.
\end{equation}

It remains to bound $R_{5n}$. To handle both threshold cases together,
set $s=\operatorname{sgn}(\gamma-\gamma_0)$ and abbreviate
$a_0=a_{\gamma_0}$ and $c_0=c_{\gamma_0}$. Then
\[
a_\gamma=a_0+sb,\quad c_\gamma=c_0-sb,\quad
A_\gamma=A_{\gamma_0}+sB,\quad C_\gamma=C_{\gamma_0}-sB.
\]
Applying the inverse-matrix identity to the last two relations gives
\begin{equation}\label{eq:oa-decomp-inverses}
\begin{aligned}
A_\gamma^{-1}-A_{\gamma_0}^{-1}
 &=-sA_{\gamma_0}^{-1}BA_\gamma^{-1},\\
C_\gamma^{-1}-C_{\gamma_0}^{-1}
 &=sC_{\gamma_0}^{-1}BC_\gamma^{-1}.
\end{aligned}
\end{equation}
We now substitute the score relations into the two quadratic forms in
$u'P_\gamma u$ and use \eqref{eq:oa-decomp-inverses} to compare them
with their values at $\gamma_0$. This gives
\begin{equation}\label{eq:oa-decomp-R5-expansion}
\begin{aligned}
R_{5n}={}&2s b'A_\gamma^{-1}a_0+b'A_\gamma^{-1}b
 -s a_0'A_{\gamma_0}^{-1}BA_\gamma^{-1}a_0\\
 &-2s b'C_\gamma^{-1}c_0+b'C_\gamma^{-1}b
 +s c_0'C_{\gamma_0}^{-1}BC_\gamma^{-1}c_0.
\end{aligned}
\end{equation}
For each regime, the sum of the absolute values of its three
contributions in \eqref{eq:oa-decomp-R5-expansion} is bounded by
$2\varpi_n\varkappa_n\|b\|
+\varpi_n\|b\|^2
+\varpi_n^2\varkappa_n^2\|B\|$.
Adding the two regime bounds yields
\begin{equation}\label{eq:oa-decomp-R5-bound}
|R_{5n}|
\le4\varpi_n\varkappa_n\|b\|
 +2\varpi_n\|b\|^2
 +2\varpi_n^2\varkappa_n^2\|B\|.
\end{equation}
Together, \eqref{eq:oa-decomp-R24} and \eqref{eq:oa-decomp-R5-bound}
establish part (b).

Finally, we prove part (c) by establishing the orders of the quantities
appearing in these bounds. Lemma~\ref{lem:local}(a) gives, uniformly
over $\Gamma$,
\[
n^{-1}A_\gamma\convp\mu D,\qquad
n^{-1}C_\gamma\convp(1-\mu)D.
\]
Both limits are positive definite almost surely, implying
$\varpi_n=O_p(n^{-1})$. The regime counts are also of order $n$.
Together with the uniform positive definiteness, this ensures that the
fixed minimum-size and rank restrictions hold throughout $\Gamma$
with probability tending to one.

We next bound the regime scores. Conditional on $\mathcal G$, ordering
observations by $q_t$ gives independent centred score increments
$x_tu_t$. Doob's inequality bounds the conditional second moment of
their maximum partial sum by $4\sigma^2\sum_t\|x_t\|^2$. Since this sum
is $O_p(n)$, we obtain $\varkappa_n=O_p(n^{1/2})$. Applying the same
argument only to observations with $q_t\in\Gamma$ gives
\begin{equation}\label{eq:oa-decomp-window-score}
\E\!\left[
 \sup_{\gamma\in\Gamma}\|b(\gamma)\|^2
 \,\middle|\,\mathcal G
\right]
\le C\sigma^2\sum_t\|x_t\|^2\1\{q_t\in\Gamma\}
=O_p(n^{1/2}),
\end{equation}
where the last order follows from Lemma~\ref{lem:local}(b).
Conditional Markov's inequality applied to
\eqref{eq:oa-decomp-window-score}, localised on a bounded value of the
normalised right-hand side, yields
$\sup_\Gamma\|b\|=O_p(n^{1/4})$. The same occupation bound, together
with positive semidefiniteness of the window moment matrices, gives
$\sup_\Gamma\|B\|=O_p(n^{1/2})$.

Substituting these orders into \eqref{eq:oa-decomp-R5-bound}, its three
terms have orders $n^{-1/4}$, $n^{-1/2}$ and $n^{-1/2}$, respectively.
The factors in \eqref{eq:oa-decomp-R24} similarly satisfy
\[
\sup_\Gamma\varpi_n\|B\|=O_p(n^{-1/2}),\qquad
\sup_\Gamma\varpi_n\varkappa_n\|B\|^{1/2}=O_p(n^{-1/4}).
\]
Together with the bound for $R_{5n}$, these establish part (c).
\hfill$\square$

\subsection*{Proof of Lemma~\texorpdfstring{\ref{lem:rate}}{A.2}
(Rate of convergence)}

We follow the localisation arguments in the proof of Theorem 1 of
\citet{Hansen2000} and Theorem 3.1 of \citet{Chen2015}. Beyond the proposed
local scale, the loss from misclassification must dominate the error score
and the gain from refitting the slopes. Part (a) establishes this comparison
under condition (iv). Part (b) then verifies that condition by ordering
observations according to their threshold values and applying a maximal
inequality to the resulting error sums.

To prove part (a), first consider thresholds above $\gamma_0$ and put
$\Gamma_K^+=[\gamma_0+K/a_n,\gamma_0+\bar h]$. On the signal event in
(iii), monotonicity gives $R_{1n}(\gamma)\ge\epsilon_0K$ throughout
$\Gamma_K^+$. Combining this lower bound with
Lemma~\ref{lem:decomp}(b) and condition (ii) yields
\begin{equation}\label{eq:rate-profile-relative}
\sup_{\gamma\in\Gamma_K^+}
\frac{R_{2n}(\gamma)+|R_{4n}(\gamma)|+|R_{5n}(\gamma)|}
     {R_{1n}(\gamma)}
=o_p(1)
\end{equation}
for each fixed $K$. Thus the profiling terms are negligible relative to
the misclassification loss. Outside the error event in (iv), the criterion
decomposition in Lemma~\ref{lem:decomp}(a), together with
\eqref{eq:rate-profile-relative}, implies that
$S_n(\gamma)-S_n(\gamma_0)>0$ uniformly over $\Gamma_K^+$ with probability
tending to one.

First choose $K$ large to control the exceptional events in (iii) and
(iv), and then choose $n$ large to control the profiling terms. This
proves the uniform outward bound on the right. Repeating the argument
for thresholds below $\gamma_0$ gives the corresponding left-hand bound.
Consistency excludes estimates outside the fixed neighbourhood, while
these outward bounds exclude estimates beyond $K/a_n$ within it.
Together, they establish the stated rate.

For part (b), we verify condition (iv) by expressing the signal and error
score in the same threshold order. Order observations above $\gamma_0$
by increasing $q_t$, write their indices as $t(i)$, and define
\begin{equation}\label{eq:clockdef}
\Theta_k=\sum_{i\le k}(\delta_0'\tilde x_{t(i)})^2,\qquad
s_k=\sum_{i\le k}\delta_0'\tilde x_{t(i)}u_{t(i)}.
\end{equation}
If $k(\gamma)$ counts observations in $(\gamma_0,\gamma]$, these
definitions give
\begin{equation}\label{eq:clock}
R_{1n}(\gamma)=n^{-2\alpha}\Theta_{k(\gamma)},\qquad
|R_{3n}(\gamma)|=2n^{-\alpha}|s_{k(\gamma)}|.
\end{equation}
Equation~\eqref{eq:clock} reduces the required comparison to a bound on
$|s_k|/\Theta_k$. Conditional on $\mathcal G$, the ordering and regressor
weights are fixed, and $s_k$ has independent centred increments with
cumulative variance $\sigma^2\Theta_k$. We can therefore apply the
inequality of \citet{HajekRenyi1955} with the nonincreasing weights
$\Theta_k^{-1}$. Using
$\sum_{k>m}(\Theta_k-\Theta_{k-1})/\Theta_k^2\le\Theta_m^{-1}$ gives
\begin{equation}\label{eq:HR}
P\left(
 \sup_{k\ge m}\frac{|s_k|}{\Theta_k}\ge\epsilon
 \,\middle|\,\mathcal G
\right)
\le\frac{2\sigma^2}{\epsilon^2\Theta_m},
\qquad \Theta_m>0.
\end{equation}

Take $m=k(\gamma_0+K/a_n)$ and $\epsilon=n^{-\alpha}/4$.
By \eqref{eq:clock}, the error event in (iv) is contained in the event
on the left-hand side of \eqref{eq:HR}. On
$\{\Theta_m\ge\epsilon_0Kn^{2\alpha}\}$, its conditional probability is
therefore at most $32\sigma^2/(\epsilon_0K)$. This bound tends to zero
as $K\to\infty$, while condition (iii) controls the probability of the
complementary signal event. Hence condition (iv) holds on the right.
Ordering observations below $\gamma_0$ in the opposite direction gives
the same bound on the left, completing the proof.
\hfill$\square$

\subsection*{Proof of Lemma~\texorpdfstring{\ref{lem:local}}{A.3}
(Window occupation and local scores)}

We first establish occupation estimates for the threshold process under
Assumption~\ref{ass:q}. We then attach the stationary regressor weights
to prove parts (a)--(c). For part (d), the weighted counts identify the
conditional variance of the local score, allowing us to apply a conditional
central limit argument. The threshold-only estimates will also be used in
the two extensions.

\paragraph{Conditional density estimates.}
We begin with density bounds for the linear-process array. These bounds
describe how the conditional density spreads out and becomes locally flat
as the prediction horizon increases, and provide the moment estimates
needed below.

Fourier inversion gives boundedness and Lipschitz continuity of $f$.
To control its local envelope, the triangular area below a Lipschitz
density gives
$f^\sharp(y)\le C P(|\eta_t-y|\le R+1)^{1/2}$
for a fixed $R$ depending only on the bound and the Lipschitz constant
of $f$. Summing over unit intervals and using the $p$th moment then gives
\[
\int(1+|y|)^j f^\sharp(y)\,dy<\infty,\qquad 0\le j<(p-1)/2.
\]

We next derive the corresponding conditional bounds for the threshold
process. Conditional on $\F_s$, write
\[
q_t=r_{t,s}+\sum_{i=s+1}^t a_{t,i}\eta_i,\qquad
r_{t,s}\in\F_s,\qquad
a_{t,i}=\sum_{j=0}^{t-i}\rho_n^{t-i-j}\varphi_j.
\]
The coefficients are uniformly bounded. Since
$\sum_jj|\varphi_j|<\infty$ and
$\varphi=\sum_j\varphi_j\ne0$, all but a fixed number $m_0$ of them
are bounded away from zero for large $n$. Thus
$\sigma_{t,s}^2=\sum_{i=s+1}^t a_{t,i}^2\asymp t-s$
for sufficiently long gaps.

The characteristic function of the normalised sum
$\sigma_{t,s}^{-1}\sum_{i=s+1}^ta_{t,i}\eta_i$ is
$\prod_i\psi_\eta(a_{t,i}\lambda/\sigma_{t,s})$, where
$\psi_\eta$ is the characteristic function of $\eta_t$.
For the low-frequency part, its modulus is bounded by
$e^{-\lambda^2/C}$. In the remaining region, all but the same $m_0$
factors are bounded by some $\bar r<1$. Retaining one integrable factor
in that region yields
\[
\int(1+|\lambda|)
\left|\prod_i\psi_\eta(a_{t,i}\lambda/\sigma_{t,s})\right|d\lambda
\le C+C\bar r^{\,t-s-m_0-1}(\sigma_{t,s}+\sigma_{t,s}^2)
\int(1+|v|)|\psi_\eta(v)|dv\le C.
\]
Applying Fourier inversion to this bound and rescaling gives a conditional
density bounded by $C(t-s)^{-1/2}$, with Lipschitz constant bounded by
$C(t-s)^{-1}$. The coefficients of $z_{t+1}$ have the same form.
Integrating against these densities therefore gives, for $g\ge0$,
\begin{equation}\label{eq:densitybounds}
\E[g(q_t)\mid\F_s]\le\frac{C\|g\|_1}{\sqrt{t-s}},\qquad
\E[g(z_t)\mid\F_s]\le\frac{C\|g\|_*}{\sqrt{t-s}}.
\end{equation}
For short gaps, the same bounds follow from the bounded one-step density
for $q_t$ and the inequality $g(z_t)\le\|g\|_\infty$.
If $g$ is signed and integrable, with zero integral and finite first
absolute moment, subtracting the conditional density at zero gives
the sharper bound
\begin{equation}\label{eq:signeddensity}
|\E[g(z_t)\mid\F_s]|
\le\frac{C}{t-s}\int|y||g(y)|dy
\end{equation}
for sufficiently long gaps.

These conditional estimates also control repeated visits to a window.
Iterating \eqref{eq:densitybounds} over ordered time indices gives
\begin{equation}\label{eq:occupationmoments}
\E\Lambda_n(g)
\le C(\|g\|_1+n^{-1/2}\|g\|_\infty),\qquad
\E N(A)^k
\le C_k\{\sqrt n|A|+(\sqrt n|A|)^k\},
\end{equation}
where $g\ge0$ is bounded and integrable, $A$ is an interval and $k\ge1$.
For the second inequality, iteration first gives the factorial moments,
from which the ordinary moments follow. Applying the same argument
to products of the kernel gives
\[
\sum_{t_1<\cdots<t_j}\E\prod_{i=1}^j g(\gamma_0-z_{t_i})
\le\frac{(C\sqrt n\|g\|_*)^j}{\Gamma(j/2+1)},
\]
where $\Gamma(\cdot)$ in this display is the gamma function.
To obtain an exponential moment bound, expand
$\prod_t[1+\bar\lambda_n g(\gamma_0-z_t)]$, with
$\bar\lambda_n=(e^{\lambda\|g\|_\infty/\sqrt n}-1)/\|g\|_\infty$,
and sum the resulting convergent series. For every fixed $\lambda>0$,
this yields
\begin{equation}\label{eq:occupationexp}
\sup_{n\ge n_\lambda}\E e^{\lambda\Lambda_n(g)}<\infty.
\end{equation}
The bound is uniform over nonnegative kernels with uniformly bounded
$\|g\|_*$. These moment estimates will control interpolation between
threshold values below and the change of measure in
Lemma~\ref{lem:cox}.

\paragraph{Occupation approximation.}
We now relate the discrete occupation sums to local time, adapting the
smoothing argument for nonlinear cointegrating functionals
\citep{ParkPhillips2001,Wang2015}. Let $g$ be bounded and integrable,
and let $\vartheta$ be piecewise continuous on $[0,1]$, with finitely
many discontinuities and one-sided limits. Put
\[
\Psi_n=n^{-1/2}\sum_t\vartheta(t/n)g(\gamma_0-z_t),\qquad
F_\epsilon(X)
=\Big(\int g\Big)\int_0^1\vartheta(s)\phi_\epsilon(X(s))ds,
\]
where $\phi_\epsilon$ is the $N(0,\epsilon^2)$ density.
We first compare these quantities when $g$ is supported on $[-K,K]$.
Define
\[
\check g_n(y)
=g(\gamma_0-y)-\Big(\int g\Big)n^{-1/2}\phi_\epsilon(y/\sqrt n),
\qquad
\Delta_n=n^{-1/2}\sum_t\vartheta(t/n)\check g_n(z_t).
\]
By construction, $\check g_n$ has zero integral. It also satisfies
$\|\check g_n\|_1\le2\|g\|_1$,
$\|\check g_n\|_\infty\le\|g\|_\infty+C/(\epsilon\sqrt n)$,
and
$\int|y-\gamma_0||\check g_n(y)|dy\le C(1+\epsilon\sqrt n)$.
We can therefore combine \eqref{eq:densitybounds} and
\eqref{eq:signeddensity} to bound each long-gap conditional mean by
\[
|\E[\check g_n(z_t)\mid\F_s]|
\le C\min\{(t-s)^{-1/2},(1+\epsilon\sqrt n)/(t-s)\}.
\]
Short gaps are controlled by the bounded-density or bounded-function
estimates. Splitting the second-moment expansion at
$t-s=\epsilon^2n$ then gives
\[
\E\Delta_n^2
\le C\{(1+\log n)/\sqrt n
+\epsilon(1+\log\epsilon^{-1})\},
\qquad n\ge\epsilon^{-2}.
\]

It remains to compare the smoothed sum with $F_\epsilon(X_n)$.
Replacing $z_t$ by $q_t$ in that sum costs at most
$C\epsilon^{-2}n^{-3/2}\sum_t|\eta_t|=o_{L^1}(1)$.
The resulting sum is a Riemann sum for $F_\epsilon(X_n)$.
Combining this approximation with the bound for $\Delta_n$ yields
\begin{equation}\label{eq:occupationsmoothing}
\lim_{\epsilon\downarrow0}\limsup_n
\E|\Psi_n-F_\epsilon(X_n)|=0.
\end{equation}

We remove the compact-support restriction by truncation.
For general bounded integrable $g$, let $g_K=g\1_{[-K,K]}$.
Equation~\eqref{eq:occupationmoments} bounds the occupation error by
$C\|g-g_K\|_1+o(1)$, and the density bound gives the same error for the
smoothed term. Letting $K\to\infty$ therefore extends
\eqref{eq:occupationsmoothing} to such $g$.

With the smoothing error controlled, the functional CLT and the
occupation formula give, jointly with $X_n$,
\begin{equation}\label{eq:kerneloccupation}
\Psi_n\convp
\Big(\int g\Big)\int_0^1\vartheta(s)\Lt(ds,0).
\end{equation}
The same limit holds with $q_t$ in place of $z_t$.
For $g\ge0$ and $\vartheta=\1_{[0,r]}$, monotonicity and continuity
of local time make convergence uniform in $r$.

We next obtain the count limits needed for parts (a) and (b).
Squeezing the full-regime counts between the events
$q_t/\sqrt n\le\pm\epsilon$ and then letting $\epsilon\downarrow0$ gives
\begin{equation}\label{eq:unweightedcounts}
\sup_{\gamma\in\Gamma}
|n^{-1}N((-\infty,\gamma])-\mu|\convp0.
\end{equation}
For bounded windows, applying \eqref{eq:kerneloccupation} to interval
indicators on a finite grid gives the corresponding uniform count
limit in part (b).

\paragraph{Shrinking windows and innovation weights.}
Part (c) requires the occupation approximation to hold uniformly down
to width $h_n$. We establish this next, allowing for the innovation
weights needed in Lemma~\ref{lem:endogsums}.
Let $g\ge0$ satisfy $g(e)\le C(1+|e|^j)$,
$j<(p-1)/2$, with $gf$ continuous almost everywhere.
For $A_h=(\gamma_0,\gamma_0+h]$, set
$Y(h)=\sum_tg(\eta_t)\1_t(A_h)$ and let $k=gf$ denote the weighted
density. We first control the predictable sum,
\[
\Lambda(h)
=\sum_t\E[g(\eta_t)\1_t(A_h)\mid\F_{t-1}]
=\sqrt nh\,\Lambda_n(\bar k_h),\qquad
\bar k_h(y)=h^{-1}\int_0^hk(y+s)ds.
\]
The density estimates above give an integrable envelope for $k$.
Its local translation modulus tends to zero almost everywhere and
in $L^1$. Combining \eqref{eq:occupationmoments} with
\eqref{eq:kerneloccupation} therefore gives
$\Lambda_n(\bar k_h)\convp\lt\int k$ uniformly for small $h$.
For $h$ bounded away from zero, uniformity follows from a finite grid.

We then control the martingale difference $Y(h)-\Lambda(h)$.
Its second moment is at most $C(\sqrt nh+1)$, so Chebyshev's
inequality on the geometric grid $h_j=(1+\epsilon)^jh_n$ gives
\[
P\left(
\max_j\frac{|Y(h_j)-\Lambda(h_j)|}{\sqrt nh_j}>\epsilon
\right)
\le\frac{C_\epsilon}{\sqrt nh_n}\to0.
\]
Monotonicity brackets intermediate widths between neighbouring grid
values. Combining this grid bound with the predictable-sum limit
and letting $\epsilon\downarrow0$ proves
\begin{equation}\label{eq:innovationoccupation}
\sup_{h_n\le h\le\bar h}
\left|\frac{Y(h)}{\sqrt nh}-\lt\,\E g(\eta_t)\right|\convp0.
\end{equation}
The same argument applies to windows below $\gamma_0$.
Applying the smoothing approximation on time blocks also gives
time-weighted versions. In particular, these estimates hold uniformly
in the calendar-time endpoint.

\paragraph{Stationary weights after ordering.}
Having established the threshold-only limits, we now use
Assumption~\ref{ass:x} to attach the regressor weights.
Conditioning on $\mathcal Q$ fixes the indices selected or ordered
by $q_t$.

We first derive a bound for centred weights after such selection.
For a centred stationary sequence $\tilde\zeta_t$, independent of
$\mathcal Q$, with an exponentially decreasing mixing coefficient
and a finite $(2+\iota)$th moment, put
$S_k=\sum_{i\le k}\tilde\zeta_{t_i}$ for distinct selected indices.
The mixing covariance inequality
\citep[Section 1.2]{Doukhan1994} gives
\[
\E[(S_k-S_j)^2\mid\mathcal Q]
\le(k-j)\sum_{m\in\mathbb Z}
C\alpha_\zeta(|m|)^{\iota/(2+\iota)}
\le C(k-j).
\]
Applying the maximal inequality of \citet{Moricz1976} to this bound,
followed by a dyadic decomposition, yields
\begin{equation}\label{eq:reorderingbounds}
\E[\max_{k\le N}S_k^2\mid\mathcal Q]
\le CN(\log2N)^2,\qquad
P\left(
\sup_{k\ge m}\frac{|S_k|}{k}>\epsilon
\,\middle|\,\mathcal Q
\right)
\le\frac{C(\log2m)^2}{\epsilon^2m}.
\end{equation}
For the second inequality, sum
$C\epsilon^{-2}m^{-1}2^{-j}[\log(2m)+j]^2$
over the dyadic blocks. These bounds use the dependence in the
original time order and do not require the reordered sequence
itself to be stationary or mixing.

\emph{Parts (a)--(c).}
Write $\psi(x_t)=\E\psi(x_t)+\tilde\psi_t$.
The contribution of the mean follows from the count limits above.
For the centred contribution, \eqref{eq:reorderingbounds} gives a
full-sample supremum of $O_p(\sqrt n\log n)=o_p(n)$ and a
bounded-window supremum of $O_p(n^{1/4}\log n)=o_p(\sqrt n)$.
Together, these establish parts (a) and (b).

For part (c), the same bound controls all shrinking windows through
the count in the smallest window:
\[
\sup_{h_n\le h\le\bar h}
\frac{\|S_{N(A_h)}\|}{N(A_h)}
\le
\sup_{k\ge N(A_{h_n})}\frac{\|S_k\|}{k}\convp0,
\]
because $N(A_{h_n})/(\sqrt nh_n)\convp\lt>0$.
Combining this negligible centred contribution with
\eqref{eq:innovationoccupation} proves part (c).

Now, we show part (d). We first establish the finite-dimensional limit and then prove
tightness. Put $\nu_n=\sqrt nh_n$.
Conditional on $\mathcal G$, the score increments ordered by $q_t$
are independent and centred. Part (c), applied at width
$\varepsilon h_n$, identifies their conditional covariance uniformly
for $\varepsilon\le|v|\le\bar v$.
Monotonicity of nonnegative quadratic forms extends this result to
$|v|<\varepsilon$ as $\varepsilon\downarrow0$, giving the uniform
covariance limit $\sigma^2|v|\lt D$.

To verify the Lindeberg condition, independence from $q_t$ and
\eqref{eq:occupationmoments} give, for some $\iota>0$,
\[
\E\left[
\nu_n^{-1}\sum_t
\|x_tu_t\|^{2+\iota}\1_t(I(\bar v,h_n))
\right]\le C.
\]
The Lindeberg sum is therefore $O_p(\nu_n^{-\iota/2})$.
The conditional independent-array central limit argument then gives
the finite-dimensional limit
$\sigma\lt^{1/2}D^{1/2}\mathcal W$, jointly with $X_n$.
Its two sides are conditionally independent because they involve
disjoint sets of error increments.

It remains to establish tightness.
Conditional Rosenthal bounds and \eqref{eq:occupationmoments}
bound fourth moments of increments over an interval of length
$\Delta$ by $C(\Delta^2+\Delta/\nu_n)$.
On a grid of mesh $\nu_n^{-1}$, we apply Billingsley's maximal
inequality \citep[Theorem 12.2]{Billingsley1968}.
To control the oscillation between grid points, note that within
a cell of threshold width $n^{-1/2}$, the absolute score sum
divided by $\sqrt{\nu_n}$ has fourth moment $O(\nu_n^{-2})$.
A union bound over $O(\nu_n)$ cells completes the tightness argument.
For $h_n\equiv1$, the same argument applies, using part (b) to
identify the covariance clock.
\hfill$\square$

\subsection*{Proof of Lemma~\texorpdfstring{\ref{lem:cox}}{A.4} (Rare visits)}
At this scale, a bounded window in the rescaled threshold coordinate contains only finitely many visits in the limit. Our argument builds on the local-time-directed Poisson limits in Proposition 1.3 of \citet{BrutscheRohde2026} and Theorem 2.1 of \citet{BrutscheHahnRohde2026}, with additional arguments for the regression marks and the nearly integrated linear-process array.  Below, we first control large marks in part (b). For part (a), we integrate out the marks and prove joint convergence of the arrivals and the rescaled path using Laplace functionals. Since an innovation producing a visit also affects the future path, we use a stopped change of measure to bound the likelihood ratio’s second moment and coupling to show that the diffusion limit is unchanged. The occupation approximation gives the local-time intensity. We then remove the stop.

\emph{Part (b): large marks.} Put $\mathcal V_n=\{t:|\sqrt n(q_t-\gamma_0)|\le K\}$. Independence from the entire visit set and stationarity of the mark marginals give
\[
\E\sum_{t\in\mathcal V_n}\1\{\|x_t\|+|u_t|>M\}
=\E|\mathcal V_n|\,P(\|x_t\|+|u_t|>M)
\le CK\,P(\|x_t\|+|u_t|>M)
\]
by \eqref{eq:occupationmoments}. Markov's inequality proves (b). The count bound also gives tightness.

\emph{Part (a): joint arrivals and marks.} By Kallenberg (\citeyear{Kallenberg2002}, Theorem 16.16), it remains to show, for continuous $\chi\ge0$ with compact support in $[0,1]\times(-K,K)\times\R^{d+1}$ and bounded Lipschitz $\mathcal H\ge0$ on $\mathbb D[0,1]$,
\begin{equation}\label{eq:laplace}
\E[e^{-\mathcal N_n(\chi)}\mathcal H(X_n)]\to\E\Big[\exp\Big\{-\int_0^1\!\!\int\!\!\int(1-e^{-\chi(s,v,x,u)})P_x(dx)P_u(du)\,dv\,\Lt(ds,0)\Big\}\mathcal H(J_c)\Big].
\end{equation}
Put $v_t=\sqrt n(q_t-\gamma_0)$ and $\bar\chi(s,v)=-\log\int\!\!\int e^{-\chi(s,v,x,u)}P_x(dx)P_u(du)$. If the visits are at least $m$ periods apart, the mixing inequality gives
\[
\left|\E[e^{-\mathcal N_n(\chi)}\mid\mathcal Q]-\exp\{-\sum_{t\in\mathcal V_n}\bar\chi(t/n,v_t)\}\right|
\le4|\mathcal V_n|\alpha_x(m).
\]
Equation \eqref{eq:densitybounds} bounds the expected number of closer pairs by $CmK^2n^{-1/2}$. Since $|\mathcal V_n|=O_p(1)$, letting $m\to\infty$ slowly integrates out the marks. The remaining issue is the arrival locations: the innovation producing a visit also moves the future path.

\paragraph{A stopped change of measure.}
We control that feedback by changing the innovation law only at boundary visits. This argument uses Assumption~\ref{ass:q} alone. Its bounds will also identify the innovation-score limit in Lemma~\ref{lem:endogsums}. Let $W_n\subset[\gamma_0-1/2,\gamma_0+1/2]$ be an interval, let $\Xi_{n,t}(e)$ be predictable with $|\Xi_{n,t}(e)|\le b_n\1\{z_t+e\in W_n\}$ and $b_n\le\bar b$, and put
\[
\xi_t=\Xi_{n,t}(\eta_t),\qquad P_t=P(q_t\in W_n\mid\F_{t-1}),\qquad
\mathcal L_n^T=\prod_{t\le T}\frac{e^{\xi_t}}{\E(e^{\xi_t}\mid\F_{t-1})}.
\]
Choose a stopping time $T$ for which $b_n^2\sum_{t\le T}P_t\le\mathcal K$. Since $\xi_t$ is bounded and vanishes outside the window,
\[
\frac{\E(e^{2\xi_t}\mid\F_{t-1})}{\E(e^{\xi_t}\mid\F_{t-1})^2}\le e^{C_0b_n^2P_t}.
\]
Multiplication and Cauchy--Schwarz give, under $dQ_n^T=\mathcal L_n^T dP$ and for any random variable $Y_n$,
\begin{equation}\label{eq:tilttransfer}
\E(\mathcal L_n^T)^2\le e^{C_0\mathcal K},\qquad
Q_n^T(|Y_n|>\epsilon)\le e^{C_0\mathcal K/2}P(|Y_n|>\epsilon)^{1/2}.
\end{equation}
Here $C_0$ depends only on the fixed bound $\bar b$.

The conditional innovation density before $T$ is $\tilde f_t(e)=f(e)e^{\Xi_{n,t}(e)}/\E(e^{\xi_t}\mid\F_{t-1})$. Couple it recursively to independent innovations $\eta_t^0$ with density $f$, retaining the pre-sample innovations, and let $q_t^0$ denote the threshold path generated by $\{\eta_t^0\}$. A maximal coupling gives
\[
|\tilde f_t-f|\le Cf\{b_n\1_{W_n-z_t}+b_nP_t\},\qquad
\E(|\eta_t-\eta_t^0|\mid\text{past})\le Cb_n|W_n|\tilde g(\gamma_0-z_t),
\]
where $\tilde g(y)=(1+|y|)f^\sharp(y)$ is bounded and integrable by the density estimates. Equation \eqref{eq:tilttransfer} transfers tightness of $\Lambda_n(\tilde g)$ to $Q_n^T$, so $\sum_t|\eta_t-\eta_t^0|=O_p(b_n|W_n|\sqrt n)$. The bounded linear-process coefficients imply $\max_t|q_t-q_t^0|/\sqrt n=o_p(1)$ whenever $b_n|W_n|\to0$. Thus the path limit is unchanged. Equation \eqref{eq:tilttransfer} also transfers the smoothing approximation \eqref{eq:occupationsmoothing}. The path and kernel occupations therefore have the same joint limit under the stopped law, by the approximation theorem \citep[Theorem 4.2]{Billingsley1968}.

The stop can be removed. If $c_n>0$ is a deterministic sequence with $c_n\sqrt n|W_n|\le\bar b$ and
\[
T_{\mathcal K}=n\wedge\inf\{t\ge0:c_n\sum_{s\le t+1}P_s>\mathcal K\},
\]
then $P_t\le|W_n|f^\sharp(\gamma_0-z_t)$ and \eqref{eq:occupationexp} give, for every fixed $\lambda>0$,
\begin{equation}\label{eq:stoptail}
\limsup_nP(T_{\mathcal K}<n)\le C_\lambda e^{-\lambda\mathcal K}.
\end{equation}

\paragraph{Completing the arrival limit.}
Apply this change of measure with $W_n=[\gamma_0-K/\sqrt n,\gamma_0+K/\sqrt n]$, $\Xi_{n,t}(e)=-\bar\chi(t/n,\sqrt n(z_t+e-\gamma_0))$, $b_n=b_\chi:=\|\bar\chi\|_\infty$ and $c_n=1$. Let $T_\kappa=n\wedge\inf\{t:\sum_{s\le t+1}P_s>\kappa\}$, write $Q_n=Q_n^{T_\kappa}$, and put $p_t=1-\E(e^{\xi_t}\mid\F_{t-1})$. On $\{T_\kappa=n\}$, the likelihood identity yields
\[
\E[e^{-\sum_t\bar\chi(t/n,v_t)}\mathcal H(X_n)]
=\E_{Q_n}\Big[\prod_{t\le n}(1-p_t)\mathcal H(X_n)\Big]
+O(P(T_\kappa<n)+Q_n(T_\kappa<n)).
\]
Changing variables in the conditional arrival probability gives
\[
p_t=n^{-1/2}\Phi(t/n)f(\gamma_0-z_t)+n^{-1/2}r_t,\qquad
\Phi(s)=\int(1-e^{-\bar\chi(s,v)})dv,
\]
where $|r_t|\le2K\operatorname{osc}_{K/\sqrt n}f(\gamma_0-z_t)$ and $\operatorname{osc}_hf(y)=\sup_{|s|\le h}|f(y+s)-f(y)|$. The occupation bound makes $\E n^{-1/2}\sum_t|r_t|\to0$. Also $\max_tp_t\le Cn^{-1/2}$, so $\prod_t(1-p_t)=\exp\{-\sum_tp_t+o_p(1)\}$. The stopped occupation approximation gives
\[
(X_n,\sum_tp_t)\convd(J_c,\int_0^1\Phi(s)\Lt(ds,0))\quad\text{under }Q_n.
\]
The likelihood bound here uses $\mathcal K=b_\chi^2\kappa$. For any $\lambda>C_0b_\chi^2$, \eqref{eq:tilttransfer} and \eqref{eq:stoptail} imply
\[
Q_n(T_\kappa<n)\le e^{C_0b_\chi^2\kappa/2}P(T_\kappa<n)^{1/2}
\le C_\lambda^{1/2}e^{-(\lambda-C_0b_\chi^2)\kappa/2}\to0.
\]
The product is bounded by one. Remove the stop and use $1-e^{-\bar\chi}=\int\!\!\int(1-e^{-\chi})dP_xdP_u$ to obtain \eqref{eq:laplace}. \hfill$\square$

\subsection*{Proof of Lemma~\texorpdfstring{\ref{lem:endogsums}}{A.5} (Endogenous error sums)}
The endogenous score contains a centred fluctuation and a predictable conditional mean. The first has the same local-time variance as in the baseline model. The second cancels to first order because its kernel has zero integral. We separate these contributions before proving (a)--(c).

Let $\mu_x=\E x_t$. For an interval $A$, write
\begin{equation}\label{eq:endogdec}
\sum_tx_tu_t\1_t(A)=b^e(A)+\varrho b^1(A)+\varrho\mu_x\{m(A)+c(A)\},
\end{equation}
where
\[
\begin{aligned}
b^e(A)&=\sum_tx_te_t\1_t(A),& b^1(A)&=\sum_t(x_t-\mu_x)\eta_t\1_t(A),\\
m(A)&=\sum_t\{\eta_t\1_t(A)-\Upsilon_A(z_t)\},& c(A)&=\sum_t\Upsilon_A(z_t),
\end{aligned}
\]
and $\Upsilon_A(z)=\int_{A-z}ef(e)de$. Given $\mathcal G$, $b^e$ has independent centred increments in the $q$-order. Given $\mathcal Q$, $b^1$ does too, with variance weights $\eta_t^2(D-\mu_x\mu_x')$. In contrast, $m$ is a martingale in chronological time. The predictable term satisfies $\int\Upsilon_A=|A|\E\eta_t=0$.

\paragraph{Predictable cancellation.}
The zero-integral property removes the leading occupation term. We need a uniform bound on the remainder. Expanding the second moment of a bounded zero-integral kernel $k$ and using \eqref{eq:signeddensity} gives
\[
\E\Big[\sum_t k(z_t)\Big]^2
\le C\sqrt n\|k\|_*\Big\{\|k\|_\infty+\log(2n)\int(1+|y|)|k(y)|dy\Big\}.
\]
Diagonal and short-gap terms give the first term. For long gaps, sum $(t-s)^{-1}$ from \eqref{eq:signeddensity}, then sum $\E|k(z_s)|$ using \eqref{eq:densitybounds}.

For $k(e)=ef(e)$, the required integrability follows from the innovation moments and density estimates. For an i.i.d.\ regressor sequence independent of $\mathcal Q$, put $Z_n(\gamma)=\sum_tx_t\int_{\gamma_*-z_t}^{\gamma-z_t}ef(e)de$. It is absolutely continuous, vanishes at $\gamma_*$, and has derivative $Z_n'(a)=\sum_tx_tk(a-z_t)$. Independence and the preceding bound imply, uniformly for $a$ in a fixed compact set,
\[
\E\|Z_n'(a)\|^2
\le\E\|x_t\|^2\sum_t\E k(a-z_t)^2
+\|\E x_t\|^2\E\Big[\sum_tk(a-z_t)\Big]^2
\le C\sqrt n\log(2n).
\]
Cauchy--Schwarz therefore gives
\begin{equation}\label{eq:zeroenergy}
\E\sup_{\gamma_*<\gamma\le\gamma_*+h}\|Z_n(\gamma)\|^2
\le h\int_{\gamma_*}^{\gamma_*+h}\E\|Z_n'(a)\|^2da
\le C\sqrt n\log(2n)h^2.
\end{equation}
This is uniform in $\gamma_*\in\Gamma$ and $0<h\le\bar h$. Markov's inequality gives the corresponding probability bound. The left side is identical. Taking $x_t=1$ controls $c(A)$ below.

\emph{Part (a): uniform orders.} The design and feasibility statements are unchanged. Doob's inequality and the innovation-weighted occupation calculation \eqref{eq:innovationoccupation}, with $g(e)=e^2$, give $\sup_{A\subset\Gamma}(\|b^e(A)\|+\|b^1(A)\|)=O_p(n^{1/4})$. For $m$, Burkholder's inequality and the density and occupation bounds give
\begin{equation}\label{eq:m4}
\E m(A)^4\le C\{1+\sqrt n|A|+(\sqrt n|A|)^2\}.
\end{equation}
The conditional variance and fourth-moment kernels are $g_A(z)=\int_{A-z}e^2f(e)de$ and $\bar g_A(z)=\int_{A-z}e^4f(e)de$. The former has $\|g_A\|_*\le C|A|$, and the latter has integral $|A|\E\eta_t^4$. Finitely many short horizons contribute the constant in \eqref{eq:m4}.

Use the grid maximal argument in Lemma~\ref{lem:local}(d), with threshold cells of width $n^{-1/2}$. For a cell $A_j$ and an interval $\tilde A\subset A_j$, put $k_j(z)=\int_{A_j-z}|e|f(e)de$ and $\tilde m(A_j)=\sum_t\{|\eta_t|\1_t(A_j)-k_j(z_t)\}$. Then
\[
|m(\tilde A)|\le|\tilde m(A_j)|+2\sum_tk_j(z_t),\qquad \E\tilde m(A_j)^4\le C.
\]
A union bound gives $\max_j|\tilde m(A_j)|=o_p(n^{1/4})$, and \eqref{eq:occupationexp} gives $\max_j\sum_tk_j(z_t)=O_p(\log n)$. Hence $\sup_{A\subset\Gamma}|m(A)|=O_p(n^{1/4})$. Equation \eqref{eq:zeroenergy} gives $\sup_{A\subset\Gamma}|c(A)|=O_p(n^{1/4}\sqrt{\log n})$.

For full regimes, $b^e$ and $b^1$ are $O_p(\sqrt n)$ by conditional Doob inequalities. Write $m((-\infty,\gamma])$ as its value at $\underline\gamma$, whose second moment is at most $n\sigma_\eta^2$, plus the bounded-window increment. The full-regime compensator is $\sum_tH(z_t-\gamma)$, with $H(z)=\E[\eta_t\1\{\eta_t\le-z\}]$. Its translated supremum on compact $\Gamma$ is bounded and integrable, so \eqref{eq:occupationmoments} gives $O_p(\sqrt n)$. Thus $\varkappa_n=O_p(\sqrt n)$ and $\sup_\Gamma\|b\|=O_p(n^{1/4}\sqrt{\log n})$. Substituting in Lemma~\ref{lem:decomp}(b) proves the profiling remainder in (a).

\emph{Part (b): the local score.} Put $\nu_n=\sqrt nh_n$. Lemma~\ref{lem:local}(c) and the threshold-only innovation calculation \eqref{eq:innovationoccupation} identify the covariance clocks of $\nu_n^{-1/2}b^e$ and $\nu_n^{-1/2}b^1$ as $|v|\lt\sigma_e^2D$ and $|v|\lt\sigma_\eta^2(D-\mu_x\mu_x')$. Truncate the innovation weights in $b^1$ at $M$. For fixed $M$, the conditional CLT and grid argument of Lemma~\ref{lem:local}(d) apply to $(b^e,b_M^1)$: given $\mathcal Q$, the summand pairs are independent, centred and have zero cross-covariance. Doob's inequality bounds the truncation error by
\[
\E\sup_{|v|\le\bar v}\|\nu_n^{-1/2}(b^1-b_M^1)(I(v,h_n))\|^2
\le C\E[\eta_t^2\1\{|\eta_t|>M\}]+o(1).
\]
The density estimate and Fubini control the long horizons, and finite short horizons contribute $o(1)$.

\paragraph{The innovation part of the local score.}
For the chronological martingale $m$, take $g(e)=e\1\{|e|\le M\}$ and define
\[
R_n(v)=\frac{\operatorname{sgn}(v)}{\sqrt{\nu_n}}\sum_t\Big[g(\eta_t)\1_t(I(v,h_n))-\E\{g(\eta_t)\1_t(I(v,h_n))\mid\F_{t-1}\}\Big].
\]
The variance clock is $|v|\lt\E g(\eta_t)^2$. To establish that its Brownian driver is independent of the threshold path, use the stopped change of measure from the proof of Lemma~\ref{lem:cox}. For a finite linear combination set
\[
\zeta_t=\sum_j\theta_j\operatorname{sgn}(v_j)\nu_n^{-1/2}g(\eta_t)\1_t(I(v_j,h_n)),
\]
and choose $W_n=[\gamma_0-\bar vh_n,\gamma_0+\bar vh_n]$, $b_n=C_\theta\nu_n^{-1/2}$, and $T_K=n\wedge\inf\{t:\nu_n^{-1}\sum_{s\le t+1}P_s>K\}$. This stop has $\mathcal K=C_\theta^2K$ in \eqref{eq:tilttransfer}. If $S_t$ is the sum of the centred $\zeta_t$ and $V_t$ its predictable variance, the likelihood identity gives
\[
\E[e^{S_{T_K}}\mathcal H(X_n)]
=\E_{Q_n^{T_K}}[e^{V_{T_K}/2+o_p(1)}\mathcal H(X_n)].
\]
Here $\max_t|\zeta_t|\to0$, the stopped variance is bounded, and $\sum_t\E(\zeta_t\mid\F_{t-1})^2=o_p(1)$ by the window-density bound. The time-weighted occupation and innovation calculations in the proof of Lemma~\ref{lem:local} give, uniformly in $r$,
\[
\nu_n^{-1}\sum_{s\le[nr]}P_s\to2\bar v\Lt(r,0),\qquad
V_{[nr]}\to v(\theta)\Lt(r,0),
\]
where $v(\theta)=\E g(\eta_t)^2\sum_{j,k}\theta_j\theta_k\1\{v_jv_k>0\}\min(|v_j|,|v_k|)$. Their smoothing approximations transfer under \eqref{eq:tilttransfer}, so $V_{T_K}\convd v(\theta)\min\{\lt,K/(2\bar v)\}$. Using the likelihood identity also with $2\theta$ ensures uniform integrability. It identifies the conditional Gaussian moment generating function \citep{Curtiss1942}. Equation \eqref{eq:stoptail} then permits $K\to\infty$.

Burkholder's inequality and \eqref{eq:occupationmoments} bound fourth moments of increments by $C(\Delta^2+\Delta/\nu_n)$. The grid and within-cell argument of Lemma~\ref{lem:local}(d) gives tightness: a within-cell increment is bounded by $C\nu_n^{-1/2}\{N(A_j)+\sum_tP(q_t\in A_j\mid\F_{t-1})\}$ and has fourth moment $O(\nu_n^{-2})$. Thus $R_n\weak(\lt\E g(\eta_t)^2)^{1/2}\mathcal B$, jointly with the threshold path. Moreover, $R_n$ is $\mathcal Q$-measurable. Multiplying the conditional characteristic-function approximation for $(b^e,b_M^1)$ by $e^{i\theta'\mathbf R_n}\mathcal H(X_n)$, where $\mathbf R_n=(R_n(v_1),R_n(v_2),\ldots)'$ stacks the values of $R_n$ at the finitely many points $v_j$ appearing in $\zeta_t$, gives their joint limit with $R_n$, with the Gaussian components independent conditionally on $J_c$.

To remove the truncation from $m$, put $a_M=\|e^2\1\{|e|>M\}f(e)\|_*\to0$. The tail martingale $m_M$ satisfies
\[
\E m_M(A)^4\le C\{1+\sqrt n|A|\E[\eta_t^4\1\{|\eta_t|>M\}]+(\sqrt n|A|)^2a_M^2\}.
\]
After division by $\nu_n^2$, this gives the same grid-tightness estimate uniformly in $M$. On the $O(\nu_n)$ local cells, predictable sums have maximum $O_p(\log(2\nu_n))=o_p(\sqrt{\nu_n})$ by \eqref{eq:occupationexp}, and the fourth-moment union bound controls martingale cell maxima. Tail finite-dimensional variances vanish with $\E[\eta_t^2\1\{|\eta_t|>M\}]$. Hence the uniform tail process vanishes as $M\to\infty$. Finally, \eqref{eq:zeroenergy} bounds the scaled compensator by $O_p(\sqrt{h_n\log n})=o_p(1)$. Adding the three Gaussian covariance contributions yields
\[
|v|\lt\{\sigma_e^2D+\varrho^2\sigma_\eta^2(D-\mu_x\mu_x')+\varrho^2\sigma_\eta^2\mu_x\mu_x'\}
=|v|\lt\sigma^2D,
\]
which proves (b), including jointness with the threshold path.

\emph{Part (c): outward error control.} Let $h_n=n^{2\alpha-1/2}$ and use the information clock $\Theta_k$ in \eqref{eq:clockdef}. The window and innovation-weighted occupation calculations imply, after excluding an arbitrarily unlikely event where $\lt$ is small, uniformly for $Kh_n\le h\le\bar h$,
\begin{equation}\label{eq:endogclock}
\Theta_{k(\gamma_0+h)}\ge\epsilon_0\kappa_0\sqrt nh,\qquad
\epsilon_0\sigma_\eta^2\sqrt nh\le\sum_t\eta_t^2\1_t((\gamma_0,\gamma_0+h])\le C_\eta\Theta_{k(\gamma_0+h)}.
\end{equation}
Bound each component of \eqref{eq:endogdec} relative to $n^{-\alpha}\Theta_k$. For $b^e$ and $b^1$, the conditional H\'ajek--R\'enyi inequality, with clocks $\Theta_k$ and $\sum_{i\le k}\eta_{t(i)}^2$, bounds each outward probability by $C/(\epsilon_0K)$.

For $m$, take shells $h_j=2^jKh_n$ and thresholds $\lambda_j=\epsilon_1n^{-\alpha}\sqrt nh_j$, with $\epsilon_1>0$ a sufficiently small constant. Equation \eqref{eq:m4} and the grid maximal inequality bound the probability on shell $j$ by $C4^{-j}K^{-2}$. Cell interpolation as in (a) adds $C2^{-3j}K^{-3}n^{-2\alpha}$ for $\tilde m$ and a negligible $O_p(\log n)$ predictable sum, since $\lambda_0\asymp Kn^\alpha\to\infty$. Summing gives $C/K^2+o(1)$. Equation \eqref{eq:zeroenergy} bounds the probability for $c$ on each shell by $Cn^{2\alpha-1/2}\log n$. Summing over $O(\log n)$ shells gives $o(1)$. Both sides of the threshold obey these bounds. Taking $K$ large proves (c). \hfill$\square$

\subsection*{Proof of Lemma~\texorpdfstring{\ref{lem:localns}}{A.6} (Nonstationary slope regressors)}
We closely follow the block approximation for normalised nonstationary regressors in \citet{Chen2015} and \citet{ChenStengos2022}. The occupation measure changes here. On a representation where $n^{-1/2}x_{[n\cdot]}\to G$ uniformly, partition time into blocks of length $\epsilon$ and approximate $\mathcal X_t=n^{-1}x_tx_t'$ by $G(s_j)G(s_j)'$ on block $j$, where $s_j$ is the left end-point of the block.

(a)--(c) For full regimes use the count calculation \eqref{eq:unweightedcounts} on each block. For local windows use the time-weighted kernel occupation \eqref{eq:kerneloccupation} and the geometric-grid calculation leading to \eqref{eq:innovationoccupation}. These threshold-only arguments from Lemma~\ref{lem:local} use Assumption~\ref{ass:q} alone. After summing the blocks, the resulting limits are respectively $M_\mu$, $|\gamma_2-\gamma_1|Q_\ell$ and $Q_\ell$, with
\[
Q_\ell=\int_0^1G(s)G(s)'\,\Lt(ds,0).
\]
The error is $O_p(\omega_G(\epsilon))+o_p(1)$ uniformly in the stated windows, where $\omega_G(\epsilon)=\sup_{|s-s'|\le\epsilon}\|G(s)-G(s')\|$ is the modulus of continuity of $G$. Apply the count argument first to nonnegative quadratic forms of $\mathcal X_t$ and recover matrix entries by polarisation. Continuity of $G$ permits $\epsilon\downarrow0$.

(d) Localise to $E_{n,K}=\{\max_tn^{-1/2}\|x_t\|\le K\}$, whose complement vanishes as $K\to\infty$. Given $(x,q)$, the score has independent centred increments and covariance clock converging by (c) to $\sigma^2|v|Q_\ell$. On $E_{n,K}$ its Lindeberg sum is at most
\[
\frac{K^2N([\gamma_0-\bar vh_n,\gamma_0+\bar vh_n])}{\nu_n}
\E[u_t^2\1\{|u_t|>\epsilon\sqrt{\nu_n}/K\}]=o_p(1),\qquad\nu_n=\sqrt nh_n.
\]
The conditional CLT and grid proof of Lemma~\ref{lem:local}(d) applies: increment fourth moments are bounded by $C_K(\Delta^2+\Delta/\nu_n)$, and cell maxima vanish by \eqref{eq:occupationmoments}. The conditional characteristic function identifies the driver as independent of $(G,J_c)$. Remove $E_{n,K}$ after taking the limit. Thus no unconditional moments of $x_t$ are required.

(e) Conditional on the independent regressor path, integrate out $u_t$. The marked test function is uniformly approximated by its continuous counterpart evaluated at $G(s)$. The arrival-location and change-of-measure arguments in Lemma~\ref{lem:cox} use only Assumption~\ref{ass:q}. Applied conditionally on the independent regressor path, they hold jointly with $G$ and $J_c$, giving the stated local-time-selected calendar marks.

(f) Parts (a)--(b), positivity of both regime matrices and conditional Doob inequalities give
\[
\varpi_n=O_p(n^{-2}),\quad \varkappa_n=O_p(n),\quad
\sup_\Gamma\|B\|=O_p(n^{3/2}),\quad \sup_\Gamma\|b\|=O_p(n^{3/4}).
\]
Feasibility follows from the same regime limits. Substitution into Lemma~\ref{lem:decomp}(b) gives \eqref{eq:rho}. The inverse-matrix orders are in probability because their limiting eigenvalues are random. \hfill$\square$

\section{Endogeneity: cancellation through aggregation}\label{sec:Sendog}

Proposition~\ref{prop:endog} establishes diminishing-effect threshold and likelihood ratio limits under $u_t=\varrho\eta_t+e_t$ without an endogeneity correction. We illustrate the cancellation behind this result and compare it with two related approaches. \citet{YuPhillipsEndogeneity2018} remove a smooth endogenous conditional-mean component by local differencing, then aggregate the jump information across covariates. Here least squares is retained. The predictable error contributions cancel when aggregated over visits to a shrinking window. The zero-energy theory of \citet{WangPhillips2011} provides the technical connection to structural nonparametric cointegrating regression \citep{WangPhillips2009b}.

\subsection{A heuristic example}\label{sec:Sendog-example}

\begin{figure}[!ht]
\centering
\includegraphics[width=\textwidth]{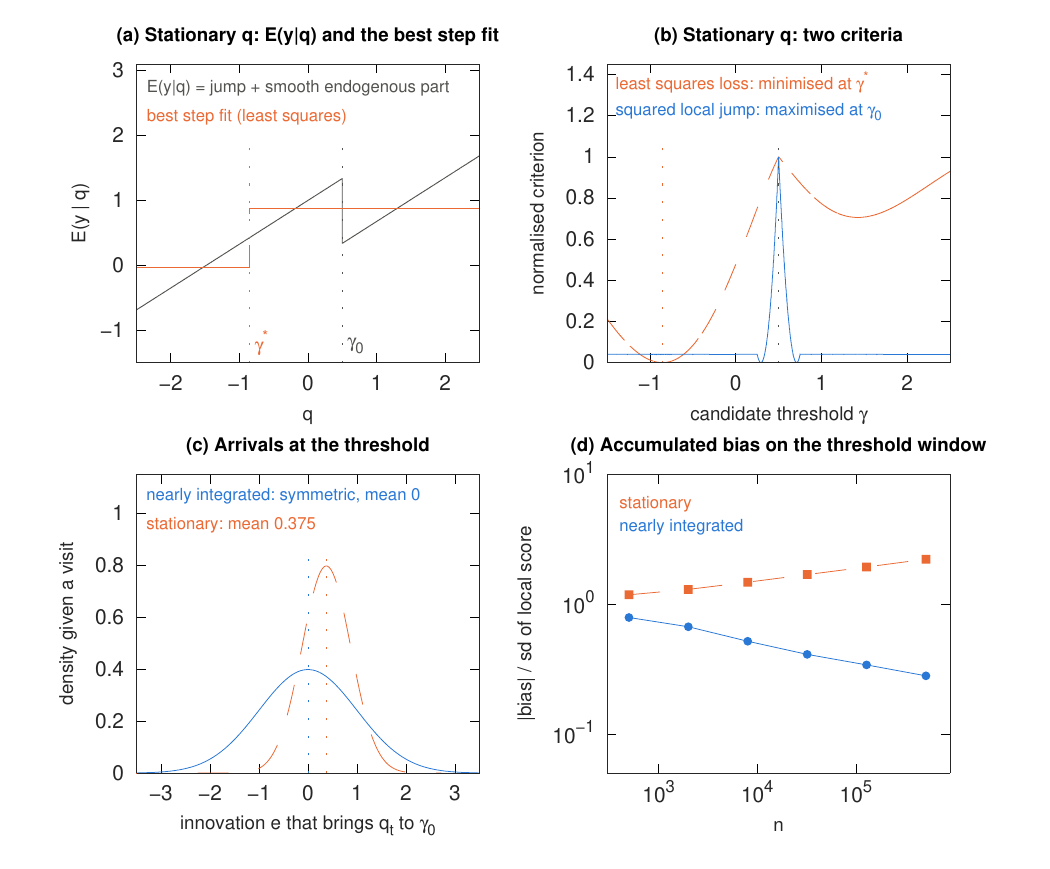}
\caption{Two forms of cancellation in model \eqref{eq:heurmodel}, with $\delta=1$, $\varrho=0.9$ and $\gamma_0=0.5$. (a) For stationary $q_t$ ($\phi=0.5$), the conditional mean is a jump plus a smooth endogenous component, and the best-fitting step splits at $\gamma^*\neq\gamma_0$. (b) Least squares loss is minimised at $\gamma^*$, whereas the squared difference of one-sided local means at bandwidth $0.25$ peaks at $\gamma_0$. (c) The conditional innovation density at a visit to $\gamma_0$ is approximately symmetric under near integration and shifted under stationarity. (d) Median ratio of accumulated predictable bias $\varrho\sum_t\Upsilon_h(z_t)$ to the standard deviation of the local score, using the limit-theory windows at $\alpha=0.1$. Results use 500 replications and log scales.}\label{fig:endog}
\end{figure}

Consider
\begin{equation}\label{eq:heurmodel}
y_t=\delta\1\{q_t\le\gamma_0\}+u_t,\qquad u_t=\varrho\eta_t+e_t,\qquad q_t=z_t+\eta_t,
\end{equation}
where $|\varrho|<1$, $\eta_t\sim N(0,1)$ and $e_t\sim N(0,1-\varrho^2)$ are independent, so that $\E u_t^2=1$, and $z_t=\phi q_{t-1}$ is known at time $t-1$. We compare $\phi=0.5$, with the process in its stationary distribution, and a unit root, $\phi=1$. Figure~\ref{fig:endog} uses $\delta=1$, $\varrho=0.9$ and $\gamma_0=0.5$.

\emph{Stationary threshold variable.} Since $\mathrm{Cov}(\eta_t,q_t)=1$ and $\mathrm{Var}(q_t)=1/(1-\phi^2)$,
\begin{equation}\label{eq:heurmean}
\E(y_t\mid q_t=q)=\delta\1\{q\le\gamma_0\}+g(q),\qquad g(q)=\varrho(1-\phi^2)q.
\end{equation}
Least squares fits a step to this conditional mean, which contains both a jump and a smooth trend. Its population split is $\gamma^*=-0.86$ in the example, far from $\gamma_0=0.5$ (Panels (a)--(b)). This illustrates the source of bias in stationary threshold regression and the motivation for control functions \citep{KourtellosStengosTan2016,YuLiaoPhillips2024}. The Monte Carlo model in Table~\ref{tab:mc4} also shifts the slope of an independent regressor. That extra threshold signal helps explain the smaller fixed-effect bias at large sample sizes in Panel C, although LR coverage remains distorted. Under a diminishing effect, the persistent bias in Panel D is more pronounced.

\citet{YuPhillipsEndogeneity2018} exploit the local nature of the jump. Because $g$ is continuous,
\[
\lim_{h\downarrow0}\left\{\E(y_t\mid q_t\in(\gamma-h,\gamma])-\E(y_t\mid q_t\in(\gamma,\gamma+h])\right\}
=\delta\1\{\gamma=\gamma_0\}.
\]
Their integrated difference kernel estimator compares one-sided conditional-mean estimates and averages squared differences over covariate values. The smooth component cancels locally, while the jump remains. Panel (b) shows the population analogue at bandwidth $0.25$: the squared local difference peaks at $\gamma_0$, whereas the least squares loss is minimised at $\gamma^*$. Under i.i.d.\ sampling their estimator is $n$-consistent, with a compound Poisson-type argmin limit.

\emph{Nearly integrated threshold variable.} For least squares, the relevant error score consists of observations in $A_h=(\gamma_0,\gamma_0+h]$. Entering this window requires $\eta_t\in A_h-z_t$, so its predictable mean is
\begin{equation}\label{eq:heurbias}
\E\big(u_t\1\{q_t\in A_h\}\mid\F_{t-1}\big)=\varrho\Upsilon_h(z_t),\qquad
\Upsilon_h(z)=\int_{\gamma_0-z}^{\gamma_0+h-z}e f(e)\,de,
\end{equation}
where $f$ is the standard normal density. A visit from below needs a positive innovation, while a visit from above needs a negative one. Conditional on landing at $\gamma_0$, the innovation density is proportional to $f(e)p_z(\gamma_0-e)$, where $p_z$ is the density of $z_t$.

For a nearly integrated state, $p_z$ becomes flat on the innovation scale.\footnote{See \eqref{eq:densitybounds} in the proof of Lemma~\ref{lem:local}.} The conditional innovation density therefore approaches $f$. Positive and negative contributions offset, giving $\E(u_t\mid\text{visit})\approx\varrho\E\eta_t=0$ (Panel (c), solid line). Under stationarity, $z_t\sim N(0,\phi^2/(1-\phi^2))$ and this density is not flat. Visits to $\gamma_0=0.5$ more often arrive from below, with $\E(\eta_t\mid q_t=\gamma_0)=(1-\phi^2)\gamma_0=0.375$ (dashed line). The bias then persists.

The formal cancellation is $\int\Upsilon_h(z)\,dz=h\E\eta_t=0$. Its accumulated contribution $\varrho\sum_t\Upsilon_h(z_t)$ has order $O_p(n^{1/4}h\sqrt{\log n})$, while the local score fluctuates at order $(\sqrt nh)^{1/2}$.\footnote{See \eqref{eq:zeroenergy} in the proof of Lemma~\ref{lem:endogsums}.} Their ratio is $O_p((h\log n)^{1/2})$, which vanishes when $h\log n\to0$, including the windows $h_n\asymp n^{2\alpha-1/2}$ for $0<\alpha<1/4$.

Panel (d) illustrates this decline at $\alpha=0.1$. It uses $h_n=10n^{2\alpha-1/2}$ for the nearly integrated variable and $h_n=10n^{2\alpha-1}$ for the stationary variable. From $n=500$ to $512{,}000$, the median bias-to-score ratio falls from 0.79 to 0.28 in the former design and rises from 1.19 to 2.23 in the latter. The log-scale slopes are close to $\alpha-1/4$ and $\alpha$, respectively.

\subsection{Relation to \citet{YuPhillipsEndogeneity2018}}\label{sec:Sendog-YP}

Both approaches use local threshold information and aggregate it, but their cancellations differ. Yu and Phillips difference out a smooth component $g(x,q)$ and aggregate the resulting jump information across covariates. Their estimator allows general endogeneity of regressors and the threshold variable under i.i.d.\ sampling. Our result instead retains ordinary least squares, allows the particular contemporaneous linear dependence of Assumption~\ref{ass:endog} with exogenous slopes, and aggregates predictable error contributions across visits of a persistent state.

This cancellation is local. Over a whole regime, the predictable error component need not vanish. The heuristic in Section~\ref{sec:endog} gives the leading contribution $-\varrho\sigma_\eta^2\lt\,\E(x_t)$ to the predictable component of $n^{-1/2}\sum_tx_tu_t\1\{q_t\le\gamma_0\}$. Thus threshold robustness does not justify conventional coefficient inference or the uncorrected sup-Wald calibration. Table~\ref{tab:mc7}, Panel B, illustrates the test distortion.

\subsection{Relation to \citet{WangPhillips2009b}}\label{sec:Sendog-WP}

The closer technical connection is to structural nonparametric cointegrating regression. For $y_t=m(x_t)+u_t$ with an integrated regressor and correlated regression errors and innovations, \citet{WangPhillips2009b} establish consistency and mixed normality of a kernel estimator, with the same limit as under exogeneity. Localisation makes the predictable error component a zero-energy functional, whose sum is smaller than the stochastic fluctuation \citep{WangPhillips2011}.

The threshold score $\sum_tx_tu_t\1\{q_t\in A_h\}$ has the same structure, using a one-sided uniform kernel of bandwidth $h$. Its compensator is $\varrho\sum_tx_t\Upsilon_h(z_t)$. The zero-energy bound and the martingale limit with a local-time clock supply the corresponding bias and fluctuation controls.\footnote{See \eqref{eq:zeroenergy} and the innovation-score argument in the proof of Lemma~\ref{lem:endogsums}.} Threshold estimation adds three requirements:
\begin{enumerate}[label=(\roman*),leftmargin=*]
\item The score must converge as a process in the local parameter $v$, jointly with the signal.\footnote{See Lemma~\ref{lem:endogsums}(b).}
\item The bias must be small relative to the signal uniformly over windows beyond $K/a_n$, requiring a maximal inequality over dyadic shells.\footnote{See Lemma~\ref{lem:endogsums}(c).}
\item The threshold effect determines the bandwidth, $h_n=n^{2\alpha-1/2}$. It shrinks precisely when $\alpha<1/4$ and contains a diverging number of visits precisely when $\alpha>0$. The averaging argument therefore covers the diminishing-effect case of Proposition~\ref{prop:endog}.
\end{enumerate}

\emph{Fixed threshold effect.} On a window of width $O(n^{-1/2})$, the same flat-density heuristic suggests that innovations and error marks retain their unconditional law. The fixed-effect results in Table~\ref{tab:mc4}, Panel A, support this conjecture. Establishing the limit would require a marked Cox convergence result allowing marks to depend on the threshold innovations, which is beyond the present analysis.

% Use the same database and author--year style as the main paper.
% Include companion papers to preserve the main paper's 2009a/b and 2018a/b labels. \nocite{WangPhillips2009a,YuPhillips2018}
\renewcommand{\thetable}{S\arabic{table}}
\section{Additional Monte Carlo evidence}\label{sec:Smc}

This section supplements Section~\ref{sec:mc} of the paper. It reports unconditional confidence-set performance, alternative inference procedures, the complete endogeneity comparison and experiments with nonstationary slope regressors. The baseline designs and reporting conventions are those of the paper unless stated otherwise. All tables report existing simulation results.

\subsection{Unconditional coverage and feasibility}\label{sec:Smc-feas}

Table~\ref{tab:mcU} evaluates $C_n^U(0.95)$ over all replications. Unlike the feasibility-conditioned summaries in Table~\ref{tab:mc1} of the paper, this construction retains infeasible threshold candidates and returns $\Gamma$ when no split is feasible. Thus both the set construction and the samples entering the coverage and length calculations differ.

The difference is most pronounced for persistent paths with few feasible splits. For example, at $c=5$, $n=250$ and $\alpha=0$, coverage is 0.990 and average length is 3.092, compared with 0.976 and 1.179 under the paper's feasibility-conditioned reporting. The parameter-space width is 4. Under weaker effects the unconditional sets can cover almost the entire search interval. These results illustrate the uncertainty associated with infeasible splits rather than an improvement in threshold precision.

\begin{table}[!ht]
\centering
\caption{Likelihood ratio confidence set without conditioning on feasibility}\label{tab:mcU}
\small
\begin{tabular}{r cc cc cc cc}
\toprule
 & \multicolumn{2}{c}{$c=-10$} & \multicolumn{2}{c}{$c=0$} & \multicolumn{2}{c}{$c=5$} & \multicolumn{2}{c}{$c=0$, dependent} \\
\cmidrule(lr){2-3}\cmidrule(lr){4-5}\cmidrule(lr){6-7}\cmidrule(lr){8-9}
$n$ & Cov & Len & Cov & Len & Cov & Len & Cov & Len \\
\midrule
% Begin output/tables/table2.tex
\multicolumn{9}{l}{Panel A: $\alpha=0.0$} \\
250 & 0.973 & 0.471 & 0.985 & 1.950 & 0.990 & 3.092 & 0.983 & 2.333 \\
500 & 0.980 & 0.290 & 0.988 & 1.450 & 0.989 & 2.607 & 0.984 & 1.868 \\
1000 & 0.989 & 0.212 & 0.986 & 1.154 & 0.990 & 2.244 & 0.988 & 1.539 \\
2000 & 0.977 & 0.146 & 0.985 & 0.854 & 0.988 & 1.730 & 0.979 & 1.189 \\
\midrule
\multicolumn{9}{l}{Panel B: $\alpha=0.1$} \\
250 & 0.962 & 1.134 & 0.978 & 2.660 & 0.989 & 3.506 & 0.979 & 3.115 \\
500 & 0.961 & 0.813 & 0.977 & 2.274 & 0.988 & 3.231 & 0.984 & 2.796 \\
1000 & 0.974 & 0.632 & 0.971 & 1.992 & 0.980 & 2.978 & 0.982 & 2.553 \\
2000 & 0.965 & 0.490 & 0.976 & 1.638 & 0.980 & 2.643 & 0.977 & 2.270 \\
\midrule
\multicolumn{9}{l}{Panel C: $\alpha=0.2$} \\
250 & 0.946 & 2.462 & 0.981 & 3.404 & 0.989 & 3.818 & 0.988 & 3.703 \\
500 & 0.957 & 2.216 & 0.976 & 3.311 & 0.987 & 3.732 & 0.991 & 3.650 \\
1000 & 0.962 & 2.011 & 0.972 & 3.172 & 0.987 & 3.683 & 0.983 & 3.576 \\
2000 & 0.964 & 1.826 & 0.970 & 3.022 & 0.987 & 3.598 & 0.988 & 3.524 \\
% End output/tables/table2.tex
\bottomrule
\end{tabular}
\vspace{0.5ex}
\parbox{0.95\textwidth}{\footnotesize Notes: The designs are those of Table~\ref{tab:mc1} in the paper. Values of $\gamma$ leaving fewer than ten observations in a regime are not rejected, and the set is $\Gamma$ when no split is feasible. Cov and Len are coverage and average Lebesgue measure of $C_n^U(0.95)$ over all 2{,}000 replications. The width of $\Gamma=[-1.5,2.5]$ is 4.}
\end{table}

\subsection{Alternative critical values and the bootstrap}\label{sec:Smc-inference}

Table~\ref{tab:mc6}, Panel A, examines hull calibration and the heteroskedasticity correction under $\alpha=0.1$. With homoskedastic errors, the hull critical value $c_{0.95}^{\mathrm{hull}}=5.99$ yields coverage near 0.96 and slightly shorter sets than the union set using $c_{0.95}=7.35$. These results are consistent with the calibration in Theorem~\ref{thm:LR} of the paper.

In the heteroskedastic design, $u_t=(0.5+0.5x_{2t}^2)^{1/2}\epsilon_t$ and the variance ratio in Remark~\ref{rem:het} is $\vartheta^2=1.5$. The uncorrected critical value therefore gives limiting coverage $P(\xi\le7.35/1.5)\approx0.83$. Finite-sample coverage is 0.90--0.93. Rescaling by $\hat\vartheta^2$ raises coverage to approximately 0.96--0.98 for the union set and 0.95--0.96 for the hull.

Panel B examines a fixed-regressor residual bootstrap under a fixed effect. We hold $x_t$ and $q_t$ fixed, resample centred residuals, re-estimate the threshold and use the 0.95 quantile of $LR_n^*(\hat\gamma)$ as the critical value. Coverage is 0.93--0.95 under Gaussian and Laplace errors and 0.95--0.96 under mixture errors. The mixture requires larger critical values and longer sets than the baseline calibration.

These experiments do not establish bootstrap validity. Under a fixed effect, only finitely many observations remain informative in a shrinking neighbourhood of the threshold, and the bootstrap fixes their regressor marks. The local-time-free LR limit in Theorem~\ref{thm:LR}(c) is therefore insufficient to justify this resampling scheme. Conditional bootstrap convergence requires a separate argument, as the change-point literature makes clear \citep{SeijoSen2011,Yu2014}.

\begin{table}[!ht]
\centering
\caption{Alternative likelihood ratio confidence sets: convex hull, heteroskedasticity-robust scaling and the fixed-regressor bootstrap}\label{tab:mc6}
\footnotesize
\setlength{\tabcolsep}{4pt}
\begin{tabular}{rr cc cc c cc cc}
\toprule
\multicolumn{11}{l}{Panel A: diminishing effect, $\alpha=0.1$} \\
 & & \multicolumn{4}{c}{homoskedastic errors} & \multicolumn{5}{c}{heteroskedastic errors, $\vartheta^2=1.5$} \\
\cmidrule(lr){3-6}\cmidrule(lr){7-11}
 & & \multicolumn{2}{c}{union, $c_{0.95}$} & \multicolumn{2}{c}{hull, $c^{\mathrm{hull}}_{0.95}$} & union, $c_{0.95}$ & \multicolumn{2}{c}{union, $\hat\vartheta^2c_{0.95}$} & \multicolumn{2}{c}{hull, $\hat\vartheta^2c^{\mathrm{hull}}_{0.95}$} \\
\cmidrule(lr){3-4}\cmidrule(lr){5-6}\cmidrule(lr){7-7}\cmidrule(lr){8-9}\cmidrule(lr){10-11}
$c$ & $n$ & Cov & Len & Cov & Len & Cov & Cov & Len & Cov & Len \\
\midrule

% Begin output/tables/table6_panelA.tex
-10 & 500 & 0.972 & 0.829 & 0.958 & 0.810 & 0.909 & 0.962 & 1.194 & 0.955 & 1.154 \\
-10 & 2000 & 0.971 & 0.492 & 0.960 & 0.488 & 0.899 & 0.968 & 0.740 & 0.952 & 0.726 \\
0 & 500 & 0.974 & 1.688 & 0.962 & 1.611 & 0.909 & 0.966 & 2.077 & 0.958 & 1.972 \\
0 & 2000 & 0.969 & 1.358 & 0.959 & 1.317 & 0.901 & 0.968 & 1.743 & 0.953 & 1.672 \\
5 & 500 & 0.969 & 1.857 & 0.960 & 1.776 & 0.926 & 0.977 & 2.138 & 0.962 & 2.030 \\
5 & 2000 & 0.967 & 1.868 & 0.958 & 1.802 & 0.912 & 0.965 & 2.240 & 0.960 & 2.145 \\

% End output/tables/table6_panelA.tex

\bottomrule
\end{tabular}

\medskip
\begin{tabular}{lr cc cc c}
\toprule
\multicolumn{7}{l}{Panel B: fixed effect, $\alpha=0$} \\
 & & \multicolumn{2}{c}{$c_{0.95}=7.35$} & \multicolumn{2}{c}{bootstrap} & \\
\cmidrule(lr){3-4}\cmidrule(lr){5-6}
Design & $n$ & Cov & Len & Cov & Len & median $c^*$ \\
\midrule

% Begin output/tables/table6_panelB.tex
$x_t=(1,x_{2t})'$, Gaussian, $c=-10$ & 500 & 0.987 & 0.290 & 0.943 & 0.228 & 5.26 \\
$x_t=(1,x_{2t})'$, Gaussian, $c=0$ & 500 & 0.976 & 0.795 & 0.932 & 0.636 & 5.21 \\
$x_t=(1,x_{2t})'$, Gaussian, $c=5$ & 500 & 0.976 & 1.064 & 0.927 & 0.864 & 4.85 \\
$x_t=1$, Gaussian, $c=0$ & 250 & 0.996 & 0.338 & 0.948 & 0.260 & 2.77 \\
$x_t=1$, Gaussian, $c=0$ & 1000 & 0.989 & 0.198 & 0.942 & 0.156 & 2.70 \\
$x_t=1$, Laplace, $c=0$ & 250 & 0.966 & 0.300 & 0.936 & 0.257 & 3.87 \\
$x_t=1$, Laplace, $c=0$ & 1000 & 0.971 & 0.199 & 0.947 & 0.170 & 4.21 \\
$x_t=1$, mixture, $c=0$ & 250 & 0.907 & 0.282 & 0.946 & 0.349 & 9.17 \\
$x_t=1$, mixture, $c=0$ & 1000 & 0.910 & 0.160 & 0.956 & 0.211 & 9.65 \\

% End output/tables/table6_panelB.tex

\bottomrule
\end{tabular}

\vspace{0.5ex}
\parbox{0.95\textwidth}{\footnotesize Notes: Panel A uses the design of Table~\ref{tab:mc1} with $\alpha=0.1$. In the heteroskedastic design $u_t=(0.5+0.5x_{2t}^2)^{1/2}\epsilon_t$ with $\epsilon_t$ \textit{i.i.d.}\ $N(0,1)$, and $\hat\vartheta^2$ is the estimate of Remark~\ref{rem:het} at $\hat\gamma$. The hull is the convex hull of the union set computed with the stated critical value, $c_{0.95}=7.35$ and $c_{0.95}^{\mathrm{hull}}=5.99$. 2{,}000 replications, with common random numbers across the two error designs. Panel B uses the designs of Tables~\ref{tab:mc1} and \ref{tab:mc3} with $\alpha=0$. The bootstrap holds $x_t$ and $q_t$ fixed, resamples the centred residuals of the threshold fit with replacement, re-estimates the threshold in each of 199 bootstrap samples and takes the $0.95$ quantile of $LR_n^*(\hat\gamma)$ as the critical value $c^*$. The set is $\{\gamma:LR_n(\gamma)\le c^*\}$. 1{,}000 replications. Cov conditions on a feasible true split and Len on a feasible fit.}
\end{table}

\subsection{Additional endogeneity results}\label{sec:Smc-endog}

Table~\ref{tab:mc4full} gives the complete comparison behind Table~\ref{tab:mc4} of the paper. Panels A and B use a unit-root threshold variable under fixed and diminishing effects, respectively. Median errors and coverage remain similar across the contemporaneous correlations considered. Panel B supports Proposition~\ref{prop:endog}. Panel A provides numerical evidence for a fixed effect, whose limiting distribution under this dependence remains unproved.

Panels C and D use a stationary threshold variable. Under a fixed effect, strong correlation lowers coverage, although performance improves with $n$. Under a diminishing effect, bias persists and coverage deteriorates sharply. The contrast with the unit-root designs concerns the specified contemporaneous dependence and exogenous slope regressors. It does not imply that endogeneity is harmless more generally.

\begin{table}[!ht]
\centering
\caption{Endogenous threshold variable: $\mathrm{corr}(u_t,w_t)=\rho_{uw}$}\label{tab:mc4full}
\small
\begin{tabular}{r ccc ccc ccc}
\toprule
 & \multicolumn{3}{c}{$\rho_{uw}=0$} & \multicolumn{3}{c}{$\rho_{uw}=0.5$} & \multicolumn{3}{c}{$\rho_{uw}=0.9$} \\
\cmidrule(lr){2-4}\cmidrule(lr){5-7}\cmidrule(lr){8-10}
$n$ & MdB & MdAE & Cov & MdB & MdAE & Cov & MdB & MdAE & Cov \\
\midrule

% Begin output/tables/table8.tex
\multicolumn{10}{l}{Panel A: unit-root $q_t$, $\alpha=0.0$} \\
250 & -0.050 & 0.164 & 0.972 & -0.049 & 0.168 & 0.978 & -0.045 & 0.172 & 0.983 \\
500 & -0.035 & 0.111 & 0.981 & -0.035 & 0.111 & 0.984 & -0.026 & 0.117 & 0.988 \\
1000 & -0.024 & 0.080 & 0.986 & -0.022 & 0.079 & 0.983 & -0.022 & 0.078 & 0.979 \\
2000 & -0.016 & 0.054 & 0.979 & -0.013 & 0.051 & 0.983 & -0.014 & 0.050 & 0.985 \\
\midrule
\multicolumn{10}{l}{Panel B: unit-root $q_t$, $\alpha=0.1$} \\
250 & -0.058 & 0.349 & 0.968 & -0.061 & 0.358 & 0.973 & -0.048 & 0.385 & 0.985 \\
500 & -0.032 & 0.234 & 0.970 & -0.024 & 0.258 & 0.983 & -0.020 & 0.277 & 0.984 \\
1000 & -0.026 & 0.185 & 0.976 & -0.021 & 0.188 & 0.980 & -0.021 & 0.195 & 0.983 \\
2000 & -0.008 & 0.157 & 0.969 & -0.006 & 0.158 & 0.972 & -0.008 & 0.162 & 0.978 \\
\midrule
\multicolumn{10}{l}{Panel C: stationary $q_t$ (AR(1), $\phi=0.5$), $\alpha=0.0$} \\
250 & -0.008 & 0.021 & 0.975 & -0.007 & 0.024 & 0.980 & -0.060 & 0.095 & 0.683 \\
500 & -0.004 & 0.010 & 0.979 & -0.004 & 0.012 & 0.985 & -0.021 & 0.033 & 0.747 \\
1000 & -0.002 & 0.005 & 0.986 & -0.002 & 0.006 & 0.996 & -0.007 & 0.012 & 0.823 \\
2000 & -0.001 & 0.003 & 0.983 & -0.001 & 0.003 & 0.988 & -0.003 & 0.006 & 0.867 \\
\midrule
\multicolumn{10}{l}{Panel D: stationary $q_t$ (AR(1), $\phi=0.5$), $\alpha=0.1$} \\
250 & -0.006 & 0.041 & 0.967 & -0.188 & 0.372 & 0.631 & -0.932 & 0.932 & 0.025 \\
500 & -0.004 & 0.024 & 0.963 & -0.270 & 0.399 & 0.529 & -0.901 & 0.901 & 0.004 \\
1000 & -0.002 & 0.011 & 0.978 & -0.517 & 0.542 & 0.381 & -0.886 & 0.886 & 0.000 \\
2000 & -0.001 & 0.007 & 0.970 & -0.770 & 0.774 & 0.195 & -0.871 & 0.871 & 0.000 \\

% End output/tables/table8.tex

\bottomrule
\end{tabular}

\vspace{0.5ex}
\parbox{0.95\textwidth}{\footnotesize Notes: The design of Table~\ref{tab:mc1} in the paper is used with $u_t=\rho_{uw}w_t+(1-\rho_{uw}^2)^{1/2}\nu_t$, where $\nu_t$ is an independent \textit{i.i.d.}\ $N(0,1)$ sequence. In Panels A--B, $q_t$ is a Gaussian random walk. In Panels C--D, $q_t=0.5q_{t-1}+w_t$. MdB and MdAE are the median signed and absolute errors over samples with a feasible fit. Cov conditions on a feasible true split. There are 2{,}000 replications. Panels B and D are reproduced as Panels A and B of Table~\ref{tab:mc4} in the paper.}
\end{table}

\subsection{Nonstationary slope regressors}\label{sec:Smc-nonstat}

Table~\ref{tab:mc5} considers independent unit-root slopes, unit-root slopes with innovations correlated with those of the threshold variable, and independent hybrid stochastic local unit root slopes. Median errors decline slowly and confidence sets remain wide. Coverage is approximately 0.98 in all three designs. The independent designs satisfy Assumption~\ref{ass:ns} of the paper. The correlated design explores a possible extension and is not covered by Theorem~\ref{thm:ns}.

\begin{table}[!ht]
\centering
\caption{Nonstationary slope regressors}\label{tab:mc5}
\small
\begin{tabular}{r ccc ccc ccc}
\toprule
 & \multicolumn{3}{c}{UR $x_t$, independent} & \multicolumn{3}{c}{UR $x_t$, $\mathrm{corr}=0.5$} & \multicolumn{3}{c}{HLSUR $x_t$} \\
\cmidrule(lr){2-4}\cmidrule(lr){5-7}\cmidrule(lr){8-10}
$n$ & MdAE & Cov & Len & MdAE & Cov & Len & MdAE & Cov & Len \\
\midrule

% Begin output/tables/table9.tex
\multicolumn{10}{l}{Panel A: $\tau=0.0$} \\
250 & 0.358 & 0.982 & 1.809 & 0.401 & 0.977 & 1.931 & 0.340 & 0.982 & 1.825 \\
500 & 0.239 & 0.983 & 1.787 & 0.300 & 0.979 & 1.893 & 0.252 & 0.981 & 1.772 \\
1000 & 0.177 & 0.980 & 1.691 & 0.214 & 0.984 & 1.816 & 0.186 & 0.979 & 1.711 \\
2000 & 0.141 & 0.977 & 1.584 & 0.171 & 0.983 & 1.735 & 0.155 & 0.977 & 1.568 \\
\midrule
\multicolumn{10}{l}{Panel B: $\tau=0.1$} \\
250 & 0.610 & 0.978 & 2.435 & 0.677 & 0.980 & 2.538 & 0.609 & 0.980 & 2.430 \\
500 & 0.501 & 0.983 & 2.533 & 0.545 & 0.985 & 2.636 & 0.502 & 0.978 & 2.520 \\
1000 & 0.491 & 0.979 & 2.512 & 0.520 & 0.984 & 2.670 & 0.442 & 0.982 & 2.533 \\
2000 & 0.395 & 0.980 & 2.547 & 0.487 & 0.978 & 2.654 & 0.400 & 0.984 & 2.461 \\

% End output/tables/table9.tex

\bottomrule
\end{tabular}

\vspace{0.5ex}
\parbox{0.95\textwidth}{\footnotesize Notes: $y_t=x_t+2n^{-1/2-\tau}x_t\1\{q_t\le0.5\}+u_t$, with $q_t$ a Gaussian random walk and $x_t=\exp(a\epsilon_t/\sqrt n)x_{t-1}+v_t$, where $a=0$ (unit root, UR) or $a=0.2$ (hybrid stochastic local unit root, HLSUR), and $\epsilon_t$ \textit{i.i.d.}\ $N(0,1)$. ``corr'' is the correlation between $v_t$ and $w_t$. MdAE and Len condition on any feasible fit. Cov conditions on a feasible true split. 2{,}000 replications.}
\end{table}

Table~\ref{tab:mc2ns} examines larger samples with an independent unit-root slope regressor. The normalised median errors increase with $n$, particularly under $\tau=0.1$. Table~\ref{tab:mcQ} compares the corresponding finite-sample quantiles with those of the simulated limit.

\begin{table}[!ht]
\centering
\caption{Nonstationary slope regressor: normalised medians of $|\hat\gamma-\gamma_0|$ ($c=0$)}\label{tab:mc2ns}
\small
\begin{tabular}{llcccc}
\toprule
 & & $n=500$ & $n=2{,}000$ & $n=8{,}000$ & $n=32{,}000$ \\
\midrule

\multicolumn{6}{l}{Unit-root slope regressor, $\delta_n=n^{-1/2-\tau}\delta_0$} \\
$\tau=0$ & $n^{0.5}\,\mathrm{med}|\hat\gamma-\gamma_0|$ & 5.42 & 6.16 & 6.34 & 7.44 \\
 & Coverage & 0.982 & 0.981 & 0.984 & 0.980 \\
$\tau=0.1$ & $n^{0.3}\,\mathrm{med}|\hat\gamma-\gamma_0|$ & 3.28 & 3.56 & 4.58 & 4.74 \\
 & Coverage & 0.985 & 0.980 & 0.978 & 0.976 \\
\bottomrule
\end{tabular}

\vspace{0.5ex}
\parbox{0.95\textwidth}{\footnotesize Notes: The model is $y_t=x_t+2n^{-1/2-\tau}x_t\1\{q_t\le0.5\}+u_t$, with $x_t$ and $q_t$ independent Gaussian random walks and independent standard Gaussian errors. The normalisation is $n^{1/2-2\tau}$. Coverage refers to $C_n^F(0.95)$ over samples with a feasible true split, with the search conventions of Section~\ref{sec:mc} in the paper. There are 1{,}000 replications.}
\end{table}

At $\tau=0.1$, the lower quartile is already close to its limiting value, while the median moves towards about 6. The upper quantiles remain substantially smaller than their limiting counterparts. Visits near the threshold can occur when the slope regressor is close to zero, making $Q_\ell$ small and the limiting scale large. In addition, the compact search space bounds the finite-sample errors. These features explain the slow tail approximation. The numerical construction and grid checks are given in Section~\ref{sec:Snumerical}.

\begin{table}[!ht]
\centering
\caption{Nonstationary slope regressor, $\tau=0.1$: quantiles of $n^{0.3}|\hat\gamma-\gamma_0|$ and of the limit $|\lambda_Q\mathcal T|$}\label{tab:mcQ}
\small
\begin{tabular}{lcccccc}
\toprule
 & \multicolumn{4}{c}{finite sample} & \multicolumn{2}{c}{limit} \\
\cmidrule(lr){2-5}\cmidrule(lr){6-7}
quantile & $n=500$ & $n=2{,}000$ & $n=8{,}000$ & $n=32{,}000$ & simulated & 95\% interval \\
\midrule

% Begin output/tables/table4.tex
25\% & 0.94 & 0.97 & 1.01 & 0.95 & 0.99 & [0.97, 1.02] \\
50\% & 3.28 & 3.56 & 4.58 & 4.74 & 6.03 & [5.91, 6.15] \\
75\% & 8.13 & 11.19 & 16.74 & 18.99 & 50.9 & [49.4, 52.5] \\
90\% & 11.61 & 17.04 & 26.72 & 36.08 & 703 & [663, 745] \\
\midrule
upper bound & 12.9 & 19.6 & 29.6 & 44.9 & $\infty$ & \\

% End output/tables/table4.tex

\bottomrule
\end{tabular}

\vspace{0.5ex}
\parbox{0.95\textwidth}{\footnotesize Notes: The finite-sample columns use the design of Table~\ref{tab:mc2ns} with $\tau=0.1$ and 1{,}000 replications. The limit uses independent Brownian slope and threshold paths, $\sigma=1$, $\delta_0=2$, and 100{,}000 simulated draws. Intervals measure Monte Carlo uncertainty in the limiting quantiles. The last row is the bound imposed by $\Gamma$. Section~\ref{sec:Snumerical} gives the approximation, grid checks and bootstrap details.}
\end{table}

\section{Numerical approximation for the nonstationary-slope limit}\label{sec:Snumerical}

This section records the numerical details behind Table~\ref{tab:mcQ} below. These simulations diagnose approximation to the limit rather than provide an additional asymptotic result.

Finite-sample columns use Table~\ref{tab:mc2}, Panel B, with $\tau=0.1$ and 1{,}000 replications. The limit $|\lambda_Q\mathcal T|$ uses $\lambda_Q=\sigma^2/(\delta_0^2Q_\ell)$, $\sigma=1$, $\delta_0=2$ and independent Brownian motions $G,J$, with $Q_\ell=\int_0^1G^2\,dL_J(\cdot,0)$. We approximate $Q_\ell$ by $N^{-3/2}h^{-1}\sum_{t\le N}x_t^2\1\{0<q_t\le h\}$, using independent Gaussian random walks, $N=400{,}000$ and $h=3$. We draw $|\mathcal T|$ by inverting its distribution function on $[0,200]$ with mesh $0.001$. There are 100{,}000 independent limit draws and 1{,}000 bootstrap resamples for percentile intervals. The simulated mean of $Q_\ell$ is 0.2682 (SE 0.0019), compared with $\E Q_\ell=2/(3\sqrt{2\pi})\approx0.2660$. Bootstrap SEs of the four finite-sample quantile rows are at most 0.09, 0.46, 1.52 and 1.07. Independent grid checks give quantiles $(0.99,5.91,48.5,619)$ at $N=10{,}000$, $(1.00,5.98,48.6,632)$ at $N=25{,}000$, $(0.99,6.01,50.6,690)$ at $N=100{,}000$ and $(1.02,6.04,50.6,700)$ at $N=400{,}000$. Lower quantiles are stable across grids, while coarse grids understate upper quantiles. The last row bounds the normalised error by $n^{0.3}\max_{\gamma\in\Gamma}|\gamma-\gamma_0|$.

\end{document}